\documentclass[10pt,onecolumn]{article}
\usepackage{amsmath}
\usepackage{amsfonts}
\usepackage{amssymb}
\usepackage{graphicx}
\usepackage{hyperref}
\graphicspath{{FIGS/}}
\usepackage{authblk}
\usepackage{color}
\usepackage{latexsym}
\usepackage{lscape}

\newcommand{\ignore}[1]{}

\renewcommand{\epsilon}{\varepsilon}
{\begingroup

 \begin{enumerate}}
 {\end{enumerate}\endgroup}

\newcommand{\ms}{M_s}

\newcommand{\mss}{M_{ss}}

\newcommand{\subfigimg}[3][,]{%
  \setbox1=\hbox{\includegraphics[#1]{#3}}% Store image in box
  \leavevmode\rlap{\usebox1}% Print image
  \rlap{\hspace*{-5pt}\raisebox{\dimexpr\ht1-2\baselineskip}{#2}}% Print label
  \phantom{\usebox1}% Insert appropriate spacing
}

\title{Dynamical mechanisms of flexible phase-locking in cortical theta oscillators}

\author[1]{Yangyang Wang}
\author[2]{Benjamin R. Pittman-Polletta}

\affil[1]{\small Department of Mathematics, Volen National Center for Complex Systems, Brandeis University, 415 South St, Waltham, MA 02453.
Email: yangyangwang@brandeis.edu}

\affil[2]{\small Department of Mathematics and Statistics, Mount Holyoke College, 50 College St, South Hadley, MA 01075.
Email: benpolletta@gmail.com}

\date{}

\begin{document}
\maketitle	
\abstract{
Oscillatory activity in auditory cortex is thought to play a central role in auditory and speech processing by synchronizing neural rhythms to external acoustic features of the speech stream. To support this function, cortical oscillators must flexibly phase-lock to inputs spanning a wide range of timescales, including rhythms substantially slower than their intrinsic frequency. Here we identify a general dynamical mechanism by which intrinsic inhibitory currents operating on multiple timescales enable such flexible phase-locking. Using tools from dynamical systems theory, including geometric singular perturbation theory and bifurcation analysis, we show that interactions between slow and superslow inhibitory processes generate prolonged post-input recovery delays through delayed Hopf phenomena, thereby substantially expanding the frequency range over which entrainment can occur. We demonstrate this mechanisms in a biophysically grounded cortical theta oscillator model for speech segmentation. Specifically, we show that both a theta-timescale (4-8 Hz) inhibitory current $I_m$ and a slower delta-timescale (1-4 Hz) inhibitory potassium current $I_{\rm K_{SS}}$ are crucial for entrainment flexibility. Their interaction creates a three-timescale structure that gives rise to pronounced delay phenomena associated with a delayed Hopf bifurcation (DHB). Interestingly, the superslow $I_{\rm K_{SS}}$ and the associated DHB play little role in the unforced oscillatory dynamics, but are recruited to support phase locking under external forcing. Moreover, the intermediate-timescale current $I_m$, rather than being redundant, further expands the phase-locking range by prolonging delayed recovery along the superslow manifold. Together, these results suggest that coordination among intrinsic inhibitory currents operating on multiple timescales may represent a key mechanism supporting flexible phase locking to rhythmic inputs in the brain.}

% \keywords{
% Phase Locking, Theta rhythm,  Geometric Singular Perturbation Theory, Speech processing, Multiple timescales}

%\begin{MSCcodes}
% 37N25, 34C23, 34C60, 34E13, 34E15, 92C20
%\end{MSCcodes}

\section{Introduction}

Macroscopic neural oscillations are ubiquitous in the brain and span a temporal scale that ranges from a few to a hundred hertz \cite{buzsaki2006rhythms}. Since their discovery by Hans Berger in 1929 \cite{berger1929elektroenkephalogramm}, neural rhythms have been extensively studied for their critical roles in neural communication, cognitive function and brain health \cite{buzsaki2004neuronal,buzsaki2012origin,giilmenbrain}. They are commonly classified into the following frequency bands: delta (1-4 Hz), theta (4-8 Hz), alpha (8-13 Hz), beta (13-30 Hz), and gamma (30-70 Hz). Across these bands, oscillatory activity is thought to support effective information processing by aligning intrinsic brain rhythms with external or internally generated temporal structure, simulating the inputs from the external environment or other oscillating neural groups. 

A prominent example of this principle arises in auditory cortex, where theta-band oscillations are believed to play a central role in auditory and speech processing \cite{lakatos2005oscillatory}. In particular, these oscillators can flexibly phase-lock to regular acoustic features of the speech stream across a wide range of frequencies, thereby facilitating syllabic segmentation and marking syllable boundaries \cite{Ghitza2009,ghitza2012role,Ghitza2014}. Importantly, speech signals exhibit hierarchical temporal organization, whose variable frequency can dip well below the intrinsic frequency of cortical theta oscillators (down to $\sim 1$ Hz) \cite{ghitza2011linking,ohala1975temporal,greenberg1999speaking,chandrasekaran2009natural,ding2017temporal,pittman21}. This raises a fundamental dynamical question: how can neural oscillators maintain robust intrinsic rhythmicity while flexibly entraining to much slower external rhythmic inputs?

We refer to this capability as \emph{flexible phase-locking}: the ability of an oscillator to flexibly synchronize its phase to rhythmic input across a broad frequency range, including frequencies substantially lower than its natural oscillation frequency. Many existing models of theta oscillators are paced by synaptic inhibition \cite{hyafil2015speech,hovsepyan2020combining}. Such inhibition-based rhythms exhibit rather limited entrainment bandwidth and especially fail to lock to input frequencies lower than their intrinsic frequency \cite{cannon2015leaky,sherfey2018flexible,pittman21}. 
Recent experimental and modeling studies suggest that intrinsic membrane currents operating on distinct timescales play an important role in organizing cortical theta rhythms and substantially extending their phase locking to slower external inputs \cite{Carracedo2013,pittman21}. However, the dynamical mechanisms by which such multiscale inhibitory structure could enable flexible entrainment have not been systematically characterized. 

In this work, we identify a general dynamical mechanism by which intrinsic inhibitory currents operating on nested timescales enable flexible phase locking to rhythmic inputs well below the intrinsic frequency. The key insight is that entrainment to slow rhythms require a sufficiently long post-input spiking recovery delay (the time between an input pulse and the next spontaneous spike) that prevents premature resumption of intrinsic spiking. If recovery occurs too soon, spontaneous spiking appear before the next input cycle arrives, breaking 1:1 phase-locking. Using tools from dynamical systems theory including geometric singular perturbation theory (GSPT) and bifurcation analysis \cite{Fenichel1979,Nan2015}, we show that interactions between slow and superslow inhibitory processes can dynamically regulate this recovery period by generating extended post-input delays through delayed bifurcation phenomena. 

We demonstrate this mechanism using a physiologically realistic biophysical computational model of cortical theta oscillator originally developed to study speech segmentation \cite{pittman21}. This model incorporates two intrinsic outward currents operating on distinct timescales that participate in pacing the rhythms: 
% theta-timescale subshtreshold oscillations (SAOs) that result from 
an intrinsic slow (theta-timescale) hyperpolarizing $m$-current ($I_m$) with a voltage-dependent time constant of activation of $\sim$10-45 ms \cite{gutfreund1995,Carracedo2013,adams2019}, and an intrinsic superslow (delta-timescale) hyperpolarizing potassium current ($I_{\rm K_{SS}}$) with calcium-dependent rise and decay times of $\sim$100 and $\sim$500 ms \cite{Carracedo2013}. 
Previous computational work has shown that oscillators combining these intrinsic inhibitory currents exhibit unusually broad entrainment ranges under strong rhythmic forcing, including synchronization to inputs substantially slower than the intrinsic frequency, whereas models relying on synaptic inhibitions do not \cite{Rostein2005,Hyafil2015,pittman21}. To uncover the dynamical origin of this flexibility, we provide a mechanistic explanation grounded in fast–slow dynamical structure and its underlying geometric organization. This geometric perspective allows us to identify key manifolds and delayed bifurcation structures that directly control post-input recovery timing and entrainment limits, and to characterize the distinct dynamical roles of the slow M-current and the superslow $\rm K_{SS}$-current.

Our analysis reveals four central findings. First, the superslow potassium current $I_{\rm K_{SS}}$ establishes the key timescale necessary for the emergence of a delayed Hopf bifurcation (DHB) \cite{Baer1989, Neishtadt1987, Neishtadt1988, Hayes2016}, which yields sufficiently long post-input delays to support entrainment to inputs well below the intrinsic theta frequency. Second, although the slow m-current $I_m$ is not required for the bifurcation itself, it plays a critical role in enabling the DHB mechanism to express its full characteristic delay. Third, removal of either current substantially shortens post-input delays and diminishes the oscillator's ability to entrain to slower inputs, demonstrating a synergistic interaction between inhibitory processes across timescales. 
Fourth, these delayed bifurcation effects are largely absent during unforced autonomous oscillations but are selectively recruited under strong rhythmic input, allowing the system to dynamically adapt its dynamics in response to external inputs that are significantly slower than its intrinsic frequency. Together, these results suggest that coordination among these intrinsic inhibitory currents operating on distinct timescales may provide a general mechanism for flexible entrainment in rhythm-generating neural circuits. Rather than relying on a single mechanism, biological oscillators can exploit multiscale inhibitory structure to decouple intrinsic rhythm generation from recovery dynamics, enabling flexible entrainment across a broad range of input frequencies. 

Phase-locking in neural oscillators to external inputs or to inputs from other oscillating populations has been extensively studied \cite{ermentrout1981n,ermentrout1996type,kopell2002mechanisms,achuthan2009phase,canavier2010pulse,perez2020phase,reyner2022phase}. The present work differs in that we consider a physiologically relevant strong forcing regime, in which input pulses are sufficiently strong to elicit spiking or bursting, rendering traditional phase-reduction approaches inapplicable \cite{schwemmer2012theory,park2017utility,wang2021shape,ermentrout2010mathematical}. We focus in particular on addressing how oscillators entrain to rhythmic inputs that are much slower than the intrinsic frequency of the target oscillator, for which general mechanistic understanding remains limited. Moreover, relatively few studies have examined oscillators that exhibit intrinsic outward currents operating on two distinct timescales and analyzed how such multiscale structure shapes phase-locking behaviors. Closest to our work is \cite{zhou2018m}, which employs fast-slow decomposition and bifurcation analysis to characterize the dynamical mechanisms underlying the role of the M-current in phase-locking. Here, however, we show that M-current alone is not sufficient; rather, it is the multiscale interaction between the M-current and the superslow $\rm K_{SS}$-current that is essential for enabling flexible entrainment to slow incoming rhythmic inputs.

The remainder of the paper is organized as follows. In Section \ref{sec:model}, we introduce the full theta oscillator model, perform geometric singular perturbation theory analysis, and quantify entrainment behaviors under strong rhythmic forcing. Sections \ref{sec:m} and \ref{sec:kca} analyze reduced models lacking either the superslow or slow inhibitory components in order to isolate their individual contributions.  In Section \ref{sec:both}, we characterize the full multiscale mechanism and demonstrate how the delayed Hopf mechanism emerges from the interaction of intrinsic currents to support a remarkable phase-locking to slower inputs. The paper ends with a discussion in Section \ref{sec:discussion}.

% The remainder of the paper is organized as follows. In Section \ref{sec:model}, we describe the full theta oscillator model, present a dimensionless analysis of the full system to reveal the presence of different timescales, define subsystems for the geometric singular perturbation theory (GSPT) analysis, introduce the two reduced models with either $I_{\rm K_{SS}}$ or $I_m$ being removed, and examine their phase-locking flexibility under strong inputs relative to the full system. In Sections \ref{sec:m} and \ref{sec:kca}, we analyze the $\rm K_{SS}^{-}$- and $\rm M^{-}$- systems, respectively, to explain why each exhibits inflexible entrainment to slow inputs.  In Section \ref{sec:both}, we investigate how the interaction $I_m$ and $I_{\rm K_{SS}}$ yields the greatest phase-locking flexibility in the intact MS model. The paper ends with a discussion in Section \ref{sec:discussion}. 

 % Analyze MIS as well? Why the flexibility of phase locking improves as the strength of inputs get stronger in MIS? What's the role of the inhibitory current that has similar timescale as the m-current. Instead of considering MIS, maybe consider self-inhibition to make the system simpler. 

\section{Cortical theta oscillator model and phase-locking to strong inputs}\label{sec:model}

The dynamics of the full theta oscillator model \cite{pittman21} is given by the following equations: 

\begin{equation}\label{eq:full}
\begin{array}{rcl}
     C\frac{dV}{dt}&=&  I_{\rm app}- I_{\rm Na}-I_{\rm K_{DR}}-I_{\rm leak}-I_m -I_{\rm NaP} \\
     && -I_{\rm Ca}-I_{\rm K_{SS}}+I_{\rm PP}(t)\\
     \frac{dn}{dt}&=& (n_{\infty}(V)-n)/\tau_n(V),\\
     \frac{dm_{\rm NaP}}{dt}&=& (m_{\infty}(V)-m_{\rm NaP})/\tau_{m},\\
     \frac{ds}{dt}&=& (1-s)\alpha_s-s\beta_s,\\
     \frac{dm_{\rm K_{DR}}}{dt}&=& \tau_{\rm fast}((1-m_{\rm K_{DR}})\alpha_{m}-{m_{\rm K_{DR}}}\beta_{m}),\\
     \frac{dh}{dt}&=& \tau_{\rm fast}((1-h)\alpha_h-h\beta_h),\\
     \frac{d\rm Ca_i}{dt}&=& -F_{\rm Ca} I_{\rm Ca}(V)-\rm Ca_i/\tau_{\rm Ca}\\
     \frac{dq}{dt}&=& (1-q)\alpha_q( {\rm Ca_i})-q\beta_q
\end{array}    
\end{equation}
with 
\begin{equation}\label{eq:currents}
    \begin{array}{rcl}
       I_{\rm Na}  &=& g_{\rm Na}m_{\rm Na}(V)^3h(V-E_{\rm Na})  \\
       m_{\rm Na}(V)&=&\alpha_{m_{\rm Na}}(V)/(\alpha_{m_{\rm Na}}(V)+\beta_{m_{\rm Na}}(V))\\
       I_{\rm K_{DR}}  &=& g_{\rm K_{DR}}m^4_{\rm K_{DR}}(V-E_K)\\
       I_{\rm leak}&=& g_{\rm leak} (V-E_{\rm leak})\\
       I_m &=& g_m n(V-E_K)\\
       I_{\rm NaP} &=& g_{\rm NaP} m_{\rm NaP}(V-E_{\rm NaP})\\
       I_{\rm Ca} &=& g_{\rm Ca}s^2(V-E_{\rm Ca})\\
       I_{\rm K_{SS}} &=& g_{\rm K_{SS}}q(V-E_K)\\
    \end{array}
\end{equation}
The gating conductances, reversal potentials, and parameters related to the applied currents are given in Table \ref{tab:para}, while the steady-state values, time constants, and rate functions for the gating variables and calcium are provided in Table \ref{tab:activation} in Appendix \ref{app:nondim}. 

\begin{table}[!htp]
\caption{Model Parameters}
\centering
\begin{tabular}{c c c c c c c c}
\hline
$g_{\rm Na}$  & $125\,\rm nS$ & $E_{\rm Na}$ & 40\,\rm mV  &$g_{\rm K_{DR}}$& $54\,\rm nS$ & $E_{\rm K}$ & $-80\,\rm mV$\\
$g_{\rm leak}$ & $0.27\,\rm nS$ & $E_{\rm leak}$ & $-65\,\rm mV$& 
$g_{m}$ &$1.4472\,\rm nS$& $g_{\rm K_{SS}}$ & $0.1512\,\rm nS$\\
$g_{\rm NaP}$& $0.4307\,\rm nS$ & $E_{\rm NaP}$ & $50\,\rm mV$&
$g_{\rm Ca}$ & $0.54\,\rm nS$ & $E_{\rm Ca}$ & $120\,\rm mV$\\
$I_{\rm app}$ & $9.8\,\rm pA$ & $I_T$ & $2000\,\rm pA$ & $d$ & $1/4$ & $C$ & $2.7\,\rm pF$\\ 
\hline
\end{tabular}
\label{tab:para}
\end{table}

The external periodic pulse inputs are given by
\begin{equation}\label{eq:Ipp}
I_{\rm PP}(t) = g_{\rm PP}\sum_{i=1}^{m} \chi_{\{t-t_i^*<=w\}}(t) 
\end{equation}
where $\chi_S(t)$ is the function that is $1$ on set $S$ and $0$ otherwise, $t_i^*=t_{\rm on}+T_s i$ for $i=1,2,...$ is the set of times at which input pulses arrive, $t_{\rm on}$ is the onset time of the first input, $T_s=1000/f$ is the input period, $f$ is the input frequency, $w=d T_s$ is is the pulse width given the duty cycle $d\in (0,1)$. $g_{\rm PP}=\frac{I_T}{m\cdot w}$ where $I_T$ is the total (integrated) input strength and $m$ is the total number of periodic inputs. In our simulations, we fix $d=1/4$. A couple of values of $I_T$ will be considered, while other parameters will be fixed at the values shown in Table \ref{tab:para}.  
Note that in \cite{pittman21} the periodic pulse input is given by
\[
I_{\rm PP}(t) = g_{\rm PP} \sum_i\chi_{\{|t_i^*|<=w(s-1)/2s\}}(t)*\exp (-(st/w)^2)
\]
where $*$ is the convolution operator and $s=25$ 
% determines how square the pulse is with higher $s$ being more square. In that work, $s=25$
so the input pulse is nearly square. In this paper, to simplify analysis we assume periodic pulses are perfectly square and hence can be rewritten as in equation \eqref{eq:Ipp}. Previous work has shown that, under strong forcing, the phase-locking flexibility characterized using periodic inputs extends to entrainment of quasi-rhythmic and speech-derived inputs \cite{pittman21}. Accordingly, we focus here on periodic forcing, which enables a mechanistic analysis using dynamical systems theory while remaining relevant to more naturalistic stimuli. 
% , as the underlying mechanism can yield important insights into the model's flexible phase locking to quasi-rhythmic and speech inputs. 

\subsection{Geometric Singular Perturbation Theory}\label{sec:nondim}

% \RED{Move to Appendix?}
% To analyze how multiscale inhibitory interactions shape entrainment dynamics, we perform dimensional analysis to explicitly identify the relevant timescales. 

The theta oscillator model \eqref{eq:full} involves at least three distinct timescales: fast voltage spiking, slow $I_m$ and superslow $I_{\rm K_{SS}}$. To analyze its dynamics and phase-locking properties, we employ the Geometric Singular Perturbation Theory (GSPT) for three-timescale problems \cite{Fenichel1979,Nan2015}. The extended GSPT has been successfully applied to study mixed mode oscillations (MMOs) \cite{Letson2017,PW2024} and complex bursting dynamics \cite{WR2016,WR2017,WR2020} in three-timescale systems.

As the first step of the GSPT approach, we perform a dimensional analysis of \eqref{eq:full} to reveal the important timescales. To do so, we introduce a dimensionless time variable $t=Q_t \tau$ and nondimensionalize the full system (see Appendix \ref{app:nondim} for details). This procedure suggests that $y=(V,m_{\rm Nap}, s, m_{\rm K_{DR}}, h, Ca_i)^{\mathrm{T}}$ evolves on fast timescales, $n$ evolves on a slow timescale, whereas $q$ evolves on a superslow timescale, leading to the following equivalent three-timescale system:
\begin{equation}\label{eq:nondim2}
\begin{array}{rcl}
     \varepsilon\frac{dy}{d\tau}&=&F(y,n,q)\\
     \frac{dn}{d\tau}&=&G(V,n),\\
     \frac{dq}{d\tau}&=& \delta Q(\mathrm{Ca_i},q).
\end{array}    
\end{equation}
Here, $F=0$ corresponds to the vanishing of all right-hand-side functions associated with the fast variables $\{V,m_{\rm Nap}, s, m_{\rm K_{DR}}, h, Ca_i\}$ in \eqref{eq:full} vanish, while $G$ and $Q$ are rescaled versions of the right-hand-side functions of the $dn/dt$ and $dq/dt$ equations. 
% The Jacobian of the fast-slow subsystem:
% \begin{equation}\label{eq:J-FS}
% \mathbf{J}=
% \begin{pmatrix}
%  f_{1V} &  f_{1m_{\mathrm {NaP}}} &  f_{1s} &  f_{1m_{\mathrm{K_{DR}}}}& f_{1h} & 0 & f_{1n} \\
%  f_{2V} &  f_{2m_{\mathrm {NaP}}} & 0 & 0 & 0 & 0 &0 \\
% f_{3V} & 0 &  f_{3s} & 0 & 0 & 0 & 0\\
% f_{4V} & 0 & 0 & f_{4m_{\mathrm {K_{DR}}}} & 0 & 0  & 0\\
%  f_{5V} & 0 & 0 & 0 & f_{5h} & 0 & 0\\
% f_{6V} & 0 & k_6  f_{6s} & 0 & 0 & f_{6 \mathrm{\mathrm{Ca_i}}} & 0 \\
% G_V & 0& 0& 0& 0& 0& G_n\\
% \end{pmatrix}.
% \end{equation}
% %%%%%%

Applying GSPT \cite{Fenichel1979,Nan2015} to this three-timescale system \eqref{eq:nondim2} results in the six-dimensional (6D) \emph{fast layer problem} (FL) describing the dynamics of all the fast variables $y$ for fixed values of the other variables, the 1D \textit{slow reduced layer problem} (SRL) that describes the dynamics of the slow variable $n$ for fixed values of $q$ and all variables restricted to the equilibrium points of the FL problem (defined below in \eqref{eq:ms}), and the 1D \textit{superslow reduced problem} (SSR) that describes the dynamics of $q$ with all variables restricted to the superslow manifold (defined below in \eqref{eq:mss}). In addition, we also define the 7D \emph{slow layer problem} (SL) describing the dynamics of the fast and slow variables, as follows: 
\begin{equation}\label{eq:slow-layer}
\begin{array}{rcl}
     \varepsilon\frac{dy}{d\tau}&=&F(y,n,q)\\
     \frac{dn}{d\tau}&=&G(V,n),
\end{array}    
\end{equation}
where the superslow variable $q$ is a parameter. Based on these singular limit subsystems, we define the \emph{critical manifold} to be the set of equilibrium points of the fast layer problem, i.e.,
   \begin{equation}\label{eq:ms}
\begin{array}{rcl}
 M_s &:=& \{(y, n, q): \, F(y,n,q) = 0 \} \\
 &=&  \{(y, n, q): \, f_i=0 \text{ for } i=1,\cdots,6 \}  
 \end{array}
\end{equation} 
The set of equilibrium points of the slow layer problem is defined as the \emph{superslow manifold}, i.e.,
\begin{equation}\label{eq:mss}
\begin{array}{rcl}
 M_{ss} &:=& \{(y, n, q): \, F(y,n,q) = G(V,n)=0 \} \subset M_s
 \end{array}
\end{equation}
A crucial step of GSPT in determining the nature of the solution is to establish the type of points that occur at the transitions between fast, slow and superslow segments of the singular orbit \cite{Nan2015}. In our model, the transitions between fast and slow segments happen along the fold curve of $M_s$. We have identified two types of transition points: jump points and folded singulariteis \cite{Desroches2012}. Singular orbits containing only jump points perturb to relaxation oscillations, whereas singular orbits containing folded singularites may perturb to more complicated oscillations such as orbits with subthreshold oscillations (i.e., MMOs). On the other hand, the transitions between fast/slow and superslow segments of the singular orbit occur at Hopf bifurcations (HB) of the fast layer problem along $M_{ss}$, another mechanism that can lead to MMOs \cite{Letson2017,PW2024}. This HB bifurcation is also known as delayed Hopf bifurcation (DHB) \cite{Neishtadt1987,Neishtadt1988,Baer1989}. In this paper, for the sake of brevity, we do not provide a detailed description of singular limit systems or the analysis needed to identify singularities on $M_s$ and $M_{ss}$, such as folded singularities and DHB points. For a comprehensive GSPT analysis of MMOs in three timescale systems, we refer readers to \cite{PW2024,Letson2017}.

\subsection{Phase locking under strong forcing}\label{sec:PLForcing}
	
% To explore the roles of intrinsic currents in determining the flexibility of phase-locking in auditory cortical theta rhythms, 

\begin{figure}[!t]
\begin{center}
\begin{tabular}{@{}p{0.3\linewidth}@{\quad}p{0.3\linewidth}@{\quad}p{0.3\linewidth}@{}}
\subfigimg[width=\linewidth]{\bf{(A)}}{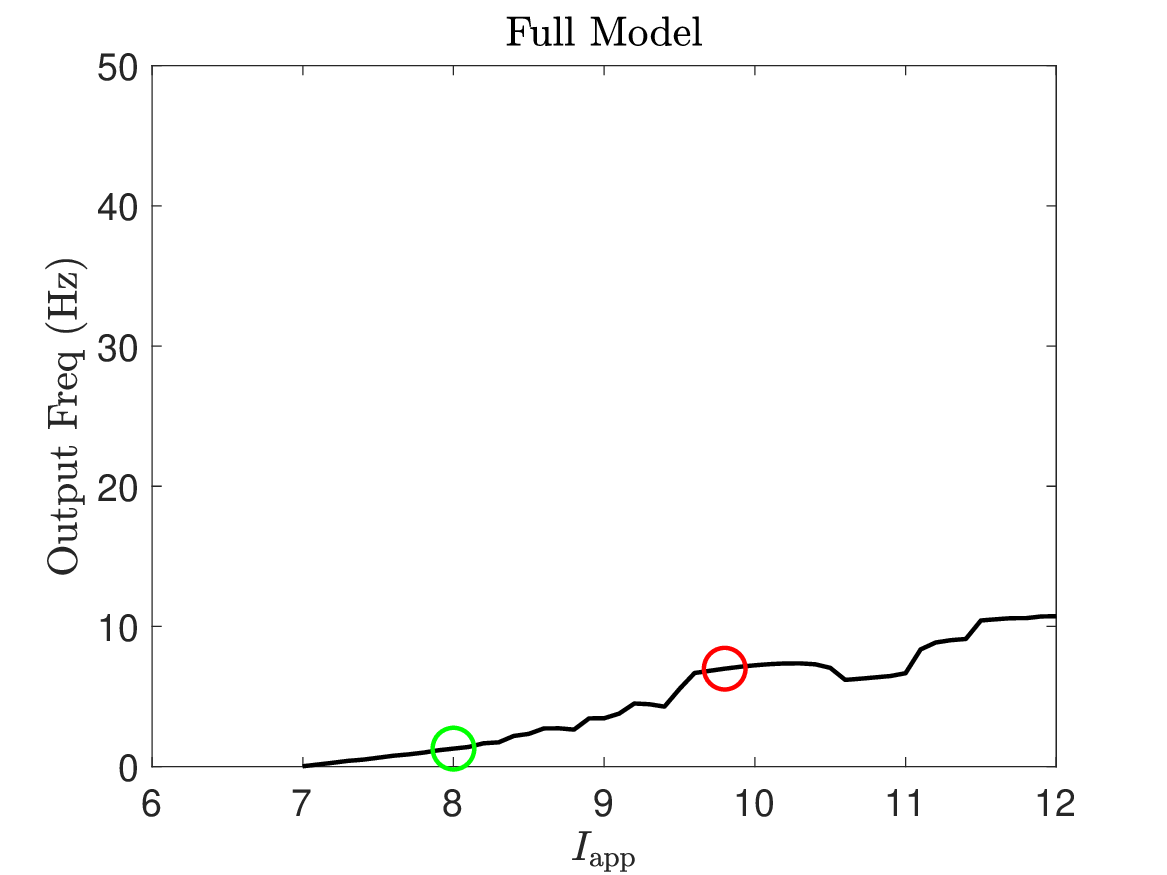}&
\subfigimg[width=\linewidth]{\bf(B)}{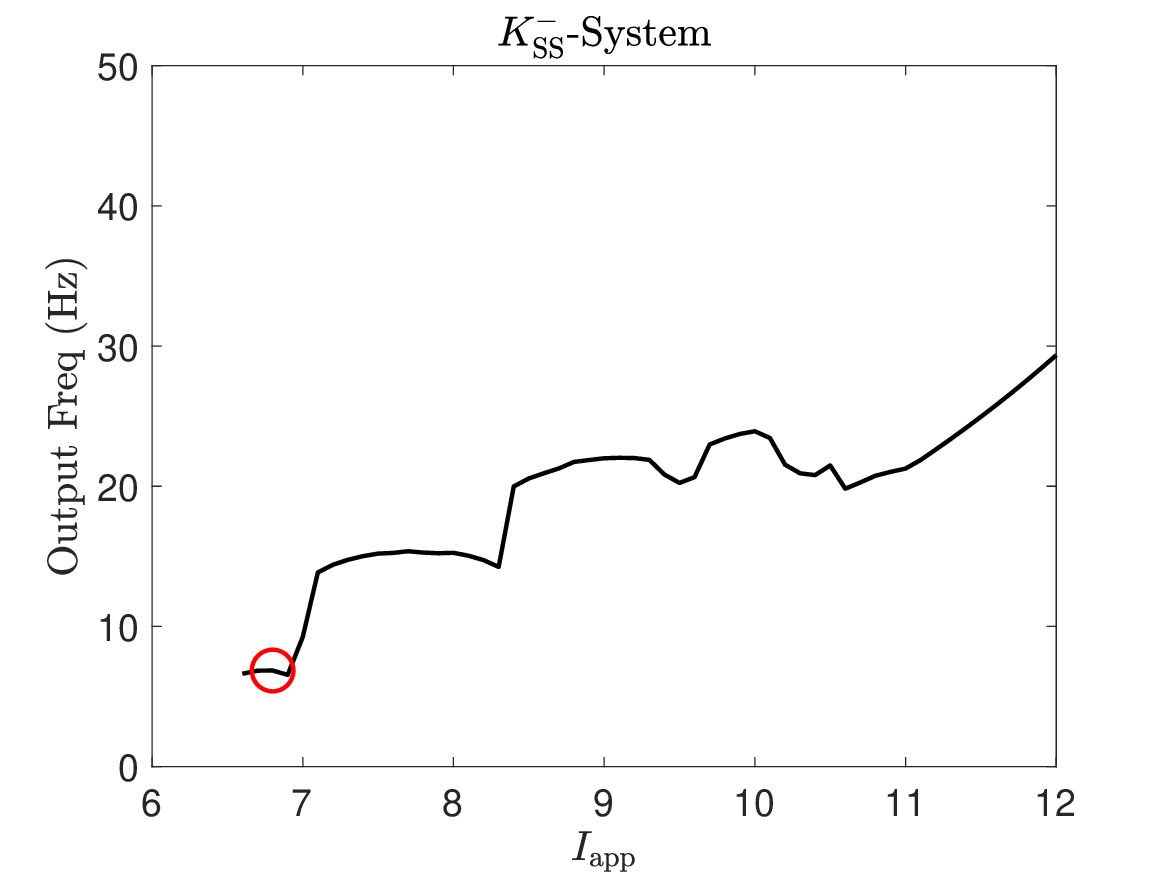}&
\subfigimg[width=\linewidth]{\bf(C)}{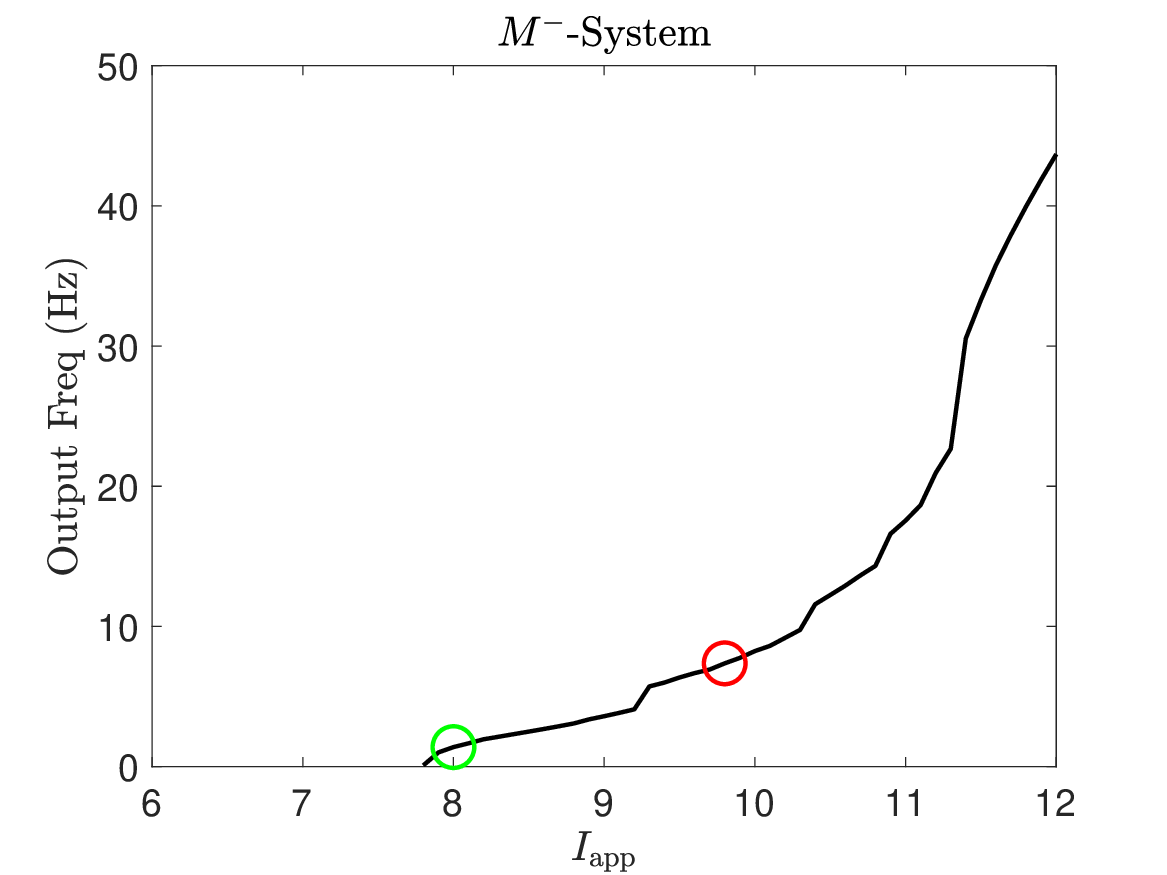}
\end{tabular}
\end{center}
\caption{\label{fig:model-diagram} FI curves show the function of output frequency as $I_{\rm app}$ varies for the (A) full model, (B) $K_{SS}^-$-system and (C) $M^-$-system. The red (resp., green) circle indicates the point on the FI curve at which $I_{\rm app}$ was fixed, to give a 7 Hz (resp., 1.4 Hz) firing rate. }
\end{figure}

To examine the roles of inhibitory currents on superslow delta- and slow theta-timescales, we construct reduced models in which either the $I_{\rm K_{SS}}$ or $I_m$ current is moved, and compare their phase locking behaviors with the intact model. For $I_{\rm PP}(t)=0$ (i.e., in the absence of periodic forcing), the frequency-current (FI) curves as a function of the applied current $I_{\rm app}$ for the full and reduced models are shown in Figure \ref{fig:model-diagram}. In the full model, $I_{\rm app}=9.8$ yields intrinsic oscillations at a frequency of $7$ Hz (Figure \ref{fig:model-diagram}A, red circle). 

%%%%%%%%%%%%%%%%%%%%%%%
%%%%%%%%%%%%%%%%%%%%%%%

We remove the superslow $\rm K_{SS}$ current from the full model by setting $g_{\rm K_{SS}} = 0$, creating a model we refer to as the \textit{$\rm K_{SS}^{-}$ system}. This is equivalent to freezing the $q$ dynamic variable at $q=0$ in \eqref{eq:full} (or equivalently \eqref{eq:nondim2}). Thus, the $\rm K_{SS}^{-}$-system can be written as
\begin{equation}\label{eq:nondimM}
\begin{array}{rcl}
     \varepsilon\frac{dy}{d\tau}&=&F(y,n,0)\\
     \frac{dn}{d\tau}&=&G(V,n),
\end{array}    
\end{equation}
which consists of 6 fast and 1 slow (6F, 1S) variables. 
In order for the $\rm K_{SS}^{-}$-system to exhibit the same natural frequency of $7\,\rm Hz$ as in the full model, we decrease its tonic input from $I_{\rm app}=9.8$ to $I_{\rm app}=6.8$ to compensate the increased excitability due to the loss of the inhibitory $\rm K_{\rm SS}$-current (Figure \ref{fig:model-diagram}B, red circle). 

\begin{figure}[!t]
\begin{center}
\begin{tabular}{@{}p{0.45\linewidth}@{\quad}p{0.45\linewidth}@{}}
\subfigimg[width=\linewidth]{\bf{\small{(A)}}}{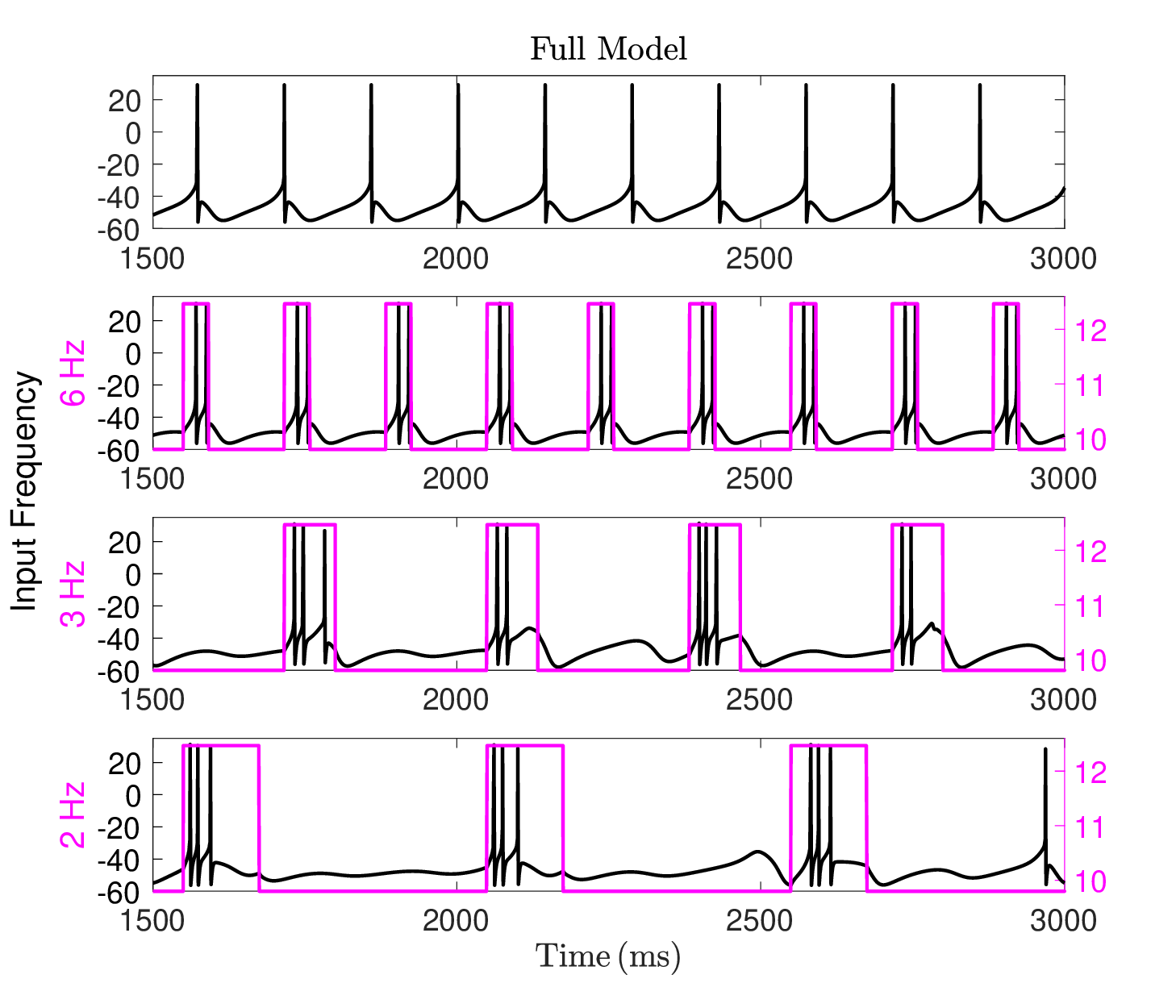}&
\subfigimg[width=\linewidth]{\bf{\small{(B)}}}{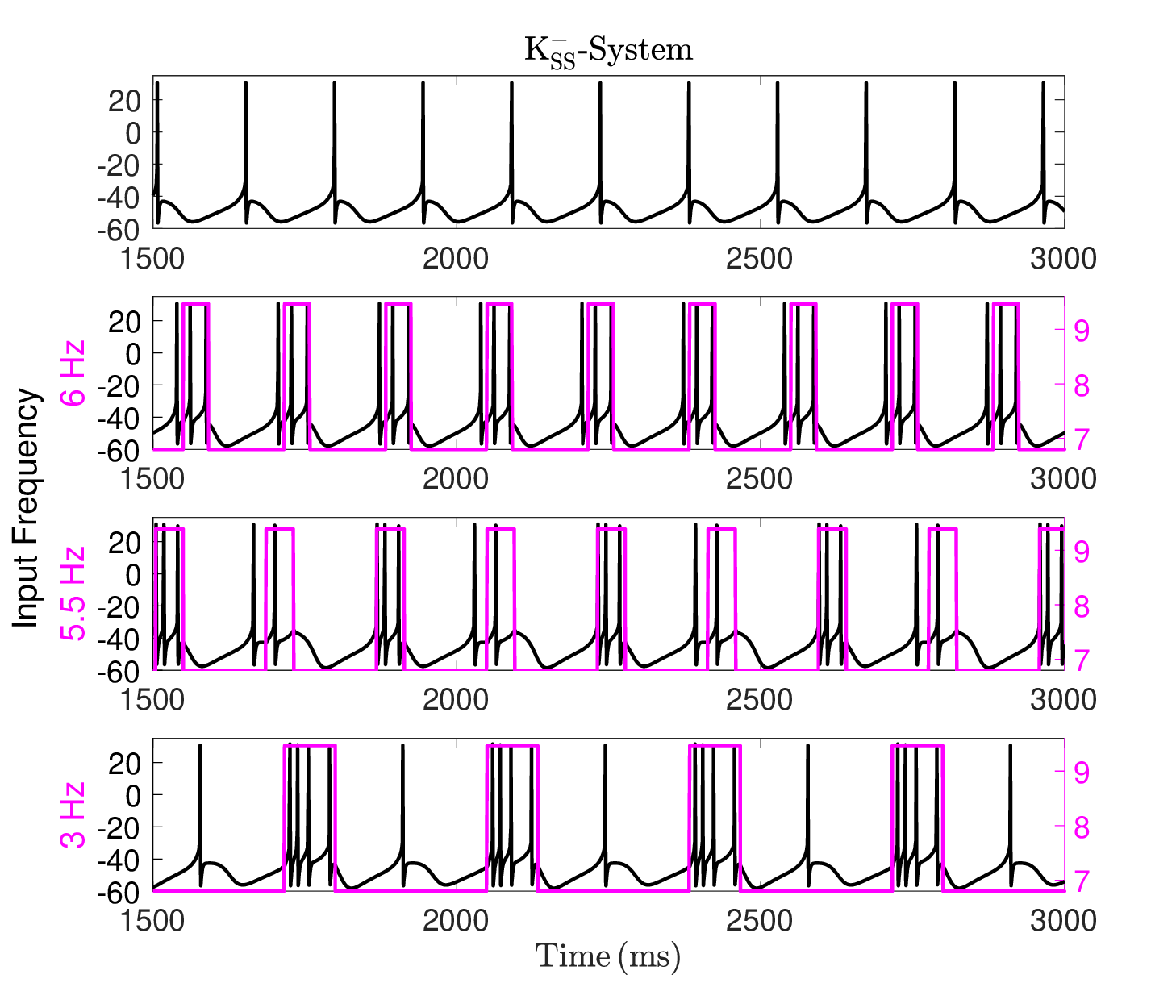}
\end{tabular}
\end{center}
\caption{\label{fig:ST-response-M} Time traces of voltage (black) for \textbf{(A)} the full model \eqref{eq:full} with $I_{\rm app}=9.8$ and \textbf{(B)} the $\rm K_{SS}^{-}$-system with $I_{\rm app}=6.8, g_{\rm K_{SS}}=0$. Other unspecified parameters in each model are given in Table \ref{tab:para}. Both models spike at an intrinsic frequency of about $7\,\rm Hz$. 
The lower three panels show their responses to 3-second slower periodic input pulses (total input current $I_T= 2000$) at different input frequencies $f$ that are lower than the intrinsic frequency. Traces of inputs whose frequency is indicated on the left are shown in magenta and the perturbed voltage traces are shown in black. The full model is able to phase lock to inputs with frequency as low as $2$ Hz, whereas the $\rm K_{SS}^{-}$-system fails at phase-locking to inputs with frequency $5.5$ Hz.}
\end{figure}

The voltage traces of the full model \eqref{eq:full} (resp., the $\rm K_{SS}^{-}$-system) without and with rhythmic inputs $I_{\rm PP}(t)$ are shown in Figure \ref{fig:ST-response-M}A (resp., Figure \ref{fig:ST-response-M}B). From Figure \ref{fig:ST-response-M}A, we can see that the full model is capable of phase-locking to rhythmic inputs at frequencies significantly lower than the intrinsic frequency (e.g., 2 Hz). 
% Here the flexible phase-locking arises from the build up of $ \rm K_{SS}$-current over each input cycle, delaying spiking until the next input arrives. 
In contrast, the $\rm K_{SS}^{-}$-system is much less flexible at phase-locking to inputs slower than its intrinsic frequency (see Figure \ref{fig:ST-response-M}B). Specifically, the top third row indicates the $\rm K_{SS}^{-}$-system fails to phase-lock to 5.5 Hz inputs, with spontaneous spikes occurring outside the input pulses. This comparison demonstrates a crucial role of $\rm K_{SS}$-current in supporting the flexible phase-locking in the full model.

%%%%%%%%%%%%%%%%%%%%%%%
%%%%%%%%%%%%%%%%%%%%%%%
	
To investigate the role of theta-timescale $m$-current in flexible phase-locking, we construct $\rm M^{-}$-system by setting $g_{m}=0$ (Figure \ref{fig:model-diagram}C).
% $g_{m}=g_{\rm NaP}=0$
This is equivalent to replacing the $n$ dynamics in \eqref{eq:nondim2} by setting $n=0$. 
Thus, the $\rm M^{-}$-system consisting of 6 fast and 1 superslow (6F, 1SS) variables can be written as  
\begin{equation}\label{eq:nondimKSS}
\begin{array}{rcl}
     \varepsilon\frac{dy}{d\tau}&=&F(y,0,q)\\
     \frac{dq}{d\tau}&=&\delta Q(\mathrm{Ca_i},q).
\end{array}    
\end{equation}
To maintain spiking activity in the $\rm M^{-}$-system system, we decrease $g_{\rm leak}$ to $0.16$. While $I_{\rm app}=9.8$ yields the same natural frequency of 7 Hz as in the full system (Figure \ref{fig:model-diagram}C, red circle), the $\rm M^{-}$-system is highly sensitive to further increase in $I_{\rm app}$, as indicated by the relatively steep slope of its FI curve near the red circle. As a result, although the two systems share the same natural frequency, the induced spiking frequency during input forcing is much higher in the M$^-$-system than in the intact model, making a direct comparison of their phase-locking properties problematic. To address this,  
we instead set $I_{\rm app}=8$ in both the $\rm M^{-}$-system and the intact model (Figure \ref{fig:model-diagram}A and C, green circles), so that they not only share the same intrinsic frequency of $1.4\,\rm Hz$ but also exhibit comparable sensitivity to increase in $I_{\rm app}$, as reflected in their FI curves. 

\begin{figure}[!t]
\begin{center}
\begin{tabular}{@{}p{0.45\linewidth}@{\quad}p{0.45\linewidth}@{}}
\subfigimg[width=\linewidth]{\bf{\small{(A)}}}{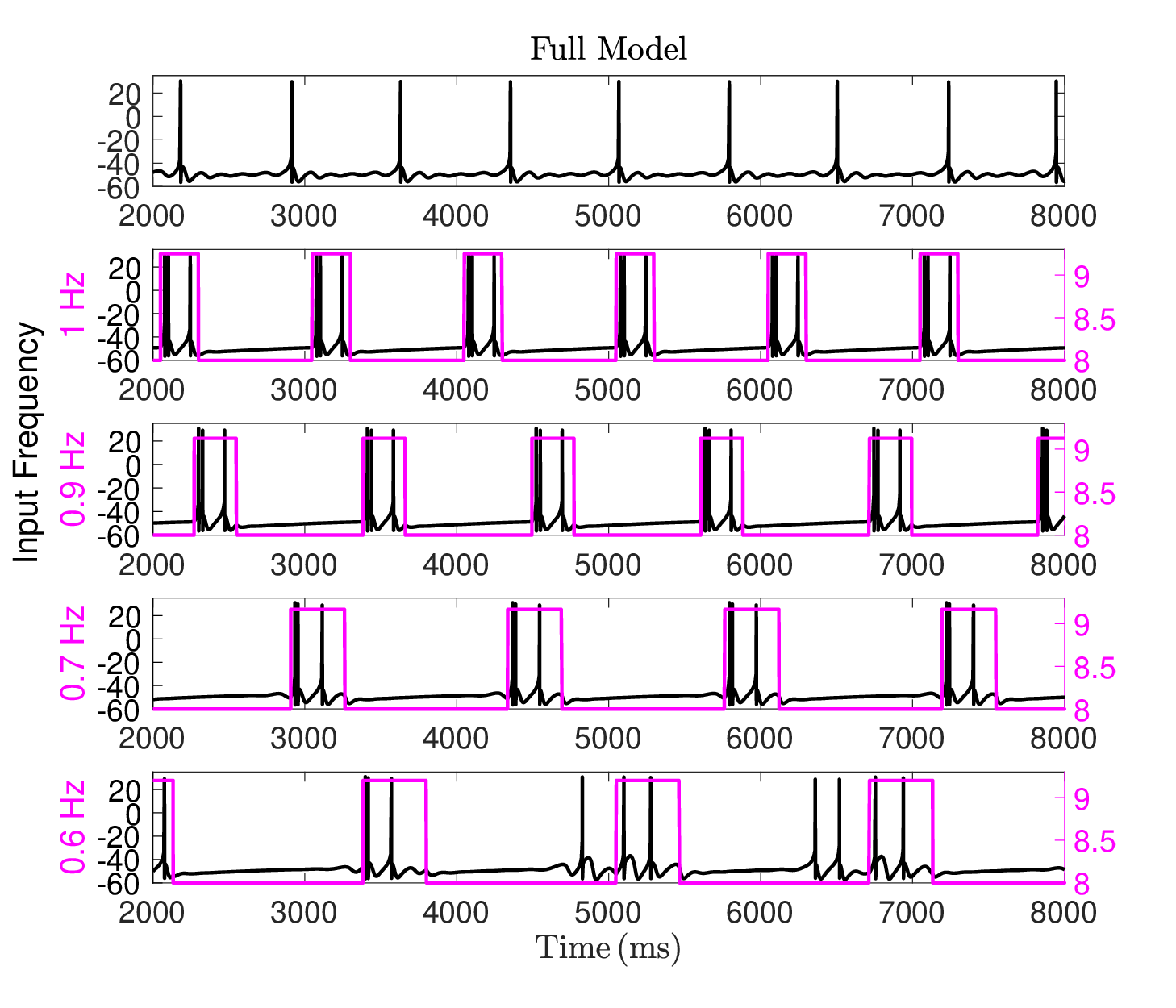}&
\subfigimg[width=\linewidth]{\bf{\small{(B)}}}{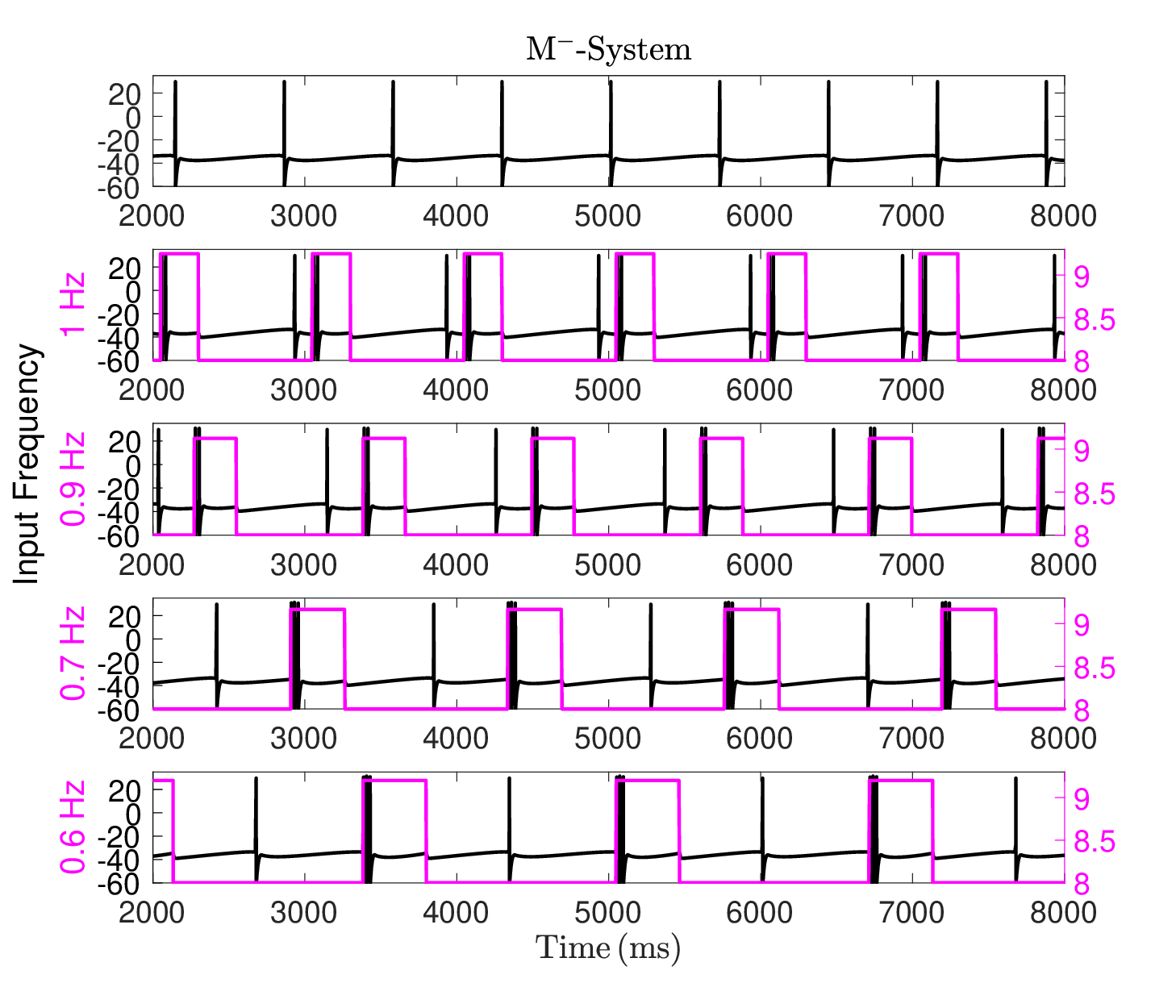}
\end{tabular}
\end{center}
\caption{\label{fig:ST-response-KCa} Time traces of voltage (black) for (A) the full model \eqref{eq:full} with $I_{\rm app}=8$ and (B) the $\rm M^{-}$-system with $I_{\rm app}=8, g_{m}=0, g_{\rm leak}=1.6$. Other unspecified parameters for each model are given in Table \ref{tab:para}. Both models spike at an intrinsic frequency of about $1.4\,\rm Hz$. The lower three panels show their responses to 8-second periodic input pulses (duty cycle =1/4 cycle, total input current = 2500) at different frequencies that are lower than the intrinsic frequency. Traces of inputs whose frequencies are indicated on the left are shown in magenta and the perturbed voltage traces are shown in black.}
\end{figure}

With this matching in place, we compare the phase-locking properties of the two models to assess the role of $I_m$ in enabling flexible phase locking to slower inputs. A comparison of Figures \ref{fig:ST-response-KCa}A and \ref{fig:ST-response-KCa}B shows that the full model exhibits a much greater phase-locking flexibility than the $\rm M^{-}$-system, highlighting the important role of $I_m$ in potentiating flexible phase-locking. In particular, the full model is able to phase-lock to periodic inputs at frequencies as low as half of its intrinsic frequency (i.e., $0.7$ Hz), whereas removal of $I_m$ leads to a substantial loss of phase-locking capability in the $\rm M^{-}$-system. 

%%%%%%%%%%%%%%%%%%%%%%%
%%%%%%%%%%%%%%%%%%%%%%%

\subsection{Post-input spiking delay}

To investigate the phase-locking characteristics of these systems, we examine the \textit{post-input spiking delay} (denoted as $D$ with a unit in $\rm s$) \cite{pittman21}. This delay represents the time until the next spontaneous spike following a single input pulse. Such delays are observed in all three systems since each input pulse initiates a burst of spiking, which activates the hyperpolarizing currents $I_m$ and/or $I_{\rm K_{SS}}$. 

If the delay in spiking is long enough, phase-locking can be achieved since the next input pulse which we assume is strong will always cause spiking. As a result, the phase-locking to input pulses can be determined by the delay D in that the lower input frequency limit ($f^*$) of phase-locking satisfies
\begin{equation}\label{eq:f-limit}
f^*\geq 1/D.
\end{equation}
It follows that if the spiking delay is small such that $D\leq 1/f_0$ where $f_0$ is the intrinsic frequency of an oscillator, then we have $f^*\geq 1/D\geq f_0$. This means this oscillator cannot phase-lock to any inputs slower than its intrinsic frequency. In other words, phase-locking of an oscillator to inputs slower than its intrinsic frequency requires the delay to be larger than its intrinsic period, that is,
\begin{equation}\label{eq:cond-pl-to-slower}
D>1/f_0.
\end{equation}
Furthermore, the larger the delay $D$, the more flexible the phase-locking to slower inputs.

In the rest of this paper, we use methods of fast-slow decomposition and bifurcation analysis to explain the inflexible phase-locking of the $\rm K_{SS}^{-}$-system to slower inputs by examining its post-input spiking delay $D$ in \S\ref{sec:m}. We carry out a similar analysis for the $\rm M^{-}$-system in \S\ref{sec:kca}. Then in \S\ref{sec:both}, we explain why the presence of both these two intrinsic currents produces a substantially longer delay $D$, which in turn leads to a markedly more flexible phase locking in the full model.

\section{$\rm K_{SS}^{-}$-system fails to phase-lock to slower inputs}\label{sec:m}

In this section, we explain why the (6F,1S) $\rm K_{SS}^{-}$-system \eqref{eq:nondimM} struggles to phase-lock to input rhythms slower than its intrinsic frequency. To accomplish the goal, we first employ timescale decomposition and bifurcation analysis to understand the intrinsic spiking dynamics of this system. We then assess its phase-locking ability by analyzing how input pulses influence its bifurcation structures and the subsequent post-input spiking delay $D$. Our findings reveal that the limited phase-locking ability of the $\rm K_{SS}^{-}$-system to slower frequencies arises from two key factors: (1) insufficient timescale separation within the $\rm K_{SS}^{-}$-system, and (2) the lack of a delayed transition into spontaneous spiking. Consequently, although input-triggered spikes generate a larger hyperpolarizing M-current than autonomous spikes in the $K_{\rm SS}^-$-system, the accumulated current decays relatively quickly back toward baseline, allowing spiking to resume rapidly and thereby preventing the model from reliably following rhythms slower than its natural frequency.

\begin{figure}[!htp]
\begin{center}
\includegraphics[width=0.5\linewidth]{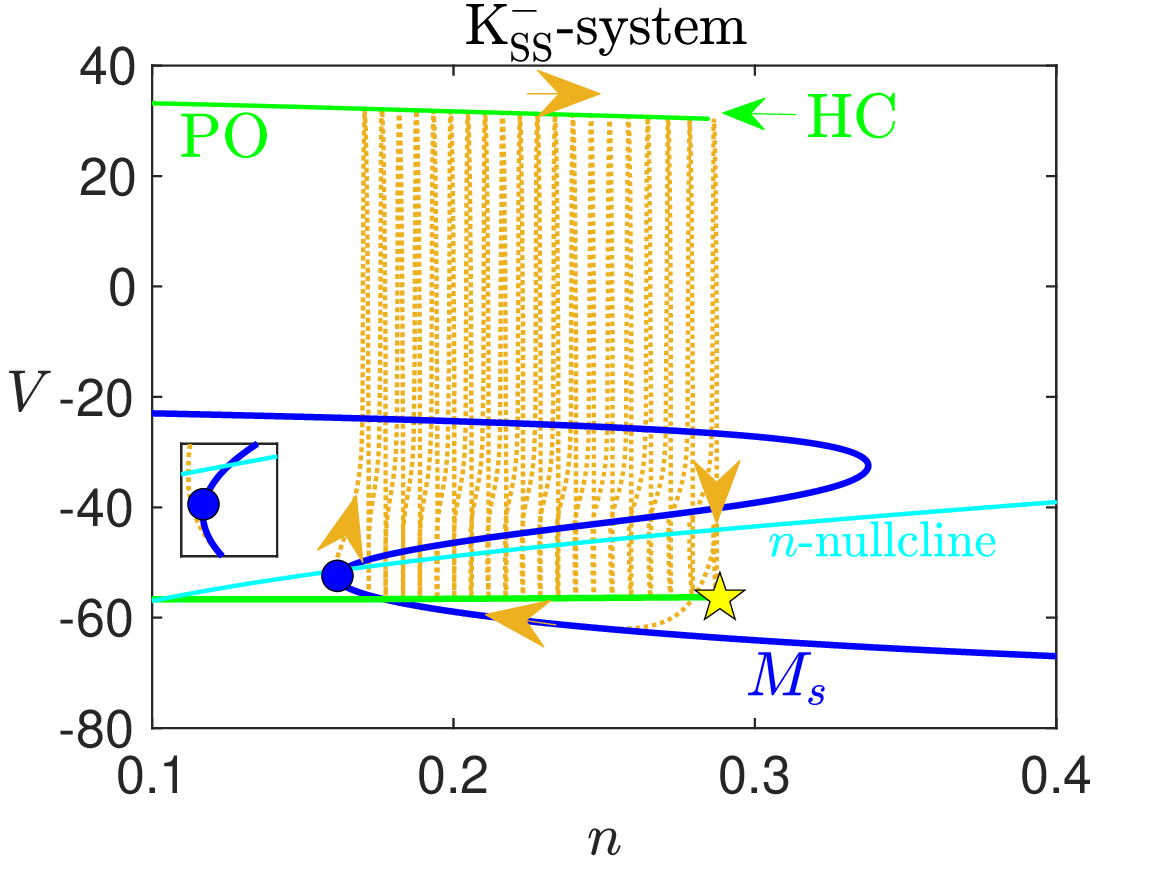}
\caption{Projection of burst trajectory (dark yellow) of the $\rm K_{SS}^{-}$-system \eqref{eq:nondimM} onto the bifurcation diagram when the timescale separation between the fast variable $y=(V,m_{\rm NaP},s, m_{\rm K_{DR}},h, \rm Ca_i$) and the slow variable $n$ is exaggerated by 10 folds. The curve $M_s$ (blue) denotes the equilibria of the fast $\rm K_{SS}^{-1}$-subsystem, and the green curve shows the maximum and minimum $V$ along the family of periodics (PO). $M_s$ and $n$-nullcline intersects at a fixed point of the $\rm K_{SS}^{-}$-system, which lies above the lower fold of $M_s$ denoted by the blue dot (see the insert for an enlarged view). HC denotes a homoclinic bifurcation in which the PO terminates. }
\label{fig:bif-n-more-sepa}      
\end{center}
\end{figure}

The fast-slow decomposition analysis of the $\rm K_{SS}^{-}$-system \eqref{eq:nondimM} is performed by treating the slow variable $n$ as a bifurcation parameter of its fast subsystem, which we refer to as the \textit{fast $\rm K_{SS}^{-}$-subsystem}. Since the $\rm K_{SS}^{-}$-system is obtained by fixing $q=0$ in the full system, the corresponding GSPT-derived subjects such as the subsystems and critical manifold are inherited from those of the full system evaluated at $q=0$ (see Subsection \ref{sec:nondim}). The bifurcation diagram of the fast $\rm K_{SS}^{-}$-subsystem (Figure \ref{fig:bif-n-more-sepa}) includes an $S$-shaped curve of equilibria, i.e., the critical manifold $M_s$ given by $F(y,n,0)=0$ (blue), and a family of stable periodic orbits PO (green) for the fast $\rm K_{SS}^{-}$-subsystem. The family PO initiates at a Hopf bifurcation at a low $n$ value (not shown here) and terminates at a homoclinic (HC) bifurcation at an $n$ value where the fast $\rm K_{SS}^{-}$-subsystem has a homoclinic point on the middle branch of $M_s$ as $t\to \pm \infty$. The dynamics of $n$ relative to this bifurcation diagram depends on the location of its nullcline (cyan curve). Here since the $n$-nullcline intersects $M_s$ on its middle branch (above the lower fold at the blue circle, see Figure \ref{fig:bif-n-more-sepa} insert), the system should exhibit a square-wave like bursting solution as long as the timescale separation between the fast and slow variables is large enough \cite{rinzel1987formal,WR2016}. This is illustrated by the yellow solution trajectory in Figure \ref{fig:bif-n-more-sepa} which is obtained when the timescale separation is exaggerated in the (6F, 1S) scheme. 

%%%%%%%%%%%%%%%%%%%%%%%
% %%%%%%%%%%%%%%%%%%%%%%%
	
\begin{figure}[!htp]
\begin{center}
\begin{tabular}{@{}p{0.45\linewidth}@{\quad}p{0.45\linewidth}@{}}
\subfigimg[width=\linewidth]{\bf{\small{(A)}}}{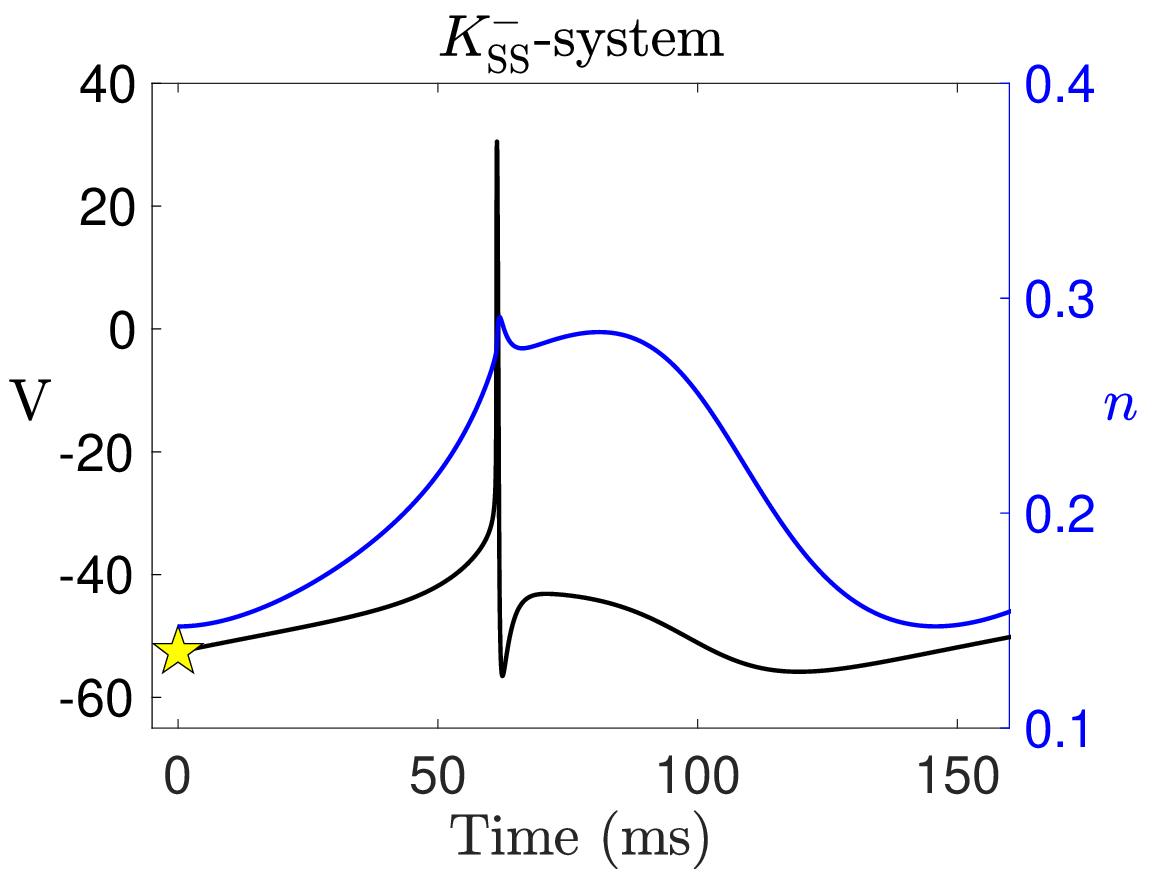} &
\subfigimg[width=\linewidth]{\bf{\small{(B)}}}{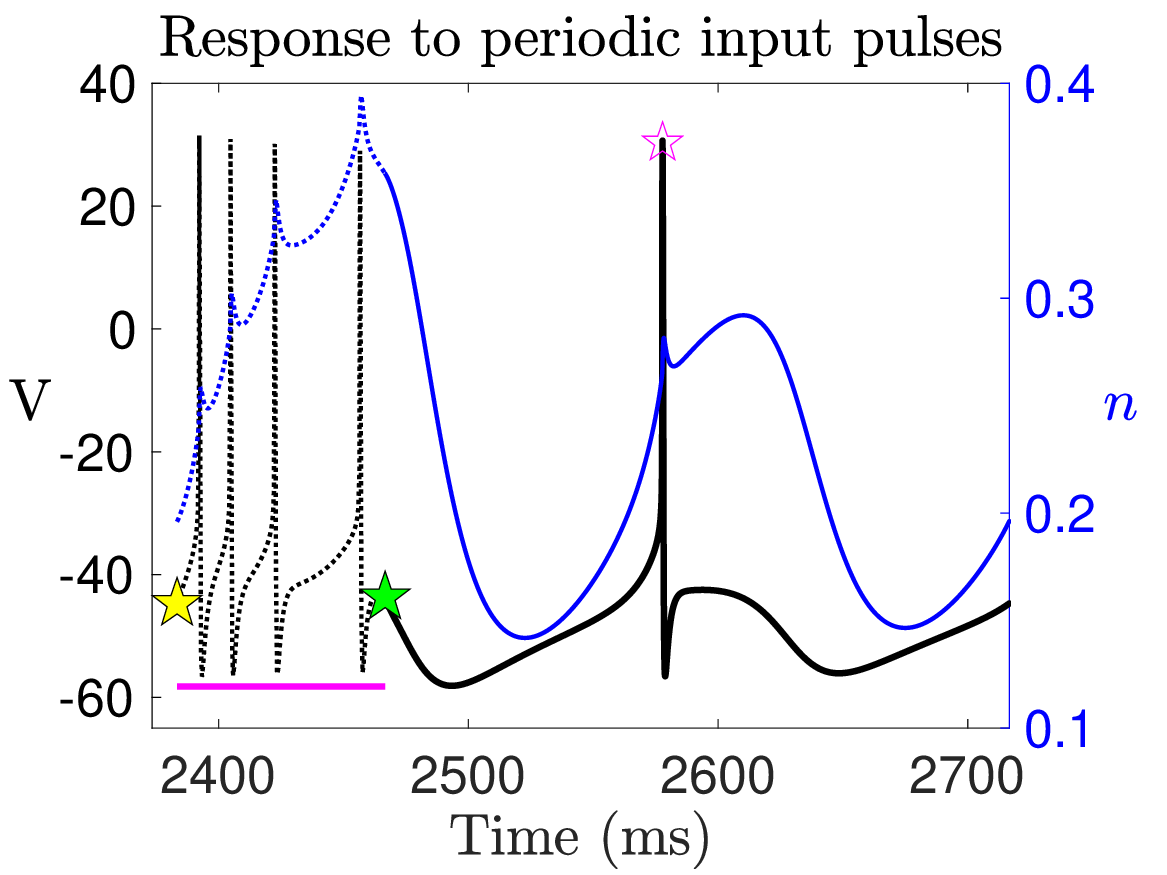} \\
\subfigimg[width=\linewidth]{\bf{\small{(C)}}}{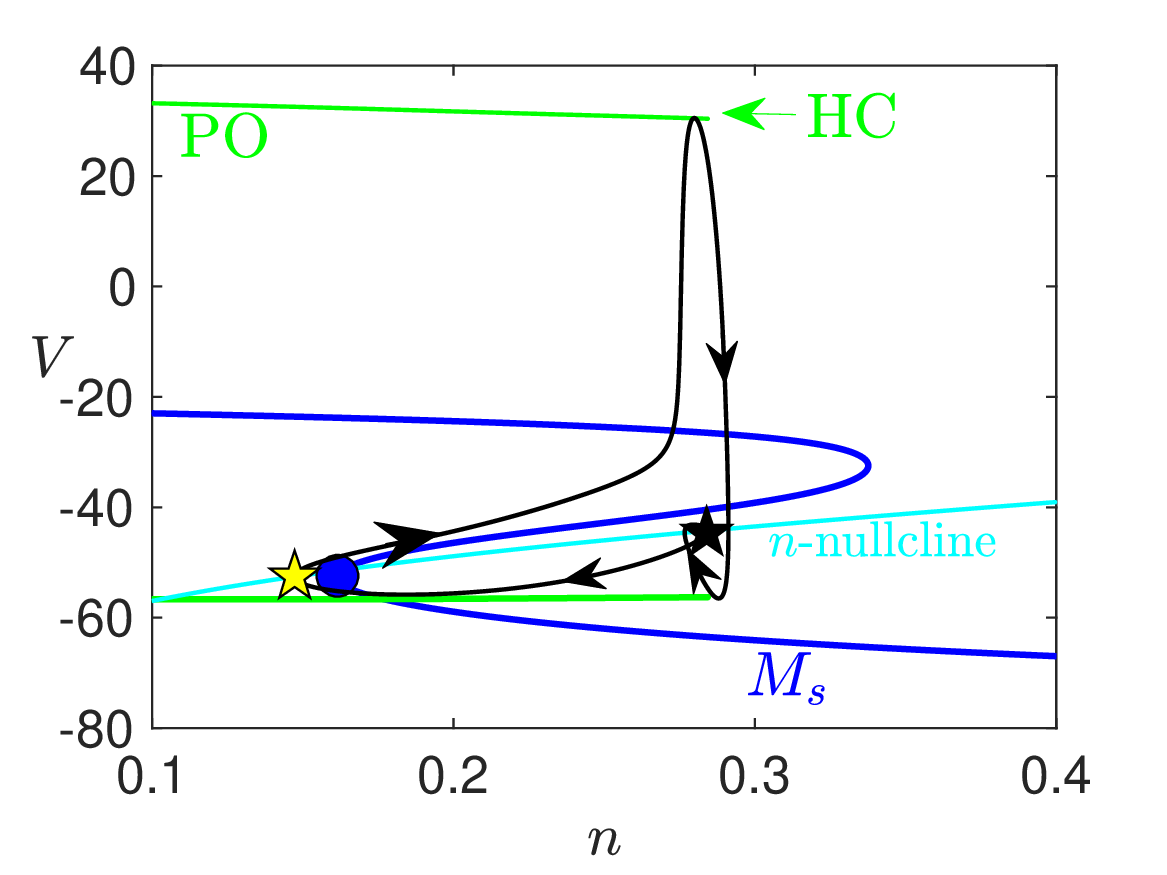} &
\subfigimg[width=\linewidth]{\bf{\small{(D)}}}{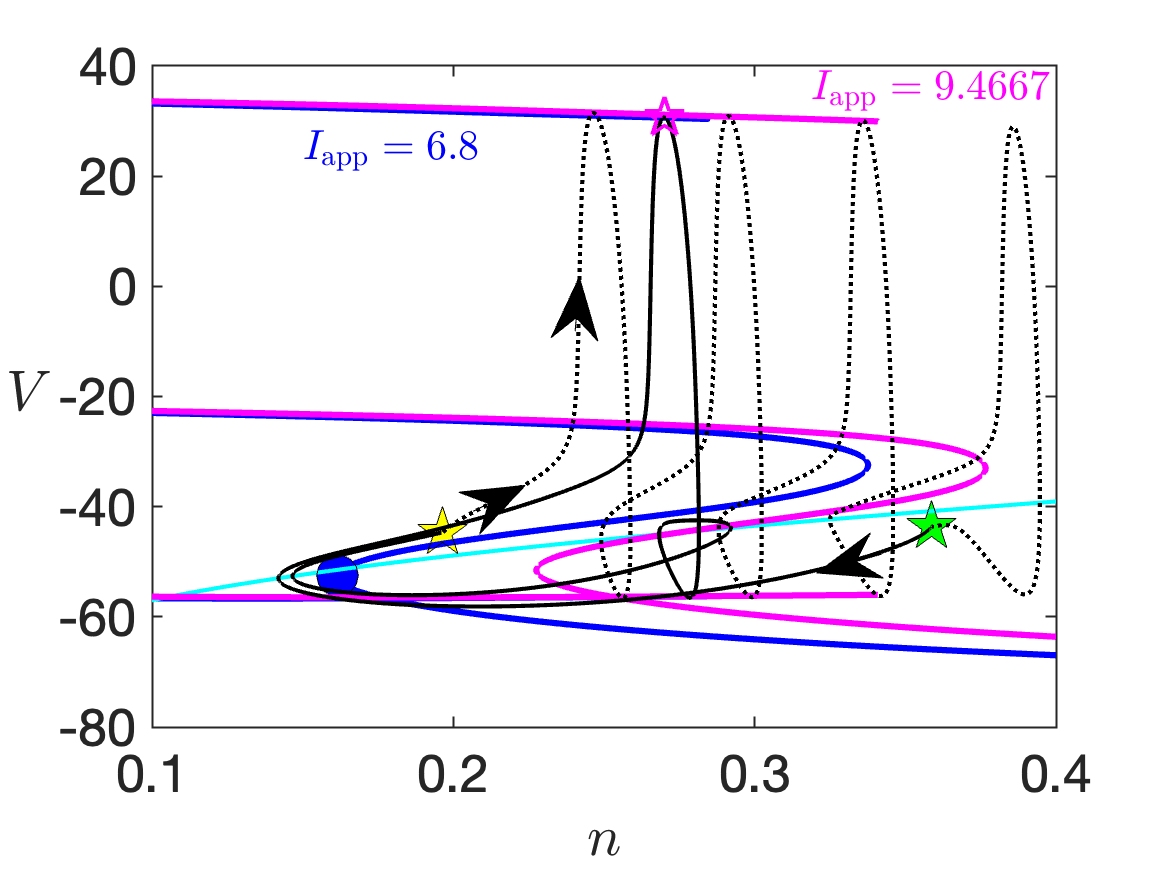}
\end{tabular}
\end{center}
\caption{ \label{fig:bif-n} Simulation of the solution of the $\rm K_{SS}^{-}$-system and its response to periodic input pulses at $3$ Hz, together with corresponding bifurcation diagrams, for $I=6.8, g_{\rm KCa}=0$ and other parameters as given in Table \ref{tab:para}. (A) One cycle of temporal evolution of $V$ (black) and $n$ (blue), as shown in Figure \ref{fig:ST-response-M}B, top panel. (B) One cycle of temporal evolution of $V$ and $n$ in response to periodic input pulses with input frequency $3$ Hz (see Figure \ref{fig:ST-response-M}B, the third panel from the top). Parameter values for the input pulses are the same as in Figure \ref{fig:ST-response-M}B. The magenta bar on the bottom indicates the timing of the input pulse, which begins at the yellow star and ends at the green star. The dotted curve indicates the solutions during the input pulse, whereas the solid lines indicate the post-input spike. (C) Projection onto $(n, V)$-space of the tonic spiking solution (black) from (A) and the 
bifurcation diagram of fast $\rm K_{SS}^{-1}$-subsystem with respect to $n$, along with the $n$-nullcline shown in cyan. Color codings of the bifurcation diagram are the same as in Figure \ref{fig:bif-n-more-sepa}. 
(D) The perturbed solution trajectory from (B) and the effect of increasing $I_{\rm app}$ on the bifurcation diagram for the fast $\rm K_{SS}^{-1}$-subsystem, projected onto $(n, V)$-space, along with the $n$-nullcline (cyan). Increasing $I_{\rm app}$ from $6.8$ (without any input pulse) to $9.4667$ (with the input pulse) results in a shift of the bifurcation diagram to the upper right (blue to magenta). }
\end{figure}

In the $\rm K_{SS}^{-}$-system with default parameters given in Table \ref{tab:para}, the timescale separation between the fast and slow variables is relatively far from the singular limit. As a result, instead of generating a square-wave bursting solution that closely follows the bifurcation structure (Figure \ref{fig:bif-n-more-sepa}), the system produces only a single spike in each cycle, as illustrated by the black solution trajectory in Figure \ref{fig:bif-n}A and C. In particular, the solution trajectory does not closely follow the lower stable branch of $\ms$ as it moves leftward toward the lower fold (blue circle), and subsequently follows the middle repelling branch for some time before generating a spike. 
% We demonstrate below that this deviation from the bifurcation structure plays a critical role in the system's inflexible phase-locking response to slow inputs. 
% As a result, the $\rm K_{SS}^{-}$-system does not spike at the minimum value of the m-current gating variable $n$ at the fold of $M_s$ but during the rising phase of its rhythm.

\begin{figure}[!htp]
\begin{center}
 \begin{tabular}{@{}p{0.45\linewidth}@{\quad}p{0.45\linewidth}@{}}
 \subfigimg[width=\linewidth]{\bf{\small{(A)}}}{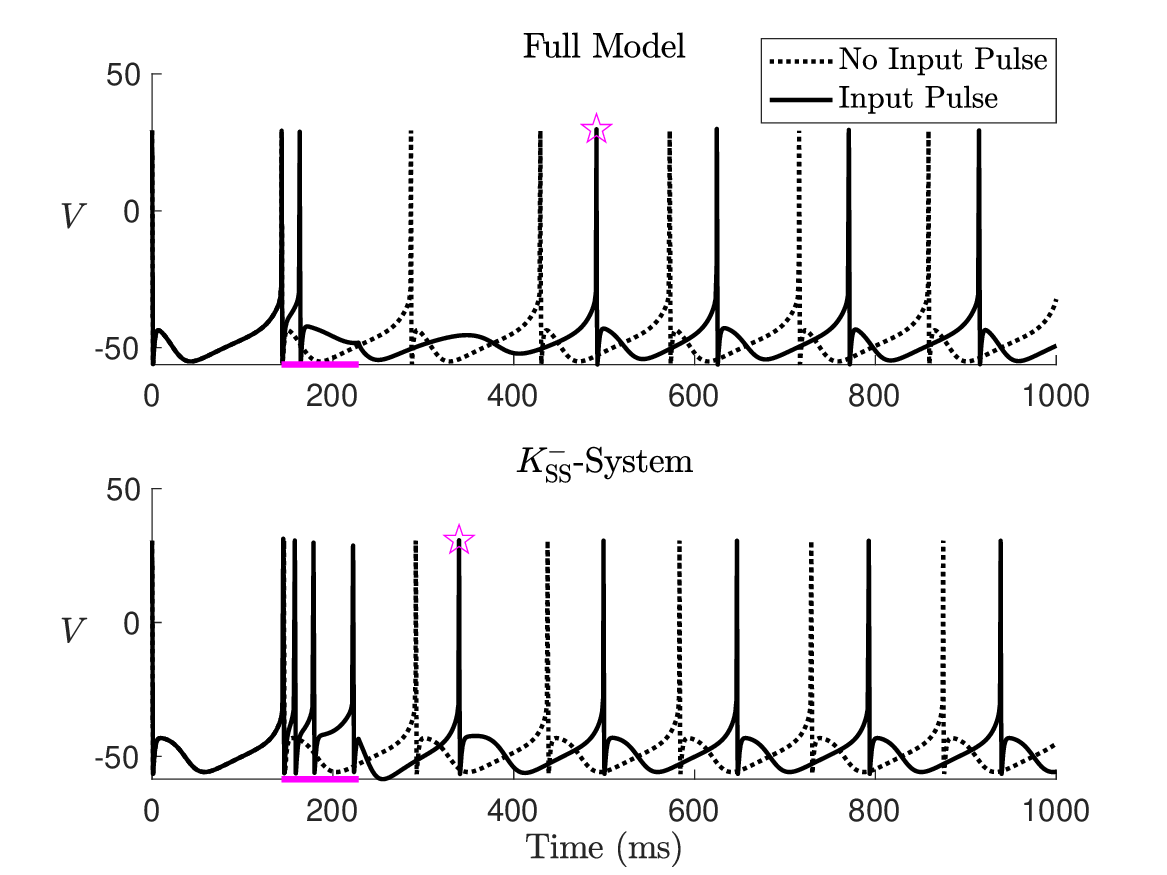}&
 \subfigimg[width=\linewidth]{\bf{\small{(B)}}}{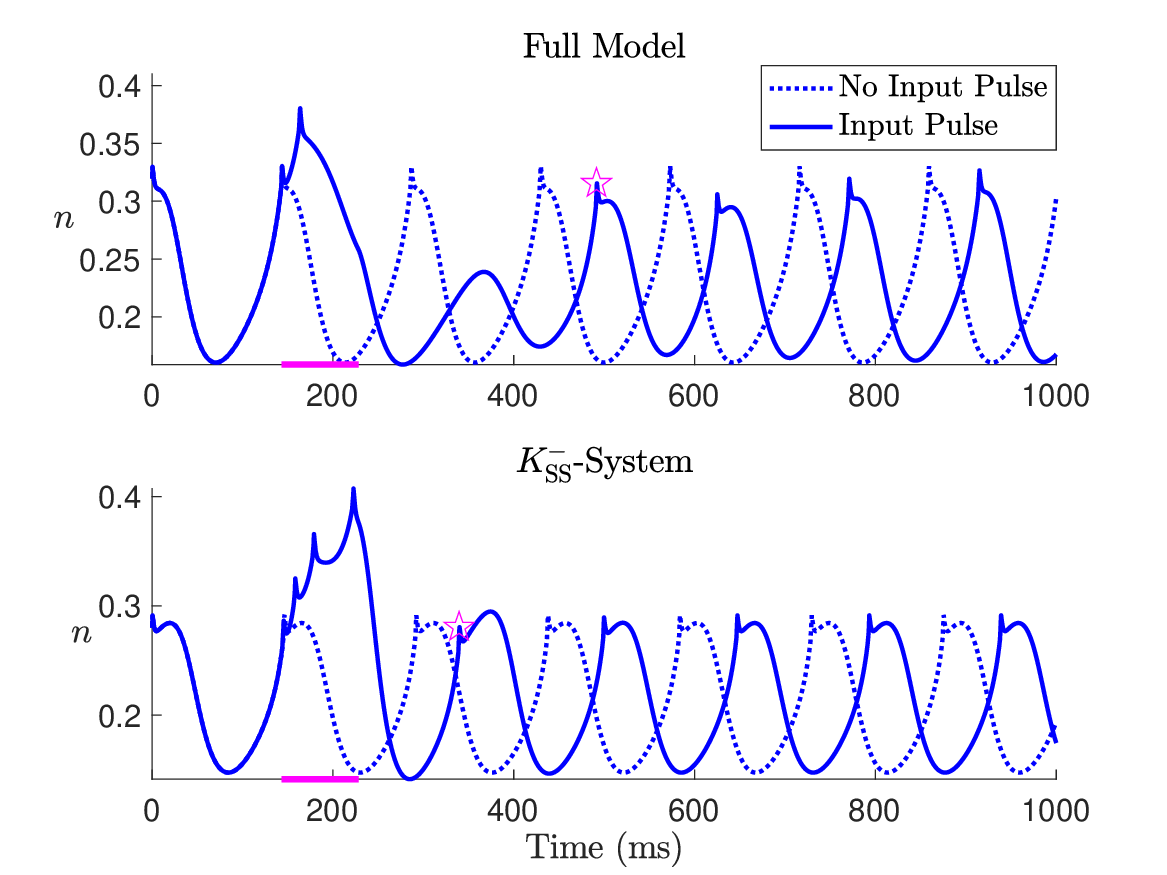}
 \end{tabular}
\end{center}
\caption{\label{fig:delay-full-Kss-} Delay of spiking $D$ of (Top Panel) the full model and (Bottom Panel) the $\rm K_{SS}^-$-system,  in response to a single pulse lasting $1/4$ of a cycle from 3 $\rm Hz$ periodic inputs. Magenta bar indicates the timing of the input pulse; magenta star indicates the first post-input spike. (A) Voltage traces with (solid lines) and without (dotted lines) an input pulse. (B) Buildup of outward $m$ current.}
\end{figure}

To investigate the phase-locking behavior of the $\rm K_{SS}^{-}$-system, we focus on the 3 Hz periodic inputs, since the $\rm K_{SS}^{-}$-system fails to phase-lock to this slower rhythm, in contrast to the full model (Figure \ref{fig:ST-response-M}). 
Figure \ref{fig:delay-full-Kss-} shows the post-input spiking delays $D$ of the full model (top panel) and the $K_{\rm SS}^-$-system (bottom panel) in response to a single input pulse whose duration and amplitude match those of the 3 Hz periodic input. Panel (A) shows the voltage dynamics, while panel (B) shows the gating variable $n$ of the inhibitory current $I_m$.
In the full model, the first post-input spontaneous spike (indicated by the magenta star in the top row) occurs after two spontaneous spikes (dashed curves), leading to phase-locking to 3 Hz periodic inputs (Figure \ref{fig:ST-response-M}A). In contrast, in the $K_{\rm SS}^-$-system, the first post-input spike follows shortly after the first spontaneous spike, preceding the arrival of the next input, which results in a failure to phase-lock (Figure \ref{fig:ST-response-M}B, bottom panel). 
% Notably, the brief delay $D$ in the $K_{\rm SS}^-$-system occurs because the system spiked for the duration of the input pulse which repeatedly reset the inhibitory current (Figure \ref{fig:delay-full-Kss-}B, bottom panel). Once the spiking stopped at the end of the input pulse, the next spike quickly followed in less than the intrinsic 7 Hz period, therby failing at phase-locking to 3 Hz inputs (see Figure \ref{fig:ST-response-M}B).
It is worth noting that in both systems, $n$ returns to a similar baseline at a comparable rate following the input pulse (see Figure \ref{fig:delay-full-Kss-}B). However, in contrast to the $K_{\rm SS}^-$-system, where spontaneous spiking resumes shortly after $n$ reaches baseline (see Figure \ref{fig:delay-full-Kss-}B, bottom panel), the full system exhibits a more prolonged delay before spiking resumes, during which $n$ oscillates (Figure \ref{fig:delay-full-Kss-}B, top panel). We will explain this delay phenomenon in greater detail in Section \ref{sec:both}. 
% The key to understanding the inflexibility of the $\rm K_{SS}^-$-system to slower inputs, therefore, lies in why it recovers spiking quickly. 

%%%%%%%%%%%%%%%%%%%%%%%
%%%%%%%%%%%%%%%%%%%%%%%

%%%%%%%%%%%%%%%%%%%%%%%
%%%%%%%%%%%%%%%%%%%%%%%

% To explore this, we define $\tau_R$ as the duration of the rising phase for $n$ (from blue circle fold to 
% roughly the black star in Figure \ref{fig:bif-n}C) and $\tau_D$ as the duration of the decaying phase (from the black star to the blue circle). The intrinsic frequency is therefore given as $f_0=1/(\tau_R+\tau_D)$. 
Figure \ref{fig:bif-n}A and B show the time series of the intrinsic dynamics ($V$ in black and $n$ in blue) of the $\rm K_{SS}^{-}$-system over one period and its response to $3\,\rm Hz$ periodic input pulses over one input cycle. 
The trajectory within an input pulse, the timing of which is indicated by the horizontal magenta line, is plotted using the dashed curve, whereas the trajectory outside the input pulse is shown by the solid curve (Figure \ref{fig:bif-n}B). The system resumes spontaneous spiking (magenta star) shortly after the input pulse terminates at the green circle, highlighting its inability to phase-lock to the 3-Hz periodic inputs. The effect of each input pulse on the bifurcation diagrams of the $\rm K_{SS}^{-}$-system over one cycle is illustrated in Figure \ref{fig:bif-n}D.
Specifically, as the input pulse begins at the yellow star, the applied current increases from a baseline of $I_{\rm app}=6.8$ to $I_{\rm app}= 9.4667$. Consequently, the bifurcation diagram shifts from the blue curve to the magenta curve. This shifts causes the trajectory at the yellow star to become further away from the perturbed HC bifurcation, allowing it to spike four times (dashed black curve in Figure \ref{fig:bif-n}D) before being terminated at a higher $n$ value. This causes the m-current gating variable $n$ to peak at about $0.38$, which is higher than it would have reached without any input. As the input terminates at the green star, the bifurcation diagram reverts from magenta to blue, and the trajectory moves in the decreasing $n$ direction to reach the unperturbed fold of $M_s$ at the blue circle. After that, an intrinsic spike occurs (solid black curve and magenta star) before the arrival of the next input pulse. That is, phase-locking is not achieved as the post-input spiking delay $D$ is insufficient to prevent spiking before the next input pulse. 

To explain why the post-input spiking delay $D$ is short, we decompose it into three components:
\begin{equation}\label{eq:D1-Msystem}
D= \frac{1}{4}\frac{1}{f}+\tilde{\tau}_D + \tilde{\tau}_R, 
\end{equation}
where the first term represents the input duty cycle, set at $1/4$ of the input period $1/f$. The second term, $\tilde{\tau}_D$, denotes the duration of the decay phase during which $n$ decreases from its value the end of the input pulse (green star in Figure \ref{fig:bif-n}D) to its baseline at the $\ms$ fold (blue circle in Figure \ref{fig:bif-n}D). The third term, $\tilde{\tau}_R$, denotes the duration of the rising phase where $n$ increases from the baseline to the first-input spike (magenta star). 
In parallel, we define $\tau_D$ and $\tau_R$ (without the tilde) as the decay and rising durations in the unperturbed system. Specifically, $\tau_D$ measures the duration from the black star in Figure \ref{fig:bif-n}C to the blue circle, and $\tau_R$ is the duration from the blue fold point to the spike. Notably, $\tilde{\tau}_R \approx \tau_R$ because, after passing the fold, the perturbed and unperturbed trajectories behave almost identically, both following the same blue bifurcation diagram. 

% If the timescale separation between $n$ and other fast variables is sufficient, $\tilde{\tau}_D$ should be larger than $\tau_D$. This expectation is due to the higher $n$ value at the end of the input pulse at the green star than the unperturbed maximum $n$ value near the black star in Figure \ref{fig:bif-n}C.
% Despite predictions that a more hyperpolarized system with a higher buildup of $n$ would take longer to resume spiking than its unperturbed counterpart, numerical simulations reveal that $\tilde{\tau}_D$, the time it takes for the perturbed system to recover its spiking behavior, is actually smaller than $\tau_D$, the recovery time of the unperturbed system. 
In contrast, $\tilde{\tau}_D\neq \tau_D$ as the initial $n$ values at the start of the decay phase differ. Specifically, the input pulse pushes $n$ to a higher level compared with the unperturbed case. Intuitively, one might expect the perturbed trajectory, starting at a higher $n$, to take longer to decay back to the same baseline, implying that $\tilde{\tau}_D$ should be longer than $\tau_D$.
However, our numerical simulations reveal the opposite: a system that is farther from the fold of $M_s$, i.e., the spiking threshold, resume spiking more quickly. This is one of the key factors underlying the short delay $D$ and hence poor phase-locking ability in the $\rm K_{SS}^-$-system. It follows from \eqref{eq:f-limit} and $\tilde{\tau}_D<\tau_D$ that the lower frequency limit $f^*$ of phase-locking for the $\rm K_{SS}^{-}$-system can be estimated as
\[ f^*\geq \frac{3/4}{\tilde{\tau}_D+\tau_R} > \frac{3/4}{\tau_D+\tau_R}\approx 5.25\,\rm Hz
\] 
%used the fact that $f*\leq f$.
Note that it is difficult to precisely estimate $f^*$ as $\tilde{\tau}_D$ implicitly depends on the input frequency $f$. Our estimate $f^*> 5.25$ implies that the $\rm K_{SS}^-$-system can, at best, phase-lock to inputs as low as $5.25$ Hz, which is consistent with our direct numerical simulation results shown in Figure \ref{fig:ST-response-M}B. 

% This is consistent with Figure \ref{fig:ST-response-M}B showing that the $\rm K_{SS}^{-}$-system fails at phase-locking to input rhythms at 5.5 Hz.  

%%%%%%%%%%%%%%%%%%%%%%%
\begin{figure}[!t]
\begin{center}
\includegraphics[width=\linewidth]{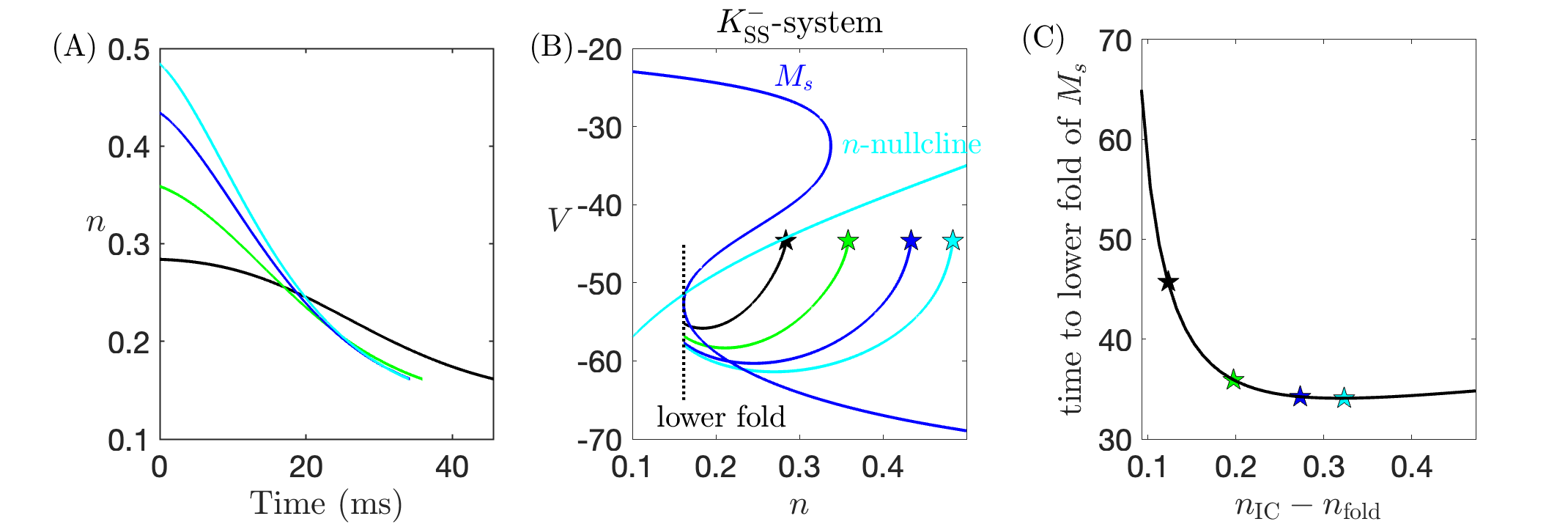}\\
\includegraphics[width=\linewidth]{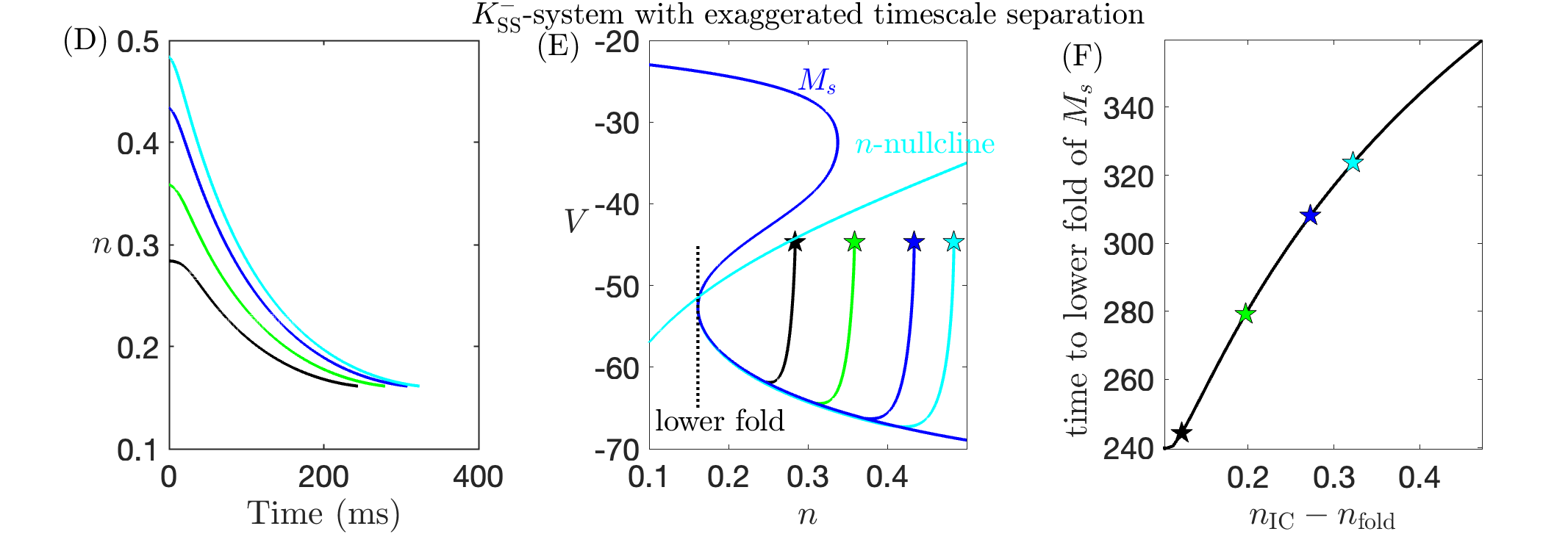}
\end{center}
\caption{ \label{fig:time-to-fold-NoKCa} Solution trajectories of the $\rm K_{SS}^{-}$-system with (Top row) default timescales and (Bottom row) exaggerated timescale separation as in Figure \ref{fig:bif-n-more-sepa}.  
(Left): Time evolution of the trajectories that start at different $n$ values until hitting the fold  and terminate upon reaching the lower fold of $M_s$. (Middle): Projection of trajectories from the left panel onto $(n,V)$-space. Stars with different colors denote the different initial conditions of the trajectories. (Right): The relationship between the difference between $n$ at the initial condition and $n$ at the fold (i.e., $n_{\rm IC}-n_{\rm fold}$) and the time for the trajectories to reach the lower fold of $M_s$.}
\end{figure}
%%%%%%%%%%%%%%%%%%%%%%%

One key factor underlying the short delay $D$ and consequently the poor phase locking in the $\rm K_{SS}^-$-system to slow inputs, as discussed above, is the relatively short $\tilde{\tau}_D$. We now explain why $\tilde{\tau}_D$ remains small despite the elevated $n$ value at the end of each input pulse. In particular, this behavior arises from the insufficient timescale separation in the $\rm K_{SS}^{-}$-system.
The top row of Figure \ref{fig:time-to-fold-NoKCa} displays the $\rm K_{SS}^{-}$-system trajectories during the decay phase, each starting with different $n$ values (denoted as $n_{\rm IC}$). The black trajectory, beginning with the same $n$ value as the black star in Figure \ref{fig:bif-n}C, takes approximately $\tau_D$ to reach the fold. In contrast, the green trajectory, starting from the $n$ value marked by the green star in Figure \ref{fig:bif-n}D, takes approximately $\tilde{\tau}_D$ to reach the fold. The plot clearly shows that trajectories beginning at higher levels of inhibition (green, blue and cyan curves) decay to the fold faster than the black trajectory. Panel (B) suggests that this faster decay for trajectories with higher $n$ values is due to their greater distance from the n-nullcline, leading to a more rapid decrease in $n$. As a result, trajectories initiated at higher $n$ values take less time to decay to the baseline at the fold and resume spiking sooner. Thus, despite the greater inhibition accumulated during the input pulse, the system recovers spontaneous spiking more quickly, leading to a shorter delay $D$ than in the absence of accumulated inhibition.

To further confirm that the small $\tilde{\tau}_D$ arises from an insufficient fast-slow timescale separation, we exaggerate this separation by making $n$ 10 times slower. As anticipated, this adjustment causes trajectories with higher initial values of $n$ to take longer to decay to the lower fold (compare Figure \ref{fig:time-to-fold-NoKCa}C and F). This occurs because, with $n$ evolving more slowly and the fast-slow $\rm K_{SS}^{-}$-system closer to its singular limit, trajectories rapidly jump down to the lower stable branch of $M_s$ during which $n$ is nearly constant, and then evolve along $M_s$ on the slow $n$ timescale (Figure \ref{fig:time-to-fold-NoKCa}E). Since $n$ decays along $M_s$ at similar rate, a higher initial $n$ value leads to a prolonged time to reach the fold, thereby extending the decay phase $\tilde{\tau}_D$ and hence the delay $D$. 
We note that the $K_{\rm SS}^-$-system with this exaggerated timescale separation exhibits bursting rather than spiking, making a direct comparison of its phase-locking capabilities with those of the full model complicated and beyond the scope of this work. Nonetheless, even under the exaggerated timescale separation, spiking resumes immediately once $n$ returns to baseline, suggesting that the system's phase-locking capability remains limited relative to the full model 

In summary, the inflexible phase-locking in the $K_{\rm SS}^-$-system arises primarily from its relatively rapid recovery of spiking due to insufficient timescale separation between fast and slow variables. Additionally, unlike the full system, which exhibits a prolonged delay after crossing the fold of $M_s$ (see Figure \ref{fig:delay-full-Kss-} top panel, between the end of the input and the magenta star), the $K_{\rm SS}^-$-system exhibits a much shorter spiking delay after passing the fold. 
% We explore the mechanism underlying the prolonged post-fold delay in the full system in Section \ref{sec:both}.

\section{$\rm M^{-}$-system cannot phase lock to significantly slower inputs}\label{sec:kca}

In the $\rm M^-$ system \eqref{eq:nondimKSS}, the gating variable $q$ for $I_{\rm K_{SS}}$ evolves on a superslow timescale whereas all other variables $y$ evolve on fast timescales. The timescale decomposition analysis can be performed by treating the superslow variable $q$ as the bifurcation parameter for the \textit{fast $\rm M^-$-subsystem} consisting of all other fast variables. As in the previous section, we consider the effect of input pulses on the bifurcation diagram of the $\rm M^-$-system and the corresponding post-input spiking delay $D$ to understand why this oscillator is unable to phase-lock to rhythmic inputs that are significantly slower than its intrinsic frequency.

\begin{figure}[!t]
\begin{center}
\begin{tabular}{@{}p{0.45\linewidth}@{\quad}p{0.45\linewidth}@{}}
\subfigimg[width=\linewidth]{\bf{\small{(A)}}}{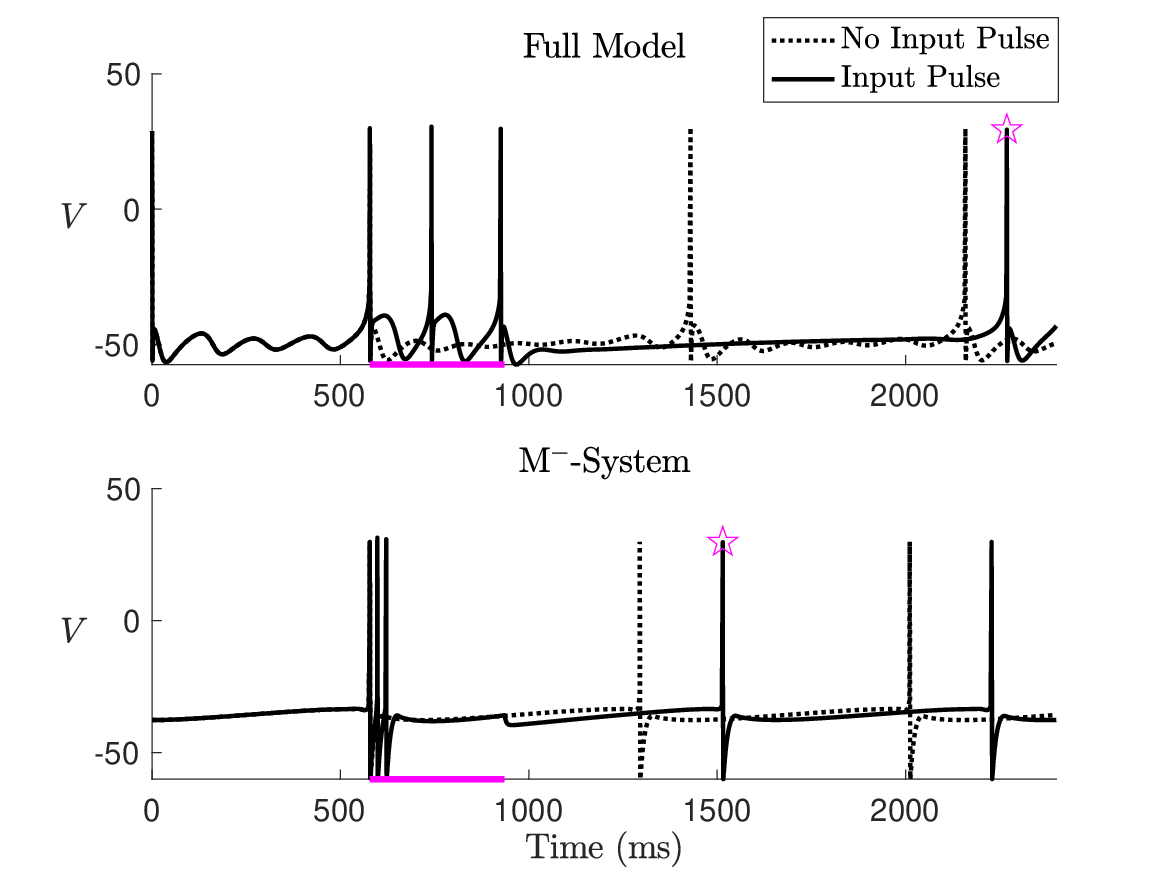}&
\subfigimg[width=\linewidth]{\bf{\small{(B)}}}{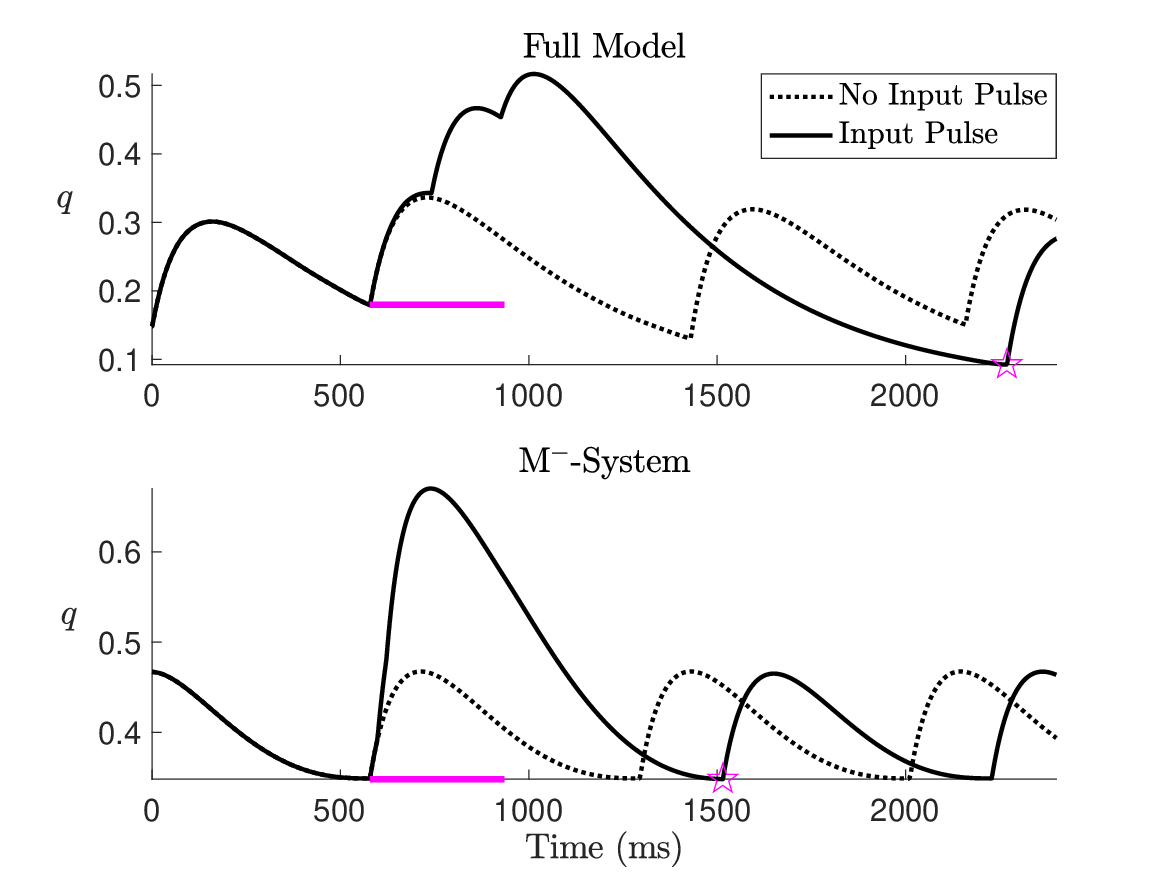}
\end{tabular}
\end{center}
\caption{\label{fig:delay-full-M-} Delay of spiking $D$ of (Top panel) the full model with $I_{\rm app}=8$ and (Bottom panel) the $\rm M^-$-system, in response to a single pulse lasting 1/4 of a cycle from $0.7$ Hz periodic inputs. Model parameters are the same as in Figure \ref{fig:ST-response-KCa}. Magenta bar indicates the timing of the input pulse; magenta star indicates the first post-input spike. (A) Voltage traces with (solid lines) and without (dotted lines) an input pulse; (B) Buildup of outward $\rm K_{SS}$ currents.}
\end{figure}

To investigate the phase-locking behavior of the $\rm M^-$ system, we focus on the $0.7$ Hz periodic inputs, as the $\rm M^-$ system fails to phase-lock to them, unlike the full system (see Figure \ref{fig:ST-response-KCa}). Figure \ref{fig:delay-full-M-} compares the spiking delays $D$ of the full model (top panel) and the $\rm M^-$-system (bottom panel), clearly showing that the full model exhibits a much longer $D$ than the $\rm M^-$-system. We highlight two major differences: First, the $\rm M^-$-system recovers spontaneous spiking as soon as $q$ returns to its baseline level, whereas in the full system, $q$ must decay to a level significantly below its baseline before the oscillator can spike again. 
This extended delay in the full system arises from a delayed Hopf bifurcation (DHB), which we discuss further in Section \ref{sec:both}. 
Interestingly, while a fast subsystem Hopf bifurcation is also present and responsible for the onset of spontaneous spiking in the $\rm M^-$-system, it does not produce a comparable delay. Second, while $q$ accumulates during the input pulse in both systems, $q$ in the full system continues to rise even after the input ends, whereas in the $\rm M^-$-system, $q$ begins to decay midway through the pulse. 
Together, these result in a much shorter delay $D$ and poorer phase-locking flexibility to slow inputs in the $\rm M^-$-system compared to the full system.

%%%%%%%%%%%%%%%%%%%%%%%
%%%%%%%%%%%%%%%%%%%%%%%

%%%%%%%%%%%%%%%%%%%%%%%
\begin{figure}[!t]
  \begin{center}
  \begin{tabular}{@{}p{0.45\linewidth}@{\quad}p{0.45\linewidth}@{}}
 \subfigimg[width=\linewidth]{\bf{\small{(A)}}}{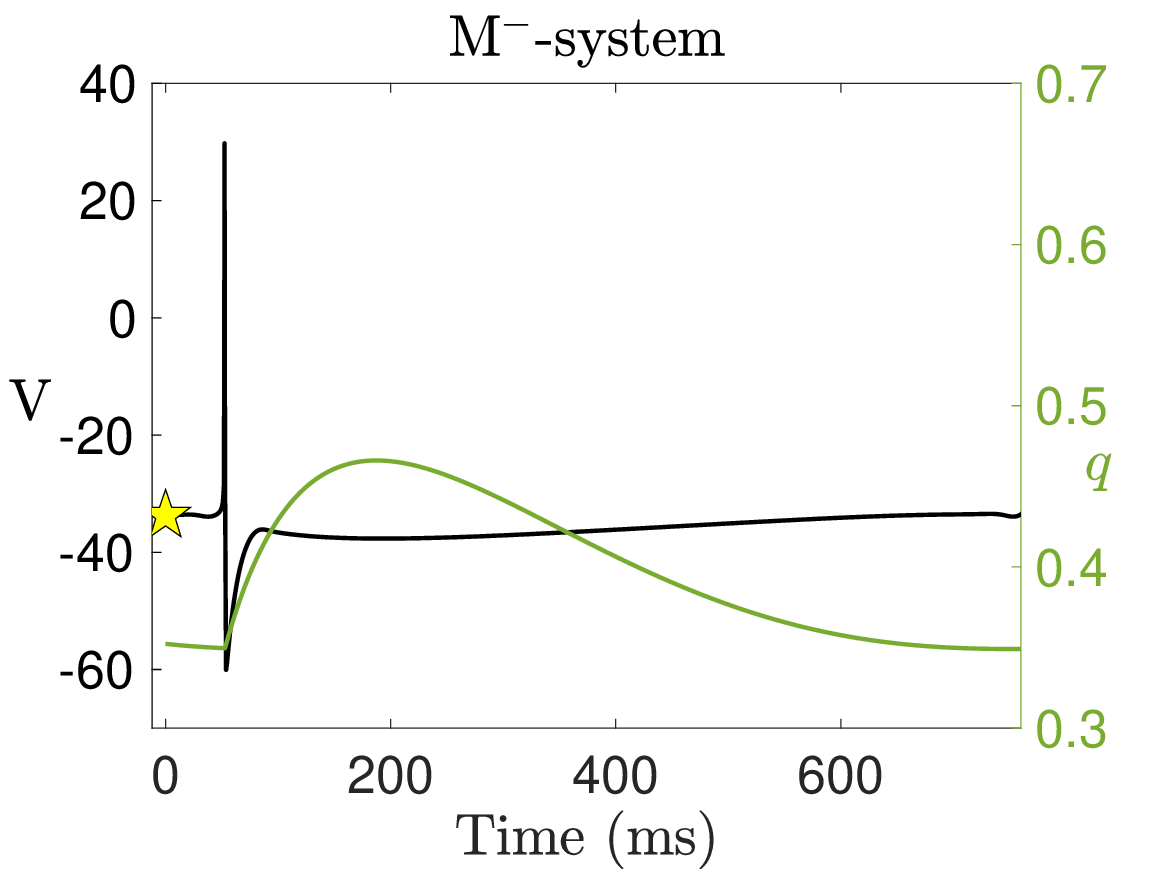}&
\subfigimg[width=\linewidth]{\bf{\small{(B)}}}{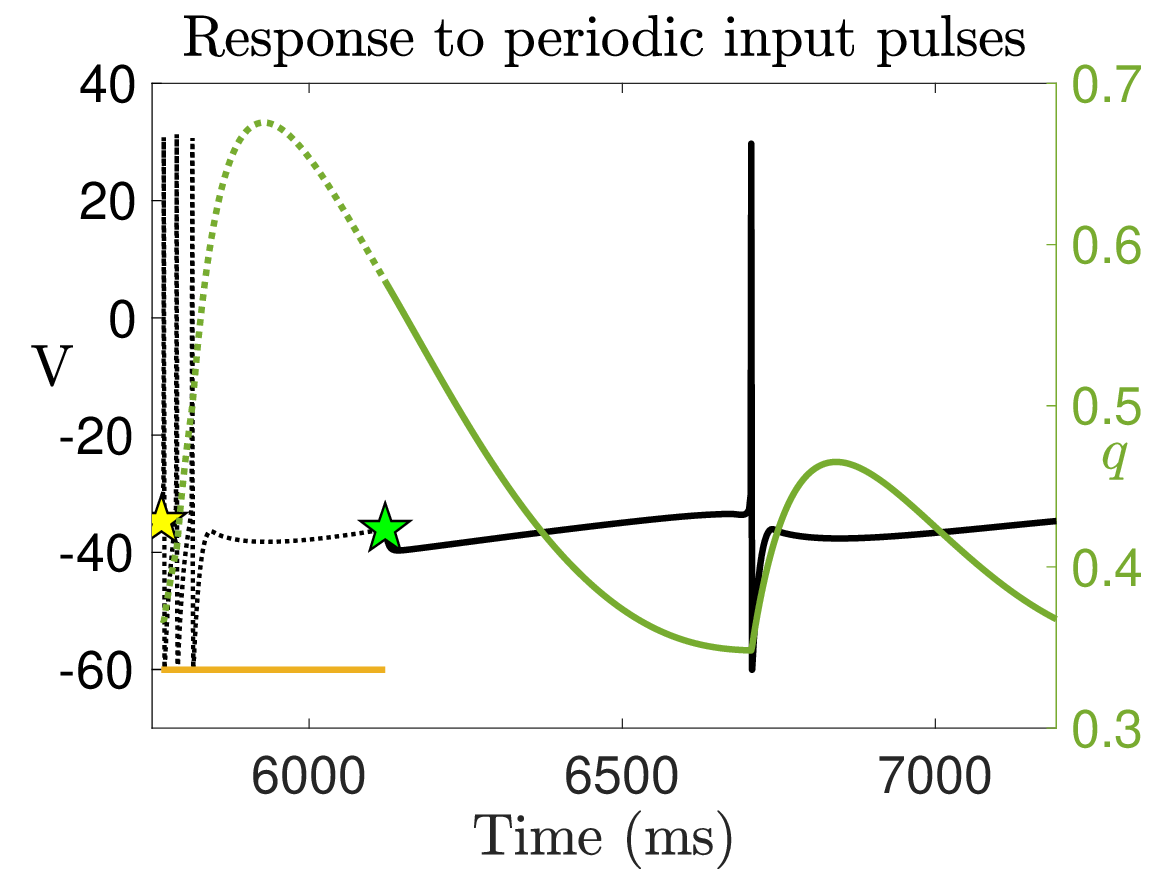}\\
\subfigimg[width=\linewidth]{\bf{\small{(C)}}}{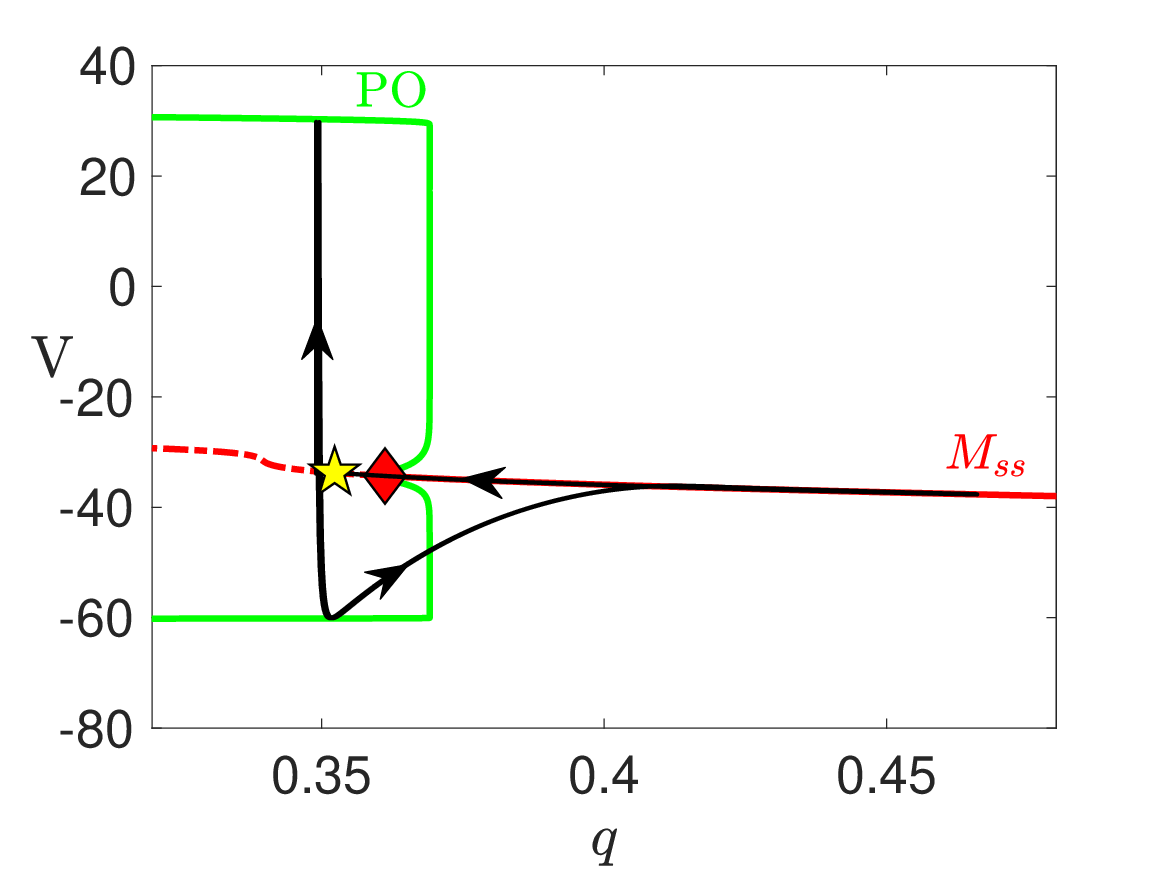}&
\subfigimg[width=\linewidth]{\bf{\small{(D)}}}{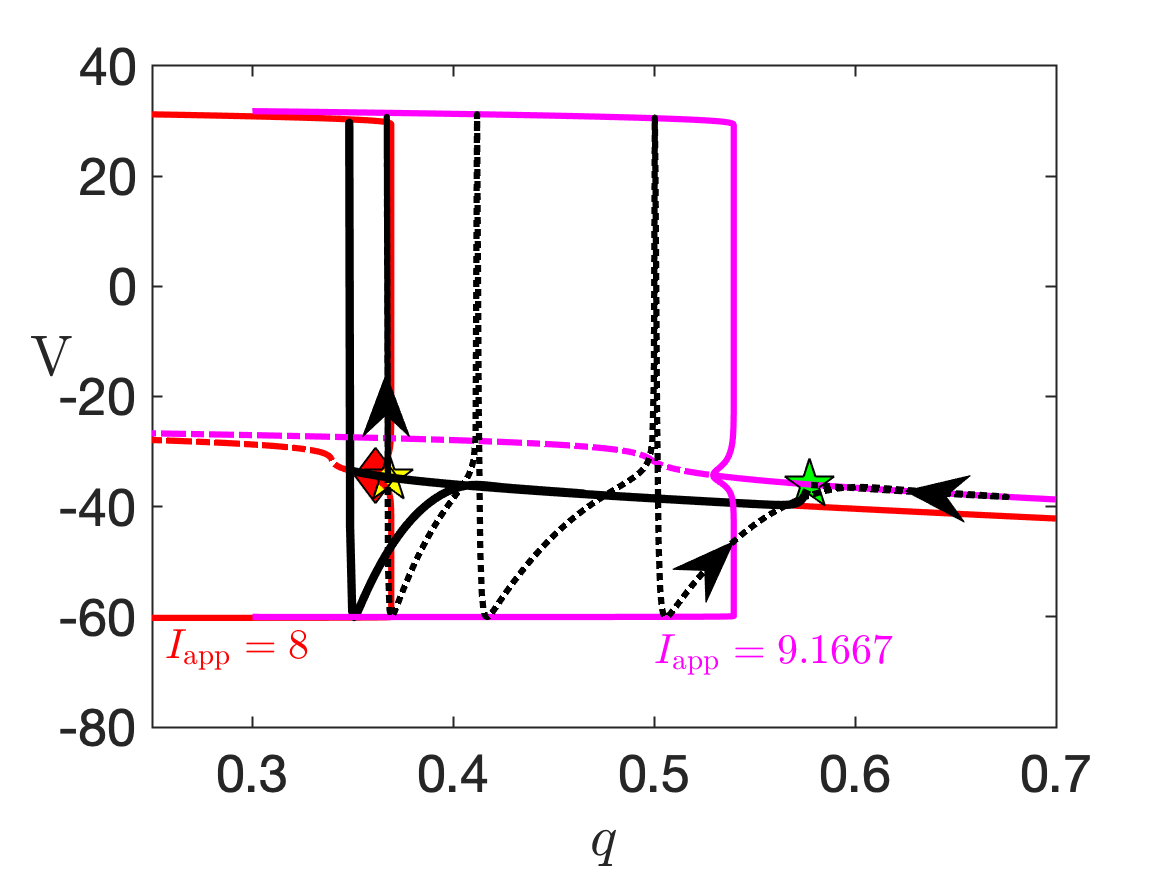}
\end{tabular}
\end{center}
\caption{\label{fig:bif-q} Simulation of the solution of the $\rm M^{-}$-system and its response to periodic input pulses at $0.7$ Hz, together with corresponding bifurcation diagrams, for $I_{\rm app}=8, g_m=0, g_l=0.16$ and other parameters as given in Table \ref{tab:para}. (A) One cycle of temporal evolution of $V$ (black) and $q$ (green), as shown in Figure \ref{fig:ST-response-KCa}B, top panel. (B) One cycle of the temporal evolution of $V$ and $q$ in response to periodic input pulses with input frequency $0.7$ Hz (see Figure \ref{fig:ST-response-KCa}B, the fourth row). The yellow bar on the bottom indicates the timing of the input pulse, which begins at the yellow star and ends at the green star. The dotted curve indicates the solutions during the input pulse, whereas the solid lines indicate the post-input spike.
(C) Projection onto $(q, V)$-space of the spiking solution (black) from panel (A) and the bifurcation diagram of fast $\rm M^{-1}$-system with $q$ taken as a constant parameter. The curve $M_{ss}$ (red) denotes the equilibria, the green curve shows the maximum and minimum $V$ along the family of periodics (PO), and the red diamond denotes the subcritical Hopf bifurcation. (D) The perturbed solution trajectory from panel (B) and the effect of $I_{\rm app}$ on the bifurcation diagram for the fast $\rm M^{-1}$-subsystem, projected onto $(q, V)$-space. Increasing $I_{\rm app}$ from $8$ to $9.1667$ results in a shift of the bifurcation diagram to the upper right (blue to magenta). Other color coding and symbols have the same meanings as in Figure \ref{fig:bif-n}.}
\end{figure}
%%%%%%%%%%%%%%%%%%%%%%%

We first examine the lack of delay associated with the Hopf bifurcation in the $\rm M^-$ system. The bifurcation diagram of the $\rm M^-$ system (Figure \ref{fig:bif-q}C) includes a curve of equilibria given by $F(y,0,q)=0$
(red curve, denoted as $M_{ss}$), and a family of periodic orbit (PO) solutions (green curve) born at a subcritical Hopf bifurcation (red diamond) for the fast $\rm M^-$-subsystem. Also shown is the projection of the intrinsic spiking solution from Figure \ref{fig:bif-q}A onto $(q, V)$-space. As the trajectory jumps near the yellow star to fire a spike, the increased $V$ triggers calcium influx and activates the $\rm K_{\rm SS}$ current. This results in a movement of the trajectory in the increasing $q$ direction until reaching its maximum. Afterward, $q$ decays superslowly as the trajectory moves leftward along the attracting branch of $M_{\rm ss}$. Note that the trajectory does not immediately jump after crossing the Hopf bifurcation at the red diamond where the stability of $M_{\rm ss}$ changes, but rather experiences a brief delay traveling along the repelling branch. However, this delay is minimal. 

\begin{figure}[!htp]
\begin{center}
\includegraphics[width=\textwidth]{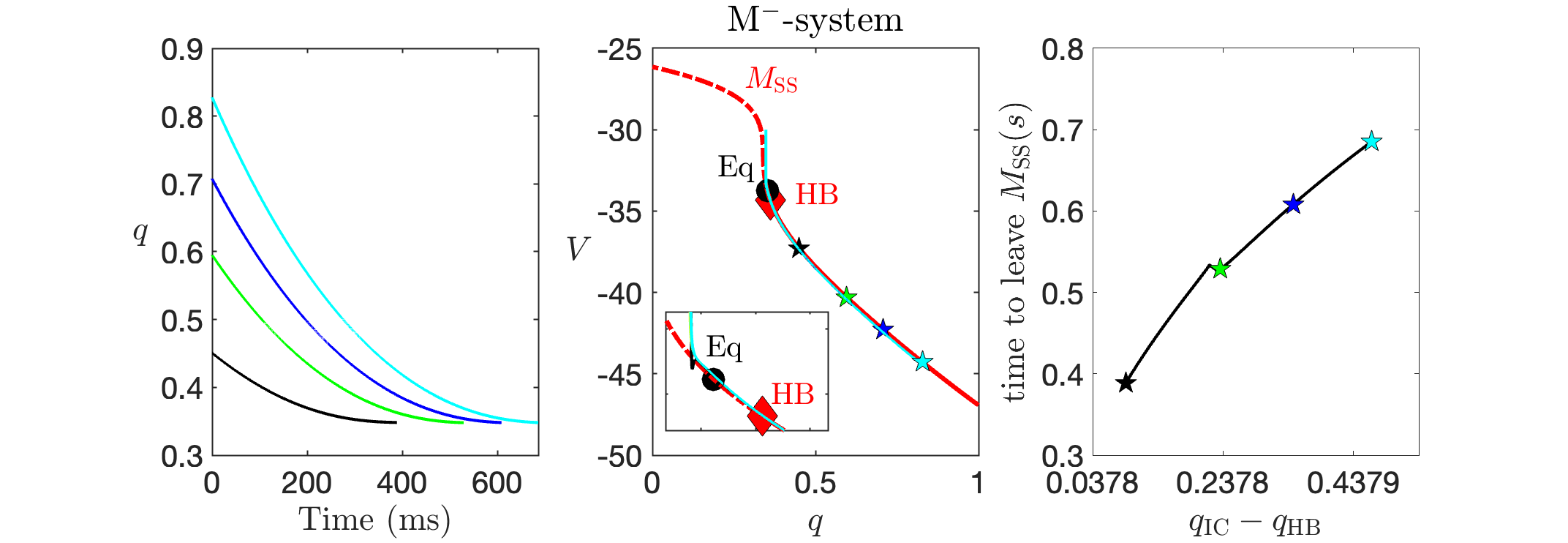}
\end{center}
\caption{ \label{fig:time-to-jump-NoSTOs} Local dynamics of the $\rm M^{-}$-system. (A) Time evolution of trajectories starting at different positions along $M_{\rm SS}$ which are denoted by stars in (B). (B) Projection of trajectories from panel (A) onto $(q,V)$-space. The red diamond denotes the Hopf bifurcation, whereas the black circle denotes the full system equilibrium which is a saddle focus. 
(C) The relationship between the difference between $q$ value at the initial condition and $q$ at the HB bifurcation ($q_{\rm IC}-q_{\rm HB}$) and the time for the trajectories to reach the HB of $M_{\rm ss}$ (i.e., $\tilde{\tau}_D$).} 
% Vertical black dashed line indicates the maximum value of $q$ for the unperturbed $\rm M^-$-system, whereas the green dashed line indicates the $q$ value at the end of the 0.7 Hz input pulse (denoted as green star in Figure \ref{fig:bif-q}B and D). 
% }
\end{figure}

To examine solution behavior near the HB more closely, we compare solutions of the $\rm M^-$ system initialized at different distances from the HB point, all chosen along the stable portion of the $M_{\rm ss}$ (see Figure \ref{fig:time-to-jump-NoSTOs}). The left panel shows the $q$-traces over time; the middle panel shows their projections onto the bifurcation diagram in $(q,V)$-space, with each star marking an initial condition colored to match its corresponding trajectory; and the right panel shows the time required for each trajectory to travel from its initial condition $q_{\rm IC}$ to leave $M_{\rm ss}$, defined as crossing $V=-30$ from below. As noted before, the trajectories experience only a very brief delay after passing the HB. In particular, the exit points of trajectories along the repelling branch of $M_{\rm ss}$ (Figure \ref{fig:time-to-jump-NoSTOs}, middle panel, red dashed curve) are nearly independent of their entry locations on the attracting side (red solid curve), in contrast to a classical delayed Hopf bifurcation, where exit locations are typically entry-dependent and organized by a buffer point \cite{Baer1989}. Our analysis suggests that this lack of delay is primarily associated with the presence of a true equilibrium of the full $M^{-1}$-system (Figure \ref{fig:time-to-jump-NoSTOs}, middle panel, black circle), located in close proximity to the HB point. This equilibrium is a saddle-focus with a two-dimensional unstable spiral manifold while all remaining eigenvalues have negative real parts. Shortly after passing the HB, trajectories are drawn toward the equilibrium along $M_{\rm ss}$ and subsequently escape along its local unstable manifold. As a result, the classical way-in-way-out analysis
% , which assumes monotone slow drift past the HB, 
no longer applies. Instead, the exit dynamics appear to be mainly governed by the local structure of the full-system equilibrium rather than by a delayed Hopf mechanism. A corresponding true equilibrium also exists in the full system when m-current is present. However, in that case it lies much further away from the HB (see Figure \ref{fig:eig-bif-q-Iapp} in Appendix \ref{app:geo}), leaving sufficient separation for the DHB to produce a pronounced delay along the repelling branch. Additional geometric and spectral features of the superslow manifold near the HB point may also limit the buildup of delay. These include its near-fold geometry, where the branch becomes nearly vertical near the HB point (see Figure \ref{fig:time-to-jump-NoSTOs}, middle panel), and the extremely narrow interval of complex eigenvalues beyond the HB. However, due to the nearby true equilibrium strongly shaping the local dynamics, it is difficult to disentangle the relative contributions of these geometric effects. A detailed characterization of these extra features is provided in Appendix \ref{app:geo}. 

Since there is almost no delay after the trajectory passes the HB point, the baseline level of inhibition in the $\rm M^-$-system, which must be fallen below for a spike to occur, can be approximated by the $q$ value at the HB point.
It is worth noting that, due to the sufficient timescale separation between $q$ and other fast variables in the $\rm M^{-}$-system, a higher buildup of $q$ at the end of each input pulse ($q_{\rm IC}$) leads to a longer time spent along the stable portion of $M_{\rm ss}$ (see Figure \ref{fig:time-to-jump-NoSTOs}, right panel), and thus a longer delay $D$. However, because $q$ begins to decay before the input ends, as we explain next, its overall accumulation during each pulse is still limited. 

% \RED{Eigenvalues at HB of the reduced system: 0.002684  +  i  0.196572; 0.002684  +  i  -0.196572, for full system: 0.000263  +  i  0.044667; 0.000263  +  i  -0.044667 }

%%%%%%%%%%%%%%%%%%%%%%%
%%%%%%%%%%%%%%%%%%%%%%%

% We now examine why $q$ begins to decay midway through the input pulse, thereby facilitating the onset of the next spontaneous spike. 
Figure \ref{fig:bif-q}B shows the time series of the perturbed $V$ (black) and $q$ (green) of the $\rm M^{-}$-system under periodic input pulses at frequency $0.7\,\rm Hz$ over one cycle. The projection of the perturbed solution onto the bifurcation diagram in $(q,V)$ space is shown in Figure \ref{fig:bif-q}D. 
As the input pulse begins at the yellow star, the bifurcation diagram in panel (D) moves rightward from the red curve to the magenta curve. The trajectory which is now further away from the perturbed HB bifurcation is able to oscillate three times between the magenta periodic orbit branch until passing the HB bifurcation and getting attracted by the stable branch of $M_{ss}$ (solid magenta curve). While the number of spikes during the input is the same as in the full model, the inter-spike interval within the induced burst are much shorter in the M$^-$-system due to the lack of the subthreshold resonance generated by the interaction of $I_m$ and $I_{\rm NaP}$ (see Figure \ref{fig:delay-full-M-}A). This leads to a rapid accumulation of $q$ to its maximum at $0.68$ during the early stage of the input perturbation, after which $q$ begins to decay even though the input is still on. By the time the input ends at the green star, $q$ has decayed to about $0.59$ and the bifurcation diagram returns to the red curve. Consequently, the trajectory jumps to the stable branch of the red $M_{ss}$ and travels along it until decaying to the original baseline level near the red diamond, after which an intrinsic spike occurs before the next input pulse arrives. 

% Similarly as discussed before, the reason why the $\rm M^-$-system is unable to phase-lock to the $0.7$ Hz inputs is that the delay $D$ is shorter than the input period $1/f$ where $f$ denotes the input frequency.

%%%%%%%%%%%%%%%%%%%%%%%
\begin{figure}[!htp]
\begin{center}
% \subfigimg[width=0.9\linewidth]{\bf{\small{(A)}}}{f_vs_tilde_tauD_KSSsystem.eps}\\\vspace{.1in}
% \subfigimg[width=0.9\linewidth]{\bf{\small{(B)}}}{f_vs_tilde_tauD_fullsystem_I_8.eps}
\begin{tabular}{@{}p{0.45\linewidth}@{\quad}p{0.45\linewidth}@{}}
\subfigimg[width=\linewidth]{\bf{\small{(A)}}}{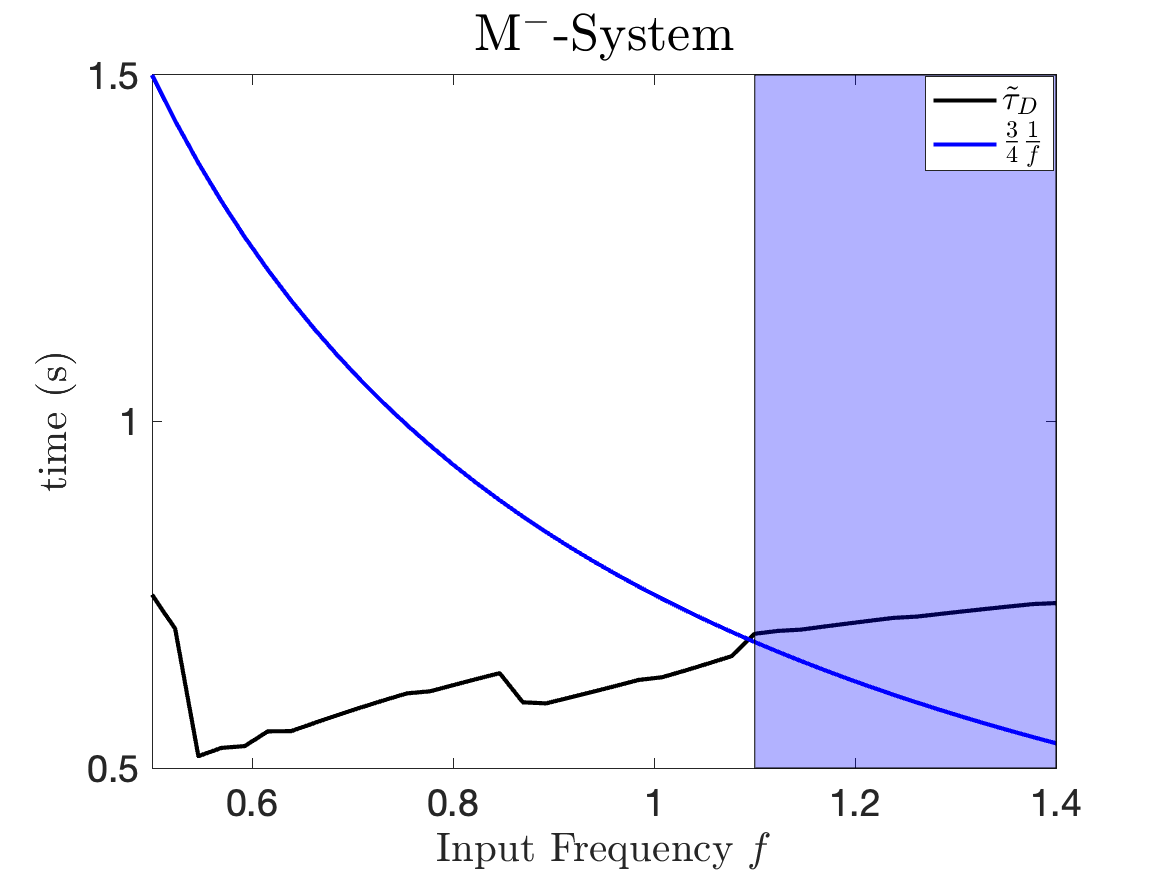} &
\subfigimg[width=\linewidth]{\bf{\small{(B)}}}{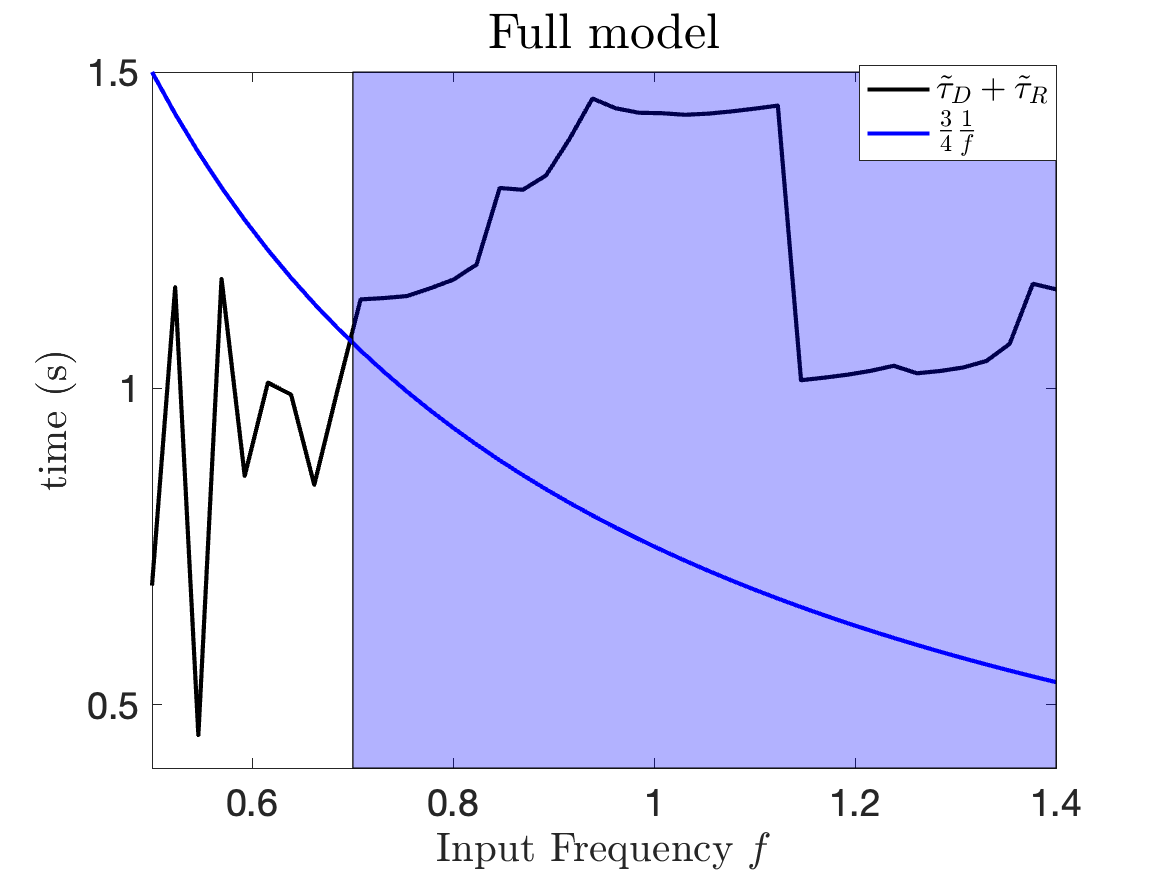}
\end{tabular}
\end{center}
\caption{\label{fig:Kss_system_f_vs_qmaxs} The relationship between input frequency $f$ that is slower than the oscillator's intrinsic frequency at $1.4\,\rm Hz$ and (A) $\tilde{\tau}_D$ for the $\rm M^-$-system and (B) $\tilde{\tau}_R+\tilde{\tau}_D$ for the full model with $I_{\rm app}=8$. Both systems fail to phase-lock to slower inputs when the black curve drops below the blue curve ($\frac{3}{4}\frac{1}{f}$), with the $\rm M^{-}$-system and the full model failing to phase-lock to inputs with frequencies below 1.1 Hz and 0.7 Hz, respectively.}
\end{figure}
%%%%%%%%%%%%%%%%%%%%%%%

In the full system, the post-input delay consists of the input duty cycle, the rising phase during which $q$ continues to increase after the input pulse ends, and the decay phase during which $q$ decreases (Figure \ref{fig:delay-full-M-} upper right):
\begin{equation}\label{eq:est_D_full}
    D^{\rm full}= \frac{1}{4} \frac{1}{f}+\tilde{\tau}_R^{\rm full} + \tilde{\tau}_D^{\rm full}.
\end{equation}
In contrast, the delay in the $\rm M^-$-system includes only two components (Figure \ref{fig:delay-full-M-} lower right):
\begin{equation}\label{eq:est_D_KSS}
    D^{\rm M^-}= \frac{1}{4} \frac{1}{f}+ \tilde{\tau}_D^{\rm M^-}.
\end{equation}
Due to the absence of a rising phase following input termination and the lack of additional delay after passing the HB point, the decay phase $\tilde{\tau}_D$ in the $\rm M^{-}$-system is significantly shorter than the combined delay $\tilde{\tau}_D + \tilde{\tau}_R$ in the full system (compare the black curves in Figure \ref{fig:Kss_system_f_vs_qmaxs}A and B). Consequently, under the same input, $D^{\rm M^-}$ is much smaller than $D^{\rm full}$, resulting in poorer entrainment to slow inputs in the $\rm M^-$-system. 
Specifically, as the input frequency $f$ decreases from the intrinsic frequency of $1.4\,\rm Hz$, the input pulse duration increases, allowing more time for $q$ in the $\rm M^-$-system to decay to lower value at the end of each pulse. This leads to a monotonic decline in $\tilde{\tau}_D$ as $f$ decreases, which soon falls below $\frac{3}{4}\frac{1}{f}$ (see Figure \ref{fig:Kss_system_f_vs_qmaxs}A), causing phase-locking to fail for frequencies below $1.1\,\rm Hz$ (represented by the white region in Figure~\ref{fig:Kss_system_f_vs_qmaxs}A). 
In contrast, in the full system, $\tilde{\tau}_D+\tilde{\tau}_R$ remains large as $f$ decreases and only falls below $\frac{3}{4}\frac{1}{f}$ at much lower frequencies (see Figure \ref{fig:Kss_system_f_vs_qmaxs}B). This allows the full system to phase-lock to much slower inputs compared with the $\rm M^-$-system. 
% Figure \ref{fig:Kss_system_f_vs_qmaxs}A shows that the maximum $\tilde{\tau}_D$ in the $\rm M^{-}$-system occurs when the input frequency $f$ equals to its intrinsic frequency $f_0=1.4\,\rm Hz$, where $q_{\rm IC} = q_{\rm max}$ (see panel A). Specifically, at $f=f_0$, $q_{\rm IC}-q_{\rm HB}\approx 0.3131$, which leads to $\tilde{\tau}_D\approx 0.584\,\rm s$ (see Figure \ref{fig:time-to-jump-NoSTOs}C). This is greater than $\frac{3}{4}\frac{1}{f} =0.536\,\rm s$, leading to 
%  $$D\approx \frac{1}{4}\frac{1}{f} +\tilde{\tau}_D> \frac{1}{f}$$ by \eqref{eq:est_D_KSS}. 
% Hence, the $\rm M^{-}$-system is capable of phase-locking to inputs at its intrinsic frequency. 

In summary, despite better timescale separation than the $\rm K_{SS}^{-}$-system, the $\rm M^-$-system still faces challenges in phase-locking to slower inputs, due to insufficient buildup of inhibition $q$ during each input pulse and a lack of delay in spiking after passing the HB.

\section{Full model with both $\rm K_{SS}$ and $\rm m$ currents exhibits the greatest phase-locking flexibility}\label{sec:both}

In this section, we examine the intrinsic oscillatory dynamics and phase-locking properties of the full model, which includes both the slow $m$-current and the superslow $\rm K_{ss}$-current and exhibits an intrinsic frequency of 7 Hz at $I_{\rm app}=9.8$. We demonstrate that a delayed Hopf bifurcation (DHB) plays a critical role in enabling the flexible phase locking to slow inputs and discuss the specific roles played by each current. A key difference between the full system and the previous reduced models is that the full system exhibits three distinct timescales, giving rise to both a critical manifold $M_s$ and a superslow manifold $M_{ss}$, which together enable an synergistic interaction between $m$- and $\rm K_{ss}$-currents. 

%%%%%%%%%%%%%%%%%%%%%%%
\begin{figure}[!htp]
	\begin{center}
	\begin{tabular}{@{}p{0.4\linewidth}@{\quad}p{0.4\linewidth}@{}}
	\subfigimg[width=\linewidth]{\bf{\small{(A)}}}{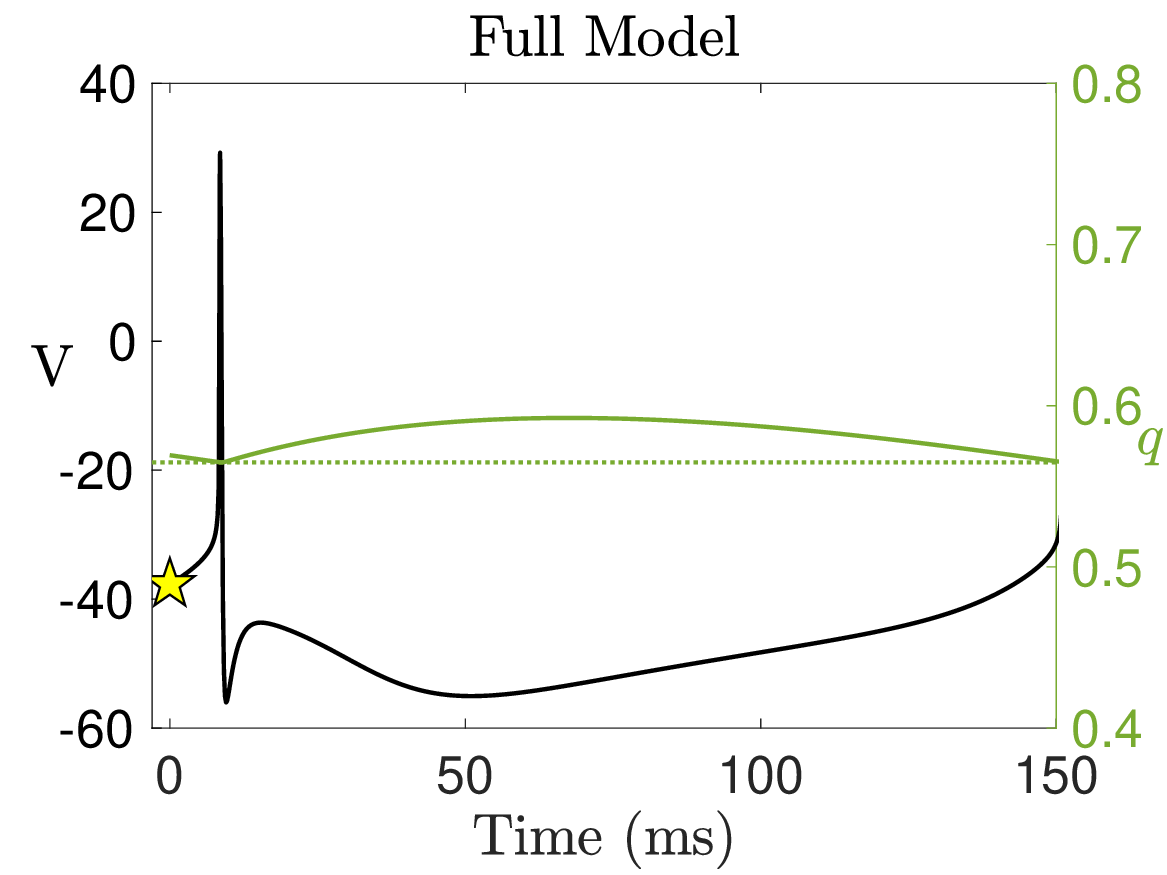}&
	\subfigimg[width=\linewidth]{\bf{\small{(B)}}}{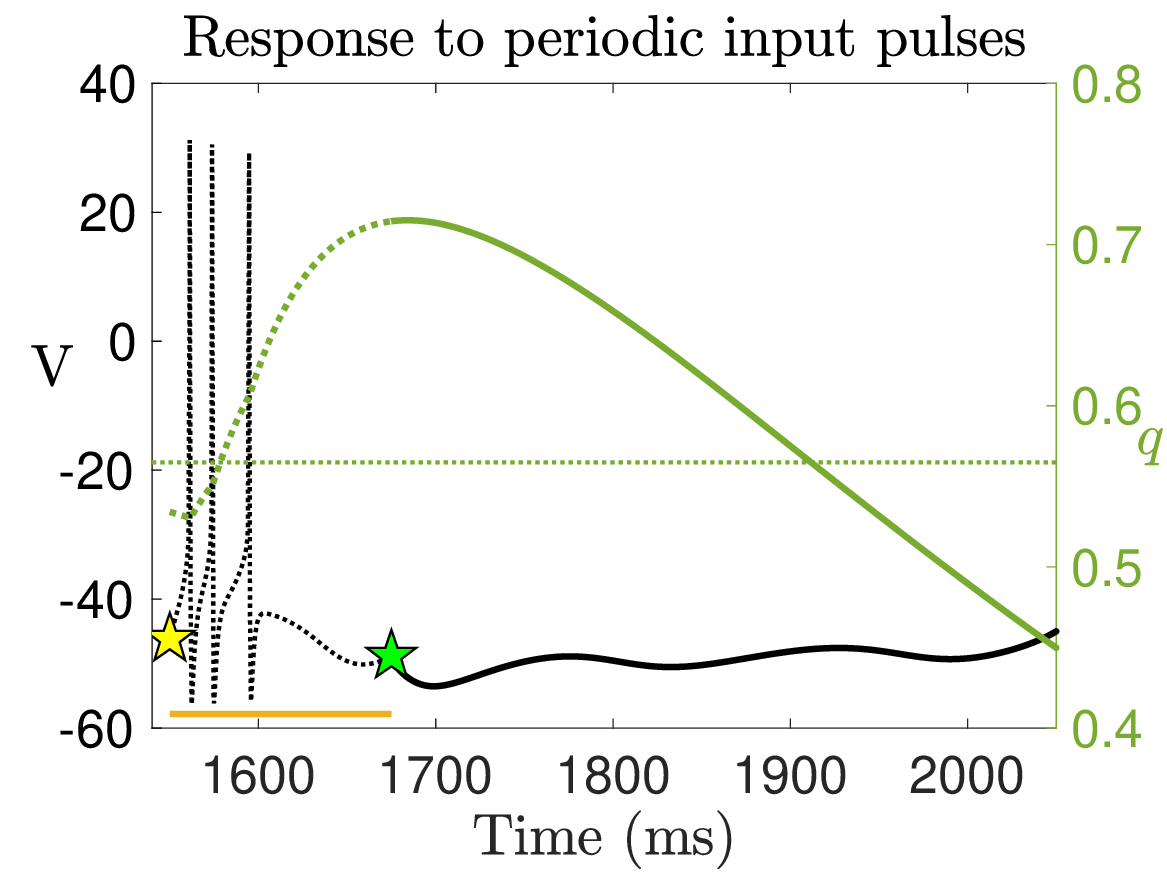}\\
	\subfigimg[width=\linewidth]{\bf{\small{(C)}}}{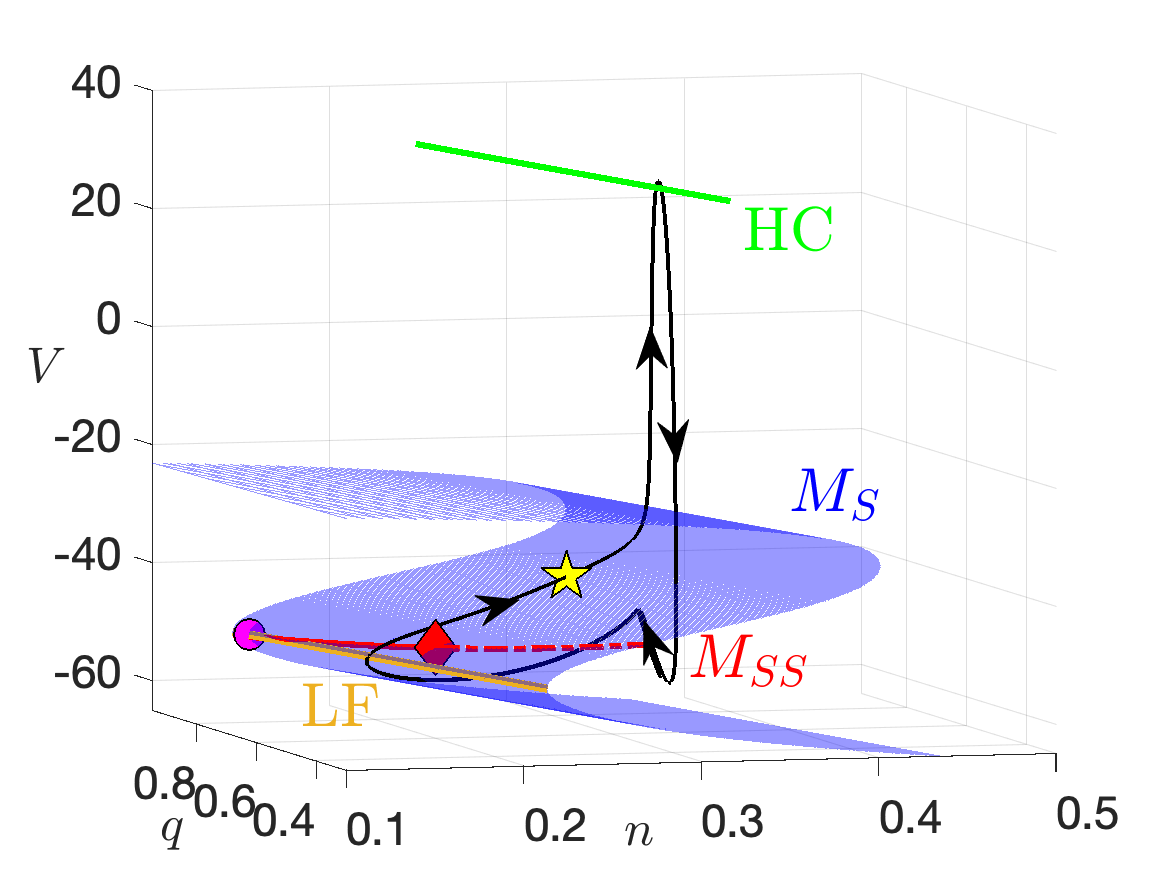}&
	\subfigimg[width=\linewidth]{\bf{\small{(D)}}}{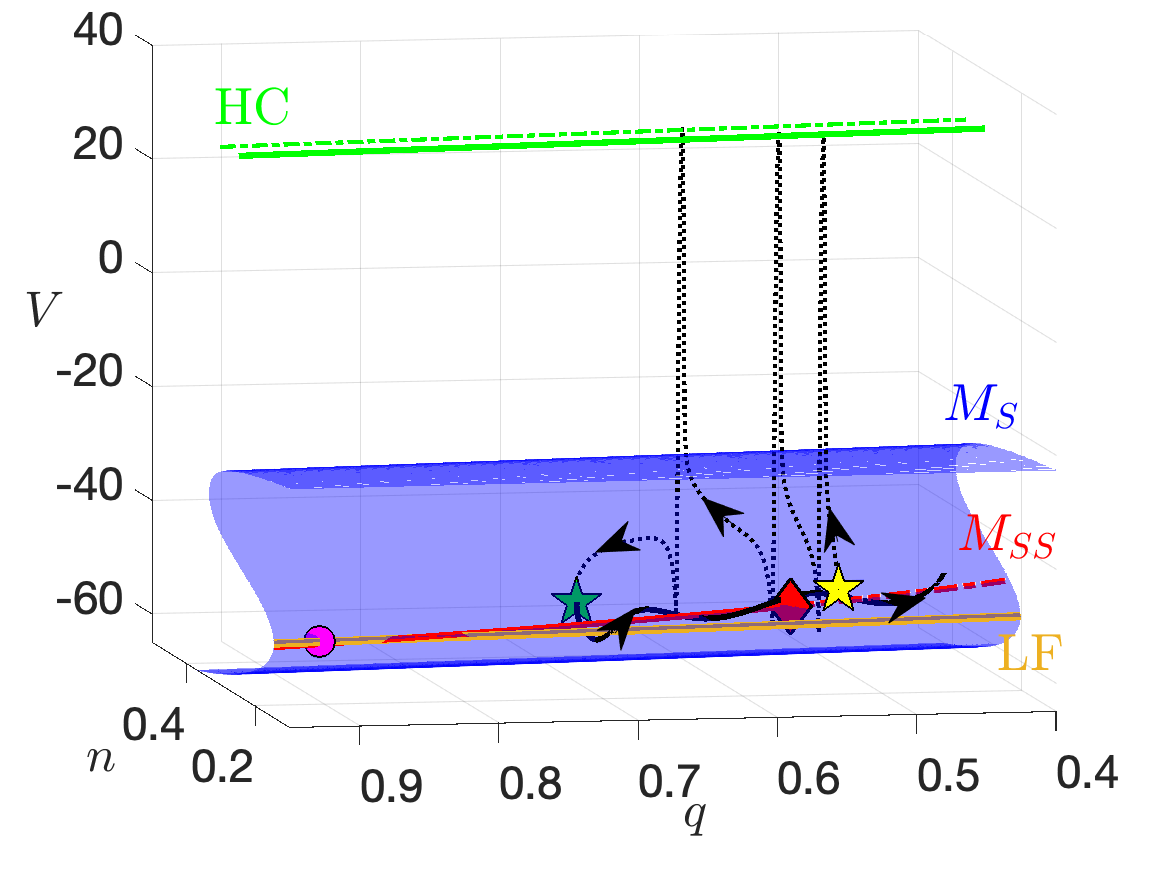}\\
	\subfigimg[width=\linewidth]{\bf{\small{(E)}}}{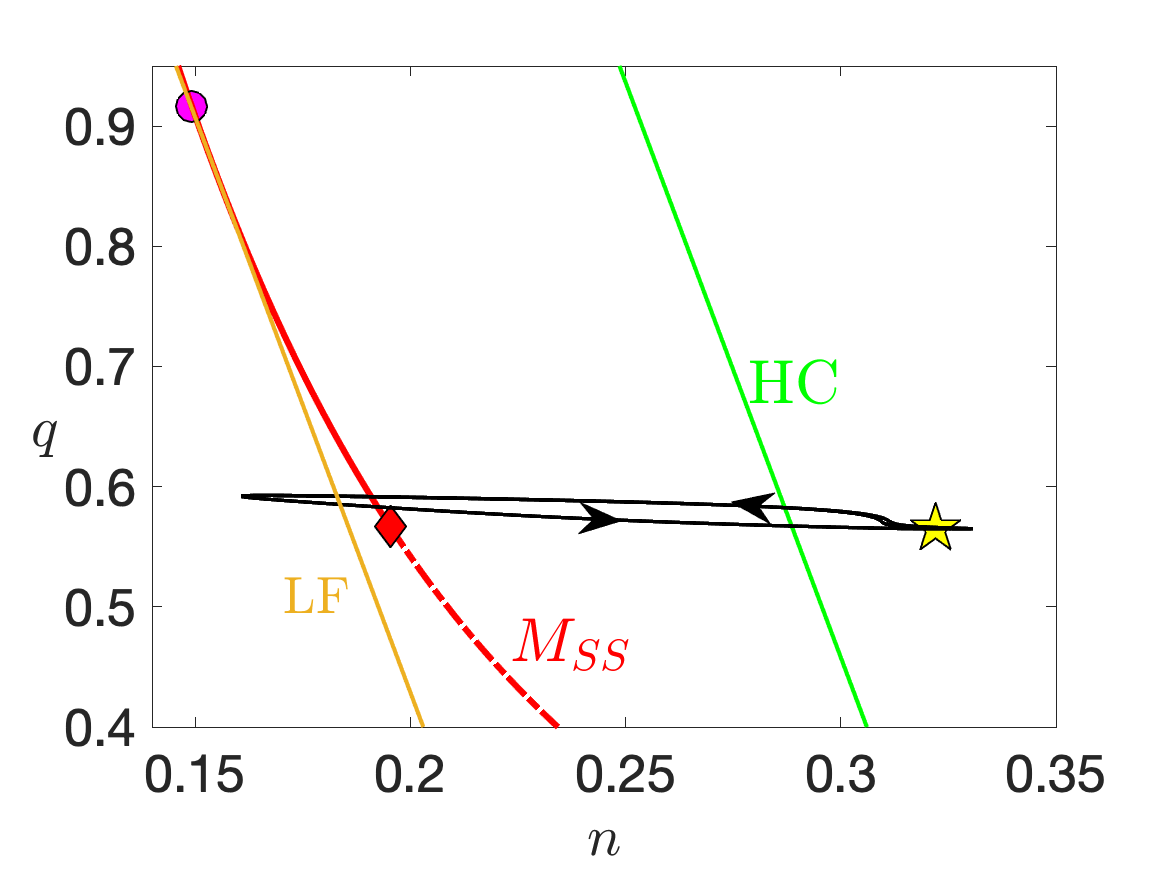}&
	\subfigimg[width=\linewidth]{\bf{\small{(F)}}}{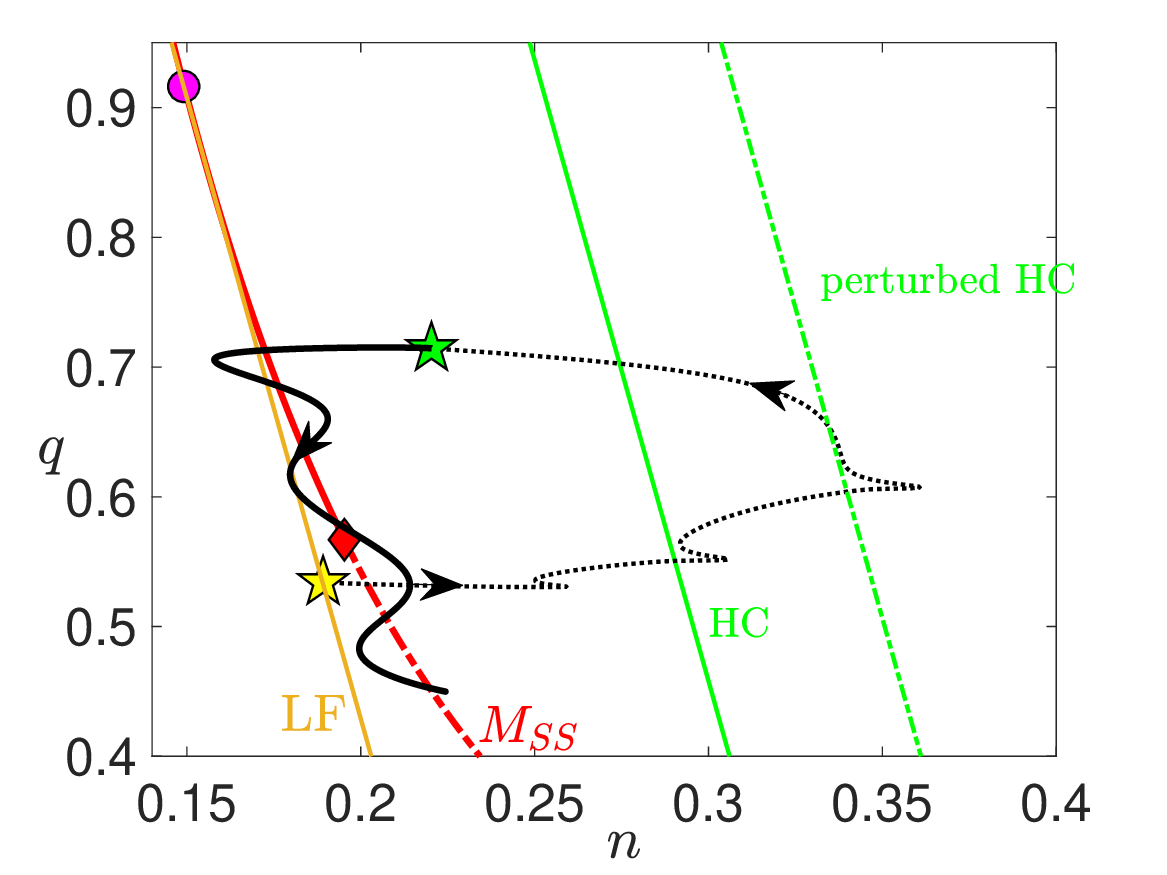}
	\end{tabular}
	\end{center}
\caption{\label{fig:ST-response-largeCm-timeseries}  Simulation of the full theta oscillator model and its response to periodic input pulses at 2 Hz, for $I_{\rm app}=9.8$ and other parameters as given in Table \ref{tab:para}. (A) One cycle of temporal evolution of $V$.
% , as shown in Figure \ref{fig:ST-response-M}A, top panel. 
(B) One cycle of temporal evolution of $V$ in response to periodic input pulses with input frequency $2$ Hz (see Figure \ref{fig:ST-response-M}, the bottom panel). The input pulse parameters are the same as in Figure \ref{fig:ST-response-M}A. Color coding and symbols have the same meanings as in Figure \ref{fig:bif-n}. (C) Projection onto $(n, q, V)$-space of the solution (black) from (A), the critical manifold ($M_s$, blue surface), the superslow manifold ($M_{ss}$, red), the curve of the homoclinic bifurcation (HC, green) and the lower fold curve (LF, blue). (D) Projection of the perturbed solution from (B) onto $(n, q_{\rm K_{SS}}, V)$-space. (E) Projection of all the curves from (A) onto $(n, q_{\rm KCa})$- space. (F) Projection of all curves from (D) onto the $(n, q_{\rm KCa})$-space, together with the HC bifurcation cure (green dashed) when $I_{\rm app}$ is increased to $12.4667$ during an input pulse.}
\end{figure}
%%%%%%%%%%%%%%%%%%%%%%% 

Figure \ref{fig:ST-response-largeCm-timeseries}A shows the unperturbed $V$ and $q$ traces over a full cycle of the system. Figure \ref{fig:ST-response-largeCm-timeseries}C and E display the corresponding trajectory (black curve) projected onto $(V,n,q)$- and $(n,q)$-space. Also shown are the projections of the critical manifold $M_s$ (blue surface), the superslow manifold $M_{ss}$ (red curve), the lower fold (LF) curve of $M_s$ (yellow curve) and the homoclinic (HC) bifurcation curve of the fast layer problem (green curve), which terminates the periodic orbit surface that originates at a HB curve on the upper branch of $M_s$ (not shown). With respect to the fast layer problem, the upper and lower sheets of $M_s$ are stable while the middle sheet is unstable. On the fold curve, there is a folded node (magenta dot), a special point that can allow trajectories to cross the fold with nonzero speed from the attracting branch to the repelling branch \cite{Desroches2012,PW2024}. The red diamond on the $M_{ss}$ represents the Hopf bifurcation of the fast-slow subsystem (slow layer problem) at which $M_{ss}$ changes its stability. This subsystem HB bifurcation is also known as delayed Hopf bifurcation (DHB). A full-system true equilibrium exists at $q=0.13385$, which is relatively far away from the HB point and not shown here as it falls outside the displayed parameter range. 

The mechanism underlying the full dynamics is similar to that of the $\rm K_{SS}^{-}$-system (Figure \ref{fig:bif-n}C): spiking initiates at the lower fold (LF) curve of the critical manifold and terminates after crossing the HC bifurcation curve (Figure \ref{fig:ST-response-largeCm-timeseries}C and E). The reason why the trajectory fails to closely follow $M_s$ is due to a lack of sufficient timescale separation between the fast and slow variables, similar to what was observed in the $\rm K_{SS}^{-}$-system (see Figure \ref{fig:bif-n}C). Although the superslow current $I_{\rm K_{SS}}$ is present, the associated superslow manifold $M_{ss}$ and the DHB do not appear to influence the intrinsic dynamics. 
% This again is due the fact that the trajectory does not remain near $M_s$, and thus never approaches $M_{ss}$. 
% This raises a question: if the intrinsic dynamics and underlying mechanisms are so similar, why does the full system not display similar phase-locking behaviors as the $\rm K_{SS}^{-}$-system (see Figure~\ref{fig:ST-response-M}A and B)?  
Nonetheless, they play a significant role in generating long post-input spiking delay $D$ in response to periodic input pulses, thereby supporting phase locking of the full oscillator model to much slower inputs.

% To answer this question, we examine the perturbed dynamics under periodic input forcing and examine the roles of $M_{s}$ and $M_{ss}$ in determining the post-input spiking delay $D$. Interestingly, while $M_{ss}$ does not influence the intrinsic dynamics, it plays a crucial role in generating a long delay $D$ in response to periodic input pulses, thereby facilitating its phase-locking flexibility. 
Figure \ref{fig:ST-response-largeCm-timeseries}B shows the time series of the perturbed $V$ (black) and $q$ (green) of the full system subject to periodic input pulses at frequency $2\,\rm Hz$, over one input cycle (also see Figure \ref{fig:ST-response-M}A, last row). From these plots, we can see that the delay $D$ is long enough to delay further spiking until the next input arrives. The projections of the perturbed solution onto the $(V,n,q)$-space and $(n,q)$-space are shown in Figure \ref{fig:ST-response-largeCm-timeseries}D and F. Below, we use these two projections to explain how $M_{ss}$ and the DHB enable the full system to generate a post-input spiking delay $D$ that is significantly longer than its intrinsic period.  
% why the full system has a much longer post-input spiking delay $D$ compared with the M- and $\rm M^{-}$-systems. 

%%%%%%%%%%%%%%%%%%%%%%%
\begin{figure}[!t]
\begin{center}
 \includegraphics[width=\linewidth]{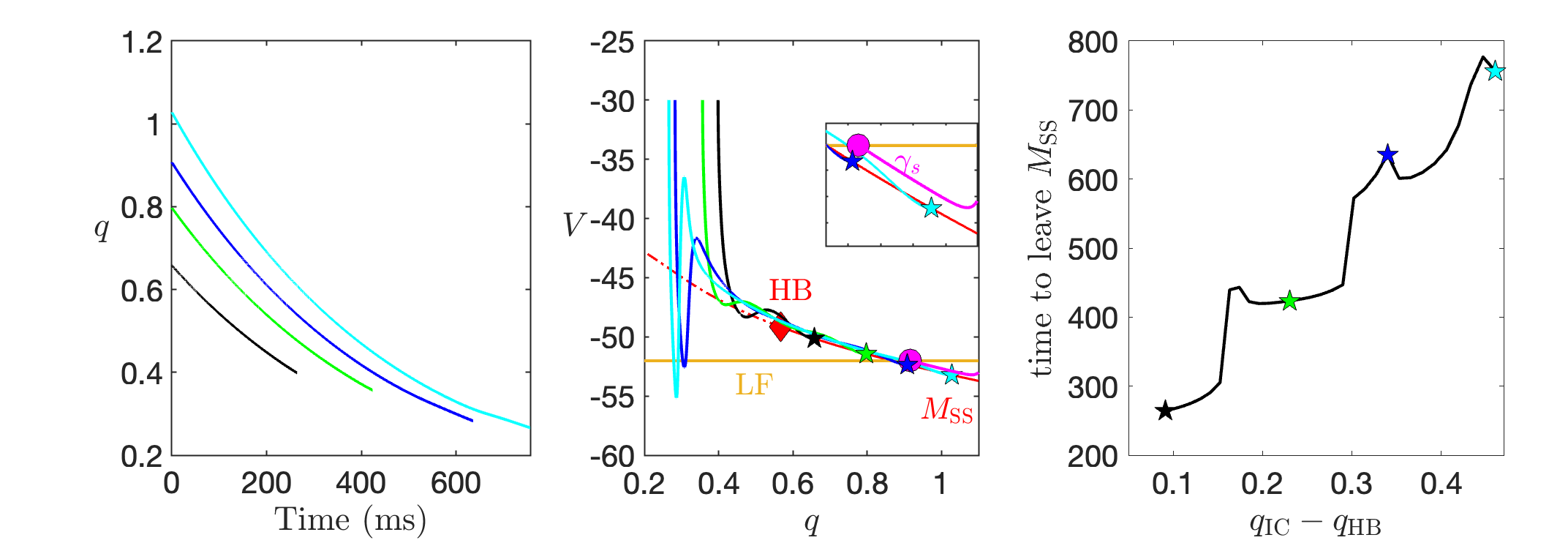}
\includegraphics[width=3in]{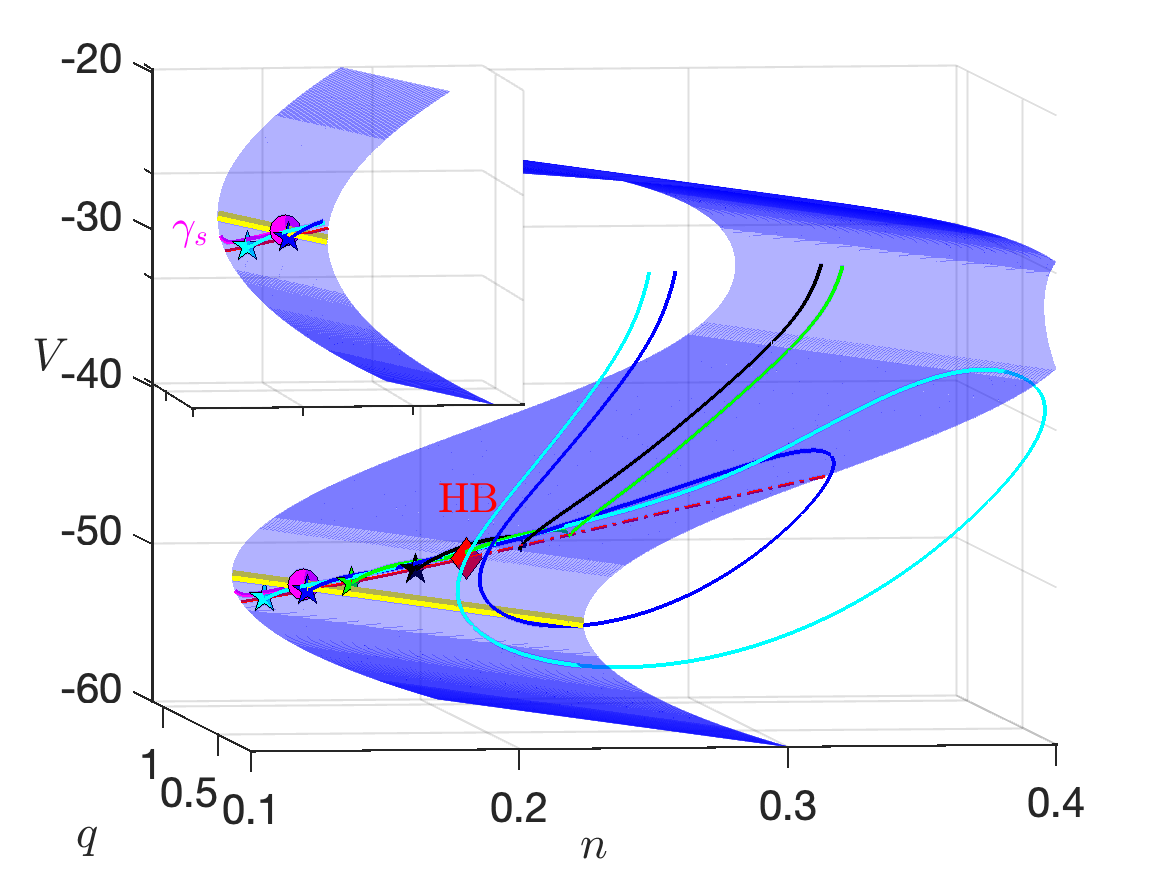}
\end{center}
\caption{ \label{fig:time-to-jump-FullModel} Local oscillations and bifurcation delay in the full system with default parameters from Table \ref{tab:para}. The top panel follows the same layout as Figure \ref{fig:time-to-jump-NoSTOs} and additionally shows the LF of $\ms$ (yellow curve), the folded node (magenta dot) and the corresponding singular strong canard $\gamma_s$ (magenta curve). The lower panel shows the projection of the trajectories from the top panel onto $(q,n,V)$-space, together with the critical manifold $\ms$ (blue surface), LF curve, $\gamma_s$ and HB point (red diamond).}
\end{figure}
%%%%%%%%%%%%%%%%%%%%%%%

As the input pulse begins at the yellow star, the elevated $I_{\rm app}$ shifts the HC curve to higher $n$ values (compare the solid and dashed green lines in panels D and F), generating additional spikes through the same mechanism described in the $\rm K_{SS}^{-}$-system. During this extended spiking phase, $q$ accumulates to a level much higher than in the absence of any input. Unlike in the $\rm M^{-}$-system where $q$ decays shortly after spiking, here $q$ continues to rise throughout the input pulse.
With increased $q$ at the end of the input (green star), the trajectory is drawn closer to the attracting side of $M_{ss}$ (red solid line in panels D and F, denoted as $\mss^a$), oscillates around $\mss^a$, passes over the DHB (red diamond) to the repelling side $\mss^r$ (red dashed lines), and then experiences a delay as it spirals outward along $\mss^r$ before jumping away, either spontaneously or upon the next input. This behavior reflects the way-in/way-out phenomenon associated with a DHB, in which the amount of delay along $\mss^r$ depends on the trajectory's entry location on $\mss^a$.

To further clarify the delay mechanism, we compare solutions of the full system initialized at different distances from the DHB along $\mss^a$ (see Figure~\ref{fig:time-to-jump-FullModel}), as in the M$^-$ system analysis. The upper left panel shows the $q$-traces over time, the upper middle panel shows their projections in $(q,V)$-space with initial conditions marked by stars, and the upper right panel shows the time required for each trajectory to travel from its initial condition $q_{\rm IC}$ to jump away from $M_{\rm ss}$, defined as crossing $V=-30$ from below. As expected from a standard DHB, trajectories initiated farther along $\mss^a$ from the DHB point experience stronger rotational attraction and exhibit more pronounced bifurcation delays \cite{Baer1989,Neishtadt1987,Neishtadt1988}.
% Such delays are absent in the $\rm K_{SS}^{-}$-system, which lacks both a superslow manifold $\mss$ and a DHB; consequently, spontaneous spiking resumes upon crossing the lower fold (LF) of the critical manifold $M_s$. In the $\rm M^{-}$-system, although a dynamic HB point is present, it does not produce a delay mainly due to a nearby true equilibrium as discussed before. 

Another potential source of delay along the repelling branch of $\ms$ is the canard mechanism due to folded node singularities (see Figures \ref{fig:ST-response-largeCm-timeseries} and \ref{fig:time-to-jump-FullModel}, magenta circle). Trajectories that land inside the trapping region (i.e., the funnel) on $\ms$, formed by the yellow fold curve and the singular strong canard $\gamma_s$ (Figure \ref{fig:time-to-jump-FullModel}, magenta curve), will converge to the folded node, thereby passing through the fold from the attracting branch to the middle repelling branch. Such solutions are so-called singular canards. The strong canard $\gamma_s$ is a special solution that locally separate those twisting around the folded node from those that do not \cite{Desroches2012}. 
As shown in Figure \ref{fig:time-to-jump-FullModel}, this funnel region is narrow, and none of the perturbed trajectories fall within it. Thus, the small-amplitude oscillations (SAOs) and the associated spiking delays are not related to the canard mechanism but arise solely from the delayed Hopf mechanism.

\section{Discussion}\label{sec:discussion}

In this paper, we investigate the phase-locking properties of a single compartment biophysical model of cortical theta oscillators under strong periodic forcing. While it is straightforward to see why a faster strong inputs could force a slower oscillator, it is not obvious how a slower forcing input can pace a faster oscillator. We therefore focus our analysis on inputs slower than the oscillator's natural frequency and examined how they enable such entrainment. The theta oscillator model includes both a theta-timescale (slow) m-current ($I_m$) and a delta-timescale (superslow) potassium current ($I_{\rm K_{SS}}$), and is forced by strong periodic inputs over a broad range of frequencies. Previous work \cite{pittman21} demonstrated that the interaction of these two intrinsic inhibitory currents on distinct timescales supports remarkably flexible phase-locking to slow inputs, enabling entrainment over a substantially broader frequency range than in models lacking either current. However, the mechanisms responsible for this flexibility have not yet been fully characterized. 

Using dynamical systems methods, including the geometric singularity perturbation theory (GSPT) and bifurcation analysis, we show that the flexible phase-locking to slow inputs in the theta oscillator model arises from a genuine multiscale synergy between its intrinsic currents $I_m$ and $I_{\rm K_{SS}}$, rather than from either current acting alone or from the sum of their separate effects. Geometrically, the slow current organizes the intrinsic dynamics around a two-dimensional critical manifold $\ms$, while the superslow current induces a one-dimensional superslow manifold $\mss\subset \ms$ that is not accessed by the intrinsic dynamics but becomes essential for entrainment to strong slow forcing. Notably, the roles of the two currents are interdependent. With only the slow current, $\ms$ exists but the absence of $\mss$ limits delays to the slow theta timescale. With only the superslow ${\rm K_{SS}}$-current, $\mss$ is present, but without the current on the slow timescale, the buildup of inhibition during forcing is only modest and the dynamic Hopf delay is lost. Thus, the presence of $I_m$ does not just add an extra delay through SAO generation, it also enables the delayed Hopf mechanism to be fully engaged. Under strong forcing, the combined action of superslow drift along $\mss$, delayed loss of $\mss$ stability via a dynamic Hopf bifurcation and the slow-timescale subthreshold oscillations substantially postpones the next spontaneous spike beyond the natural interspike interval, thereby enabling entrainment to inputs well below the oscillator's intrinsic frequency.

To investigate the role of the superslow $I_{\rm K_{SS}}$ in phase-locking, we constructed the $\rm K_{SS}^-$ system obtained by removing $I_{\rm K_{SS}}$ from the full model. Figure \ref{fig:ST-response-M} shows that removing $I_{\rm K_{SS}}$ significantly reduced the model's ability to phase-lock to inputs slower than their intrinsic frequency. In particular, the $\rm K_{SS}^-$-system can hardly entrain any slower inputs. To understand this behavior, we apply GSPT to the $\rm K_{SS}^-$-system and identify the lower fold (LF) of the critical manifold $\ms$ as the spike-initiation threshold: a spike can begin only after the $I_m$ inhibition falls below the LF level.
Paradoxically, although input-triggered spikes produce a larger $I_m$ inhibition than autonomous spikes, the trajectory decay to cross the fold LF even more quickly. As a result, the spontaneous spike following an input occurs sooner than the natural interspike interval, preventing the cell from following a rhythm slower than its natural frequency. 
Our analysis shows that this paradoxical effect is mainly due to the limited separation between fast and slow timescales. Following input-triggered spikes, trajectories with elevated inhibition (stars with higher $n$ values in Figure \ref{fig:time-to-fold-NoKCa}B) lie further from the slow $n$-nullcline. Because the fast variables do not relax fast enough to the critical manifold $\ms$, their dynamics, together with a faster decay rate of $n$ when it is further from its nullcline, pull the trajectory toward the fold of $\ms$ more rapidly. In contrast, when the fast-slow timescale separation is increased by slowing down the m-current (i.e., closer to the singular limit), elevated inhibition following an input prolongs the drift along $\ms$, producing a post-input delay longer than the natural interval (see Figure \ref{fig:time-to-fold-NoKCa}E and F).

% It was conjectured in \cite{pittman21} that flexible phase-locking in the theta oscillator model also depends critically on the subthreshold oscillations (SAOs) arising from the interplay of m- and persistent sodium currents. 
In the M$^-$-system, increased inhibition following a strong input pulse prolongs the drift along the attracting side of $\mss$, thereby extending the post-input delay. This behavior contrasts with the $\rm K_{SS}^-$-system, where greater inhibition shortens the interspike interval.
As a result, the M$^-$-system takes longer than its natural interspike interval to generate the next spike, allowing it to be able to follow some rhythms slower than its natural frequency. 
However, the overall delay remains limited for two reasons. First, the inhibition accumulates only modestly during an input due to the absence of subthreshold oscillations (SAOs) on the slow timescale. 
Second, the M$^-$-system does not display the pronounced spike-onset delay typically associated with a dynamic Hopf bifurcation \cite{Baer1989,Neishtadt1987,Neishtadt1988,engler2025delays}. There observations are consistent with previous studies showing the importance of the m-current in promoting SAOs and Hopf-mediated (Type II) excitability \cite{ermentrout2001effects,acker2003synchronization,pittman21,rotstein2014frequency,gutfreund1995}. Importantly, however, in our model the loss of delay does not arise from a transition to Type I excitability. Although removing the m-current shifts the system closer to the Type I boundary, as reflected in a nearly folded equilibrium branch, oscillation onset remains Hopf-mediated. Our analysis indicates that this suppression of delay is primarily associated with a nearby saddle-focus equilibrium located just beyond the Hopf point. Prior studies \cite{Vo2014,PW2024} have also shown that delayed Hopf dynamics can be significantly altered when the Hopf bifurcation lies close to a fold or Bogdanov-Takens bifurcation. In the M$^-$-system, although no true fold is present, the superslow manifold exhibits a near-fold structure in the HB region and lies close to a codimension-two cusp bifurcation organizing the fold branches, which may further influence the local spectral properties and delay behavior (Appendix \ref{app:geo}, Fig.~\ref{fig:bif-q-Iapp}). A more systematic investigation of how fold–cusp structure affects dynamic bifurcation delay represents an interesting direction for future work.

% Similar loss of delay has been reported in other systems where the HB lies close to an actual fold point (e.g., \cite{Vo2014,PW2024}). Our results suggest that this near-fold geometry of $\mss$ alone, even without an actual fold, may also strongly influence dynamics by suppressing the usual bifurcation delay associated with the HB. A more detailed characterization of this mechanism, the role of SAOs and its relation to cusp unfoldings represent an interesting direction for future work.  

Our analysis of the full system shows that flexible phase-locking emerges from a multiple-timescale synergy between the slow $I_m$ and superslow $I_{\rm K_{ss}}$ that cannot be reduced to the sum of their separate effects. 
In the absence of forcing, spiking in the full system is paced by the slow theta timescale $I_m$ with trajectories following $\ms$, while $\mss$ and superslow dynamics are not involved (Figure \ref{fig:ST-response-largeCm-timeseries}A,C,E). Nonetheless, periodic forcing can engage both currents, creating long delays along $\mss$ and substantially expanding the entrainment frequency range (Figure \ref{fig:ST-response-largeCm-timeseries}B,D,F). Specifically, $I_m$-mediated subthreshold oscillations space spikes during an input more widely, providing the temporal window necessary for $I_{\rm K_{SS}}$ to accumulate effectively throughout the input phase. In turn, the buildup of $I_{\rm K_{SS}}$ brings the trajectory closer to $\mss$ and causes it to land farther from the delayed Hopf bifurcation (DHB) point. The trajectory then undergoes a prolonged superslow drift along $\mss$, during which small-amplitude oscillations are observed. The combination of this superslow drift and the delayed loss of stability via a standard way-in/way-out mechanism at a DHB on $\mss$ significantly postpones the next spike beyond the natural interspike interval. 

Interestingly, although SAOs are absent in the intrinsic dynamics since $\mss$ and the DHB are not engaged without input, they emerge under forcing and contribute critically to flexible phase-locking. Thus, it is not the SAOs themselves, but the underlying delayed Hopf-mediated mechanism they reflect, that enables flexible entrainment to slower inputs. This perspective is consistent with the observation that synaptic inhibition-based theta rhythms exhibit inflexible phase-locking to input frequencies lower than their intrinsic frequency \cite{cannon2015leaky,sherfey2018flexible,pittman21}. Even when synaptic currents operate on timescales comparable to $I_m$, such systems do not exhibit SAO-associated dynamics under forcing and do not display the same degree of phase-locking flexibility. Our results thus highlight that flexibility depends not simply on the existence of multiple timescales, but on synergistic interactions among currents operating across those timescales. While the present work focuses on oscillators paced by intrinsic inhibitory currents, extending this multiple timescale analysis to oscillators incorporating both intrinsic and synaptic inhibitory currents \cite{pittman21} represents an interesting direction for future work. 
% Notably, the roles of $\ms$ and $\mss$ are interdependent. With only the slow $I_m$ current, $\ms$ exists but the absence of $\mss$ limits delays to the slow timescale. With only the superslow $I_{\rm K_{SS}}$ current, $\mss$ is present, but without the SAO mechanism on the slow timescale, buildup of inhibition is only modest and the dynamic Hopf delay is truncated by the near-fold geometry of $\mss$. Thus, the presence of $I_m$ does not just add a delay by creating small-amplitude oscillations, it also creates the right geometry for the delayed Hopf mechanism to be actually engaged. 

To analyze phase locking behavior of the theta oscillator, we focused on a geometric singular perturbation theory approach. A widely used alternative mathematical framework for studying phase locking in forced oscillators is the phase response curve (PRC), which has been developed for week forcing \cite{ermentrout1981n,ermentrout1996type,kopell2002mechanisms,schwemmer2012theory,park2017utility,wang2021shape,ermentrout2010mathematical,malerba2013phase} and extended for strong but pulsatile forcing \cite{cui2009functional,wedgwood2013phase,cannon2015leaky,castejon2013phase}. Our results lie in a dynamical regime in which the PRC theory does not directly apply, since our forcing is both strong and long-lasting such that the oscillator completes multiple cycles (i.e., bursting) during the input pulse. As a result, the phase at the end of forcing is not uniquely determined by the phase at which the forcing began. Moreover, the presence of multiple (more than two) interacting timescales makes it difficult to characterize the system state in terms of a single phase variable, since both phase and amplitude influence its dynamics. 
Recent advances in augmented phase-amplitude reduction techniques, which incorporate amplitude dynamics through isostable coordinates, extend phase-based description beyond weak inputs \cite{perez2020global,wilson2019augmented,wilson2020phase,wilson2016isostable} and to bursting dynamics \cite{wilson2025phase}, suggesting promising avenues for progress. Extending these approaches to systems studied here, which exhibits multiple interacting timescales and bursting oscillations under strong and long-lasting forcing, represents an exciting direction for future study. 
% In particular, to study properties of synaptic inhibition based rhythms under pulsatile periodic strong forcing, \cite{cannon2015leaky} used a singular phase response curve (sPRC) to characterize the response of synaptic inhibition-based oscillators to strong forcing pulse in the singular limit of well-separated fast and slow timescales, and proved that the shape of the sPRC persists away from the singular limit. It remains an open and interesting question whether this framework can be extended to capture the dynamics of oscillators with three timescales and the presence of mixed-mode oscillations, such as those observed in our model. 

Our findings echo the theme of degeneracy observed widely in biology across multiple levels of organization, where distinct parameter combinations can give rise to virtually identical baseline activity \cite{edelman2001degeneracy,prinz2004similar,goaillard2021ion,albantakis2024brain}. In many examples of degeneracy, such equivalence can mask substantial differences in the underlying mechanism. Interestingly, our results show that this degeneracy does not necessarily extend to functional flexibility under external forcing: while similar baseline dynamics can be supported by different parameter sets, the most flexible entrainment arises only when coordinated interactions of intrinsic currents across multiple timescales are present, even if some components do not participate when no forcing is applied. 
From a broader perspective, this highlights the important functional role of multiple-timescale interactions of currents in neural oscillators. Rather than serving as redundant pathways for slowing, the intermediate slow current further extends the delay by engaging the superslow manifold in a delayed Hopf regime, thereby supporting flexible phase-locking over a wide frequency range. While degeneracy can support robustness of baseline activity, interactions among mechanisms acting on multiple timescales may be especially important in cortical theta oscillators that must flexibly respond to perturbations, where speech segmentation requires entrainment to inputs that vary substantially in timescale \cite{ghitza2011linking,ohala1975temporal,greenberg1999speaking,chandrasekaran2009natural,ding2017temporal}. 
% This observation suggests that multi-timescale coordination may represent a general design principle for flexible rhythmic processing in the brain. 

% In a similar spirit, our results show that coordinated interactions among currents on different timescales reveal dynamical structure that support flexible entrainment to external inputs, even if some components do not participate when no forcing is present.  Our analysis provides a mechanistic explanation for how such robustness can emerge, through multi-timescale interactions that reshape the underlying invariant manifold geometry to enable a wide range of functional responses.

\section*{Acknowledgement}
The authors thank Nancy Kopell for helpful discussions and critical reading of the manuscript. YW acknowledges support from NIH/NIDA R01DA057767.
% Portions of original manuscript text were submitted to an AI large language model to improve readability and grammar. After review and revision, some of these passages were included. 

\section*{Conflict of Interest}
The authors have no conflicts to disclose.
% This could be because the Hopf bifurcation is close to where an actual fold in $\mss$ might develop, i.e., close to a double zero eigenvalue at a Bogdanov-Takens (BT) bifurcation in the fast $\rm M^-$-subsystem (see e.g., \cite{Baer1989,PW2024}).

% \section*{Data Availability}
% Data sharing is not applicable to this article as no new data were created or analyzed in this study.

% %%%%%%%%%%%%%%%%%%%%%%%
% \begin{figure}[!htp]
% \begin{center}
% \includegraphics[width=7in]{FIGS/time_to_jump_timescaleKCa_5_FullModel_Iapp_9p8.eps}
% \includegraphics[width=4in]{FIGS/time_to_jump_3D_timescaleKCa_5_FullModel_Iapp_9p8.eps}
% \end{center}
% \caption{ \label{fig:time-to-jump-FullModel-qslower} Time for the trajectory to jump is ... when $q$ is made 5 times slower. \RED{Switch green and cyan colors!! And fix code accordingly.}}
% \end{figure}
% %%%%%%%%%%%%%%%%%%%%%%%

% %%%%%%%%%%%%%%%%%%%%%%%
% \begin{figure}[!htp]
% \begin{center}
% \includegraphics[width=3in]{n-v-ss-bifs-Iapp-9p8-Fix-q.eps}
% \includegraphics[width=3in]{n-v-ss-bifs-Iapp-9p8-Fix-q-zoom.eps}
% \end{center}
% \caption{\label{fig:} Initially, bifurcation is at the blue position, with perturbation, it shifts to the magenta curve. As the perturbation turns off at the green star, the bifurcation curve moves leftward to the green position, which slowly moves rightward as $q_{\rm KCa}$ increases on the superslow timescale. Right: enlarged view of the lower left corner of the left figure.}
% \end{figure}

 \appendix

 \section{Nondimensionalization of the theta oscillator model \eqref{eq:full}}\label{app:nondim}

In this section, we nondimensionalize system \eqref{eq:full} in the absence of external periodic inputs $I_{\rm PP}(t)$ to identify the intrinsic timescales of all variables. For convenience, we restate the system here:
\begin{equation}\label{eq:full-app}
\begin{array}{rcl}
     C\frac{dV}{dt}&=&  I_{\rm app}- I_{\rm Na}-I_{\rm K_{DR}}-I_{\rm leak}-I_m -I_{\rm NaP} \\
     && -I_{\rm Ca}-I_{\rm K_{SS}}:=f_1\\
     \frac{dn}{dt}&=& (n_{\infty}(V)-n)/\tau_n(V),\\
     \frac{dm_{\rm NaP}}{dt}&=& (m_{\infty}(V)-m_{\rm NaP})/\tau_{m},\\
     \frac{ds}{dt}&=& (1-s)\alpha_s-s\beta_s,\\
     \frac{dm_{\rm K_{DR}}}{dt}&=& \tau_{\rm fast}((1-m_{\rm K_{DR}})\alpha_{m}-{m_{\rm K_{DR}}}\beta_{m}),\\
     \frac{dh}{dt}&=& \tau_{\rm fast}((1-h)\alpha_h-h\beta_h),\\
     \frac{d\rm Ca_i}{dt}&=& -F_{\rm Ca} I_{\rm Ca}(V)-\rm Ca_i/\tau_{\rm Ca}\\
     \frac{dq}{dt}&=& (1-q)\alpha_q( {\rm Ca_i})-q\beta_q
\end{array}    
\end{equation}
where the steady-state values, time constants, and rate functions for the gating variables and calcium are provided are provided in Table \ref{tab:activation}.

\begin{table}[!htp]
\caption{Activation variable dynamics}
\centering
\begin{tabular}{|c|c|c|}
\hline
$h$ & $\tau_{\rm fast}=5.6115$, $\alpha_h(V)=0.07\exp(-(V+30)/20)$ & $\beta_h(V)=(\exp(-V/10)+1)^{-1}$\\
$m_{\rm Na}$ & $\alpha_m(V)=-\frac{V+16}{10(\exp(-(V+16)/10)-1)}$& $\beta_m(V)=4\exp(-(V+41)/18)$\\
$m_{\rm K_{DR}}$&  $\alpha_m(V)=-0.01\frac{V+20}{\exp(-(V+20)/10)-1}$ & $\beta_m(V)=0.125\exp(-(V+30)/80)$\\
$n$ & $n_{\infty}(V)=(1+\exp(-(V+35)/10))^{-1}$ & $\tau_n(V) = \frac{81.085}{\exp((V+35/40))+\exp(-(V+35)/20)}$\\
$m_{\rm NaP}$& $m_\infty(V)=(1+\exp(-(V+40)/5))^{-1}$& $\tau_m=5\,\rm ms$\\
$s$ & $\alpha_s(V)=1.6(1+\exp(-0.072(V-65)))$ & $\beta_s(V)=0.02\frac{V-51.1}{\exp(\frac{V-51.1}{5})-1}$\\
$q$ & $\alpha_q(\mathrm{Ca_i})=\rm min(0.1\mathrm{Ca_i}, 1)$ & $\beta_q=0.002$\\
$\rm Ca_i$ & $F_{\rm Ca}= 2.2222$ & $\tau_{\rm Ca}=100$\\
\hline
\end{tabular}
\label{tab:activation}
\end{table}

We introduce a dimensionless time variable $t=Q_t \tau$ and rewrite the system in the following equivalent form: 
\begin{equation}\label{eq:nondim}
\begin{array}{rcl}
     \frac{C}{g_{\rm max}Q_t}\frac{dV}{d\tau}&=& f_1(V,n,m_{\rm NaP},s,m_{\rm K_{DR}},h,q)\\
     \frac{dn}{d\tau}&=&Q_tT_n \frac{n_{\infty}(V)-n}{\bar{\tau}_n(V)},\\
     \frac{dm_{\rm NaP}}{d\tau}&=& Q_t\frac{1}{\tau_{m}}(m_{\infty}(V)-m_{\rm NaP}),\\
     \frac{ds}{d\tau}&=& Q_tT_s((1-s)\bar{\alpha}_s-s\bar{\beta}_s),\\
     \frac{dm_{\rm K_{DR}}}{d\tau}&=& Q_t\tau_{\rm fast}T_{\rm K_{DR}}((1-m_{\rm K_{DR}})\bar{\alpha}_{m}-{m_{\rm K_{DR}}}\bar{\beta}_{m}),\\
     \frac{dh}{d\tau}&=& Q_t\tau_{\rm fast}T_h((1-h)\bar{\alpha}_h-h\bar{\beta}_h),\\
     \frac{d\rm Ca_i}{d\tau}&=&Q_tF_{\rm Ca} g_{\rm Ca}Q_v (- \bar{I}_{\rm Ca}(V)-\frac{\rm Ca_i}{\tau_{\rm Ca}g_{\rm Ca}Q_v})\\
     \frac{dq}{d\tau}&=& Q_tT_q((1-q)\bar{\alpha}_q( {\rm Ca_i})-q\bar{\beta}_q)
\end{array}    
\end{equation}
where $g_{\rm max}=125\,\rm nS$ is a conductance scale, $\bar{f}_1=\frac{f_1}{g_{\rm max}Q_t}$ with $f_1$ being the right-hand-side function of the $V$ equation in \eqref{eq:full}, $T_n=\max(1/\tau_n(V))$, $T_y=\max(\max\alpha_y,\max\beta_y)$, $\bar{\alpha}_y=\alpha_y/T_y$, and $\bar{\beta}_y=\beta_y/T_y$ for $y\in\{s, m_{\rm K_{DR}},h, q\}$. Numerical evaluations of these quantities over $V\in [-60,30]\,\rm mV$ and $\rm Ca_i\in (0,0.1)$ shows that $\frac{C}{g_{\rm max}}\approx0.02\,\rm ms$, $T_n\approx 0.0631\,\rm ms^{-1}$, $T_m=1/\tau_m\approx0.2\,\rm ms^{-1}$, $T_s=2.222\,\rm ms^{-1}$, $\tau_{\rm fast}T_{K_{\rm DR}}\approx3\,\rm ms^{-1}$, $\tau_{\rm fast}T_{h}\approx5\,\rm ms^{-1}$, $F_{\rm Ca}g_{\rm Ca}Q_v\approx120\,\rm ms^{-1}$, and $T_q\approx0.0042\,\rm ms^{-1}$.
% The $m$-current is activated at depolarized membrane potentials, with a voltage-dependent time constant of activation. 
% Note that the $I_{\rm K_{SS}}$ current exhibits a slow rise time $O(100)\,\rm ms$ and an even slower decay time $O(500)\,\rm ms$, so together the rate is $T_q=O(0.01)\,\rm ms^{-1}$. 

We choose the timescale for $n$ as our reference time, i.e., pick $Q_t = 1/T_n\approx 15.84\,\rm ms$. The above dimensionless system then becomes

\begin{equation}\label{eq:nondim}
\begin{array}{rcl}
     R_1\frac{dV}{d\tau}=\frac{C}{g_{\rm max}Q_t}\frac{dV}{d\tau}&=& \bar{f}_1\\
     R_2\frac{dm_{\rm NaP}}{d\tau} = \frac{\tau_m}{Q_t} \frac{dm_{\rm NaP}}{d\tau}&=& m_{\infty}(V)-m_{\rm NaP}:=\bar{f}_2,\\
     R_3\frac{ds}{d\tau}=\frac{1}{Q_tT_s}\frac{ds}{d\tau}&=& ((1-s)\bar{\alpha}_s-s\bar{\beta}_s):=\bar{f}_3,\\
     R_4\frac{dm_{\rm K_{DR}}}{d\tau}=\frac{1}{Q_t\tau_{\rm fast}T_{\rm K_{DR}}}\frac{dm_{\rm K_{DR}}}{d\tau}&=& ((1-m_{\rm K_{DR}})\bar{\alpha}_{m}-{m_{\rm K_{DR}}}\bar{\beta}_{m}):=\bar{f}_4,\\
     R_5\frac{dh}{d\tau}= \frac{1}{Q_t\tau_{\rm fast}T_h}\frac{dh}{d\tau}&=& ((1-h)\bar{\alpha}_h-h\bar{\beta}_h):=\bar{f}_5,\\
     R_6\frac{d\rm Ca_i}{d\tau}=\frac{1}{Q_tF_{\rm Ca} g_{\rm Ca}Q_v}\frac{d\rm Ca_i}{d\tau}&=& (- \bar{I}_{\rm Ca}(V)-\frac{\rm Ca_i}{\tau_{\rm Ca}g_{\rm Ca}Q_v}):=\bar{f}_6\\
     \frac{dn}{d\tau}&=& \frac{n_{\infty}(V)-n}{\bar{\tau}_n(V)}:=G(V,n),\\
     \frac{dq}{d\tau}&=& \delta ((1-q)\bar{\alpha}_q( {\rm Ca_i})-q\bar{\beta}_q):=\delta Q(\mathrm{Ca_i},q)
\end{array}    
\end{equation}

where $R_1=0.0014,R_2=0.3155, R_3=0.0284,R_4=0.0233,R_5=0.0118,R_6=5e-4$ and $\delta=Q_t T_q=0.0666$ are all unitless. Moreover, functions on the right-hand side ($f_i$, $G$ and $Q$) are $O(1)$ and hence the timescales are indicated by $R_i$ and $\delta$. Note that we do not rescale $V$ and $\mathrm {Ca_i}$ as they do not affect the timescales. While $m_{\rm Nap},s,m_{\rm K_{DR}}$, and $h$ are slower than $V$ and $\rm Ca_i$, it is clear that they are all relatively faster than $n$, which is 10 times faster than $q$. Hence we choose to consider $n$ as slow, to classify $q$ as superslow, and to group all the other variables as evolving on the fast timescale. For simplicity, we let $y\in \mathbb{R}^6$ denote all the fast variables $(V,m_{\rm Nap}, s, m_{\rm K_{DR}}, h, Ca_i)^{\mathrm{T}}$ and rewrite equation \eqref{eq:nondim} as 
\begin{equation}\label{eq:nondim-appen}
\begin{array}{rcl}
     \varepsilon\frac{dy}{d\tau}&=&F(y,n,q)\\
     \frac{dn}{d\tau}&=&G(V,n),\\
     \frac{dq}{d\tau}&=& \delta Q(\mathrm{Ca_i},q)
\end{array}    
\end{equation}
where $F=0$ is equivalent to all $\bar{f}_i$ in \eqref{eq:nondim} vanish. This is the system \eqref{eq:nondim2} in Section \ref{sec:model}. 

 \section{Geometric Structure near the Hopf Point in the $M^{-1}$-system}\label{app:geo}

\begin{figure}[!t]
\begin{center}
\includegraphics[width=0.5\linewidth]{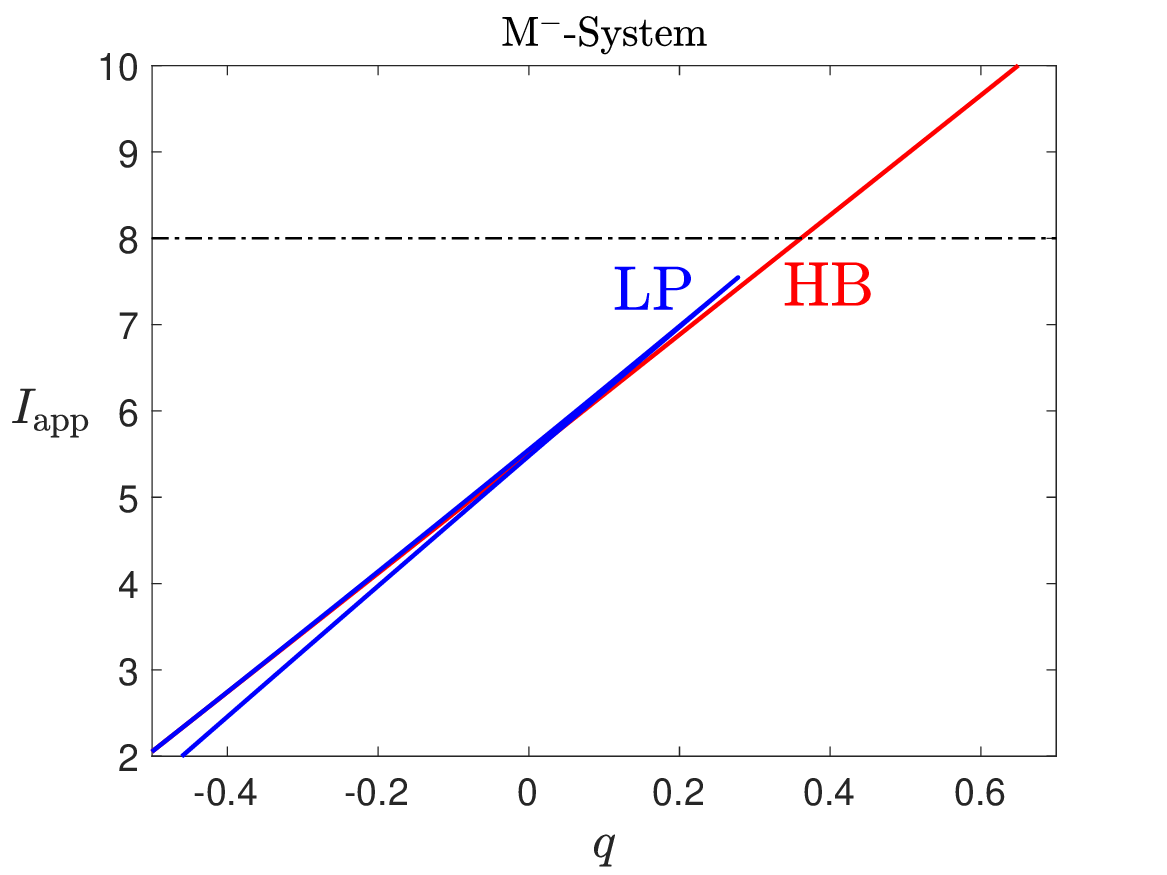}
\caption{Two-parameter bifurcations of the fast M$^-$-subsystem with respect to the superslow variable $q$ and the applied current $I_{\rm app}$. The Hopf bifurcation curve is shown in red and the curve of saddle-node bifurcations (LP) is shown in blue. Horizontal black dashed lines denote the default $I_{\rm app}$ value used in the $\rm M^-$-system.}
\label{fig:bif-q-Iapp}      
\end{center}
\end{figure}

To further characterize the local bifurcation structure underlying the absence of delay in the $M^{-1}$-system as discussed in Section \ref{sec:kca}, we examine the geometry of $M_{\rm ss}$ and its associated spectral properties near the Hopf point. Although no true fold occurs, the $M_{\rm ss}$ equilibrium branch is nearly folded: its slope becomes nearly vertical near the HB and an inflection arises without an actual fold point. This near-fold geometry suggests that the system lies close to the unfolding of a cusp bifurcation that generates saddle-node (fold) points. Indeed, by varying an additional parameter (e.g., $I_{\rm app}$), we identify a cusp bifurcation near the HB at which two fold bifurcation points (LP) are created (Figure~\ref{fig:bif-q-Iapp}). Similar unfoldings occur under variation of other parameters such as $g_m$ (data not shown). The $M^{-1}$-system parameter set lies just outside the cusp region, explaining why $M_{\rm ss}$ appears nearly folded but lacks a true fold. Thus, while no actual saddle-node bifurcation is present in the default parameter regime, the geometry of $M_{\rm ss}$ remains strongly influenced by a nearby cusp unfolding. 

%%%%%%%%%%%%%%%%%%%%%%%
% %%%%%%%%%%%%%%%%%%%%%%%\

%%%%%%%%%%%%%%%%%%%%%%%
\begin{figure}[!htp]
\begin{center}
\includegraphics[width=0.8\linewidth]{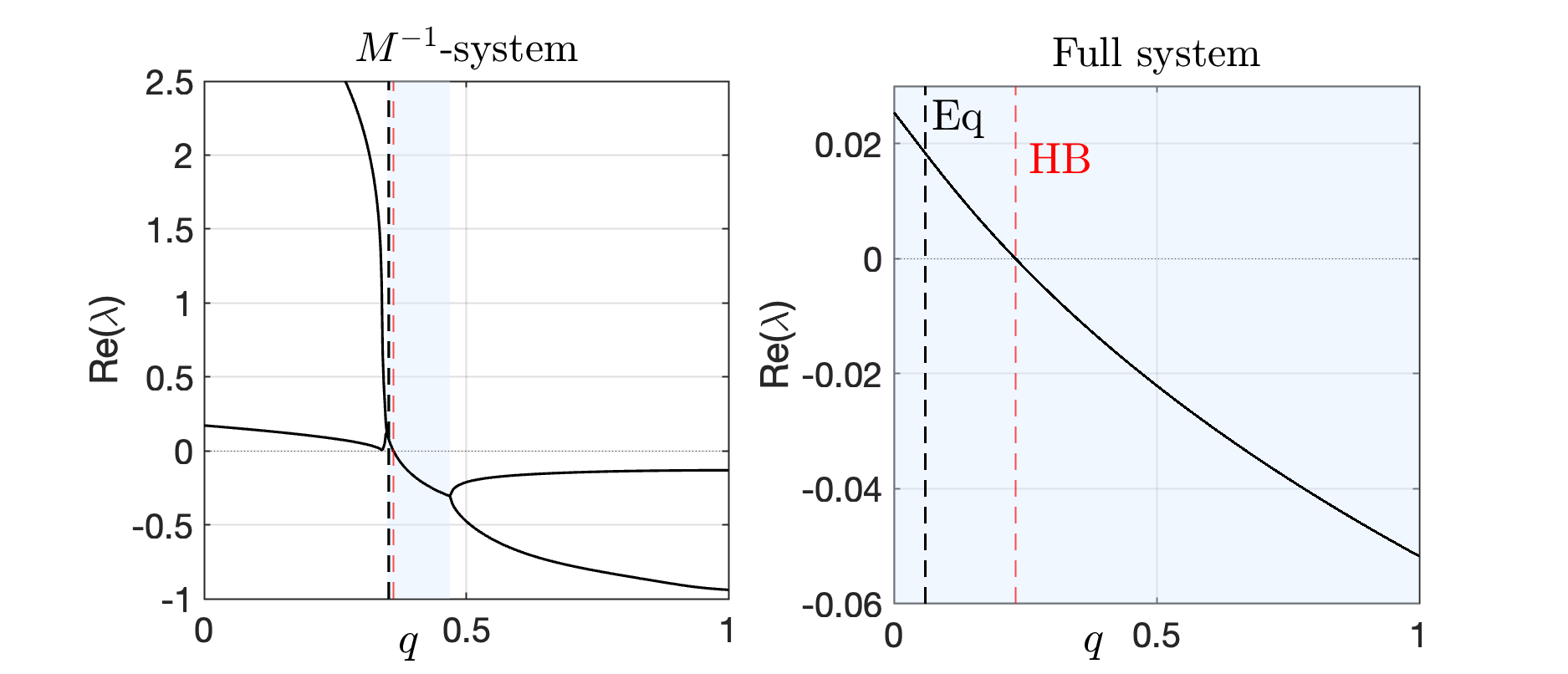}
%   \begin{tabular}{@{}p{0.45\linewidth}@{\quad}p{0.45\linewidth}@{}}
%  \subfigimg[width=\linewidth]{\bf{\small{(A)}}}{FIGS/eig_pair_3D_NoSTOs.png} &
%  \subfigimg[width=\linewidth]{\bf{\small{(B)}}}{FIGS/eig_pair_3D_allcurrents.png}
% \end{tabular}
\caption{Real part of the two eigenvalues of the fast-slow subsystem $(y, n)$ along the superslow manifold $\mss$, that switch between real and complex in $(q,\rm Re(\lambda))$-space, for (left) the $\rm M^-$-system and (right) the full model, with $I_{\rm app}=8$. Other model parameters are the same as in Figure \ref{fig:ST-response-KCa}. The eigenvalues along $\mss$ are real when there are two branches of $\rm Re(\lambda)$ and complex when there is a single branch. The HB point is indicated by the red vertical dashed line and the true equilibrium is indicated by the black vertical line. }
\label{fig:eig-bif-q-Iapp}   
\end{center}
\end{figure} 
% %%%%%%%%%%%%%%%%%%%%%%%

Proximity to a fold is known to modify delayed Hopf bifurcation behavior. Previous studies have shown that when a Hopf bifurcation occurs near an actual fold or a Bogdanov–Takens (BT) point, the classical delay can be significantly reduced or eliminated \cite{Vo2014,PW2024}. In the present case, although no true fold is present, the near-fold geometry may influence the spectral structure near the HB, thereby affecting the associated delay. Specifically, Figure \ref{fig:eig-bif-q-Iapp}A shows the real parts of the two critical eigenvalues $\lambda$ of the fast-slow subsystem evaluated along $\mss$ that transition between real and complex values, while the other eigenvalues stay real and negative for $q$ between $0$ and $1$. The eigenvalues are real when two distinct branches of Re $\lambda$ are present and become complex when these branches merge into a single curve. On the stable branch of $\mss$ ($q>q_{\rm HB}=0.3612$), the eigenvalues are initially real and negative, and become complex as $q$ approaches the HB (red dashed vertical line). Although the complex pair persists briefly past the HB, the region of complex eigenvalues is extremely narrow, and $\lambda$ quickly becomes real again, similar to the observation when a true fold is nearby \cite{Vo2014,PW2024}. In contrast, when the m-current is present, the two eigenvalues remain complex along $\mss$ for all $q\in [0, 1]$ (see Figure \ref{fig:eig-bif-q-Iapp}B). 

\bibliographystyle{plain}
\bibliography{bib-file}

\begin{thebibliography}{100}

\bibitem{achuthan2009phase}
Srisairam Achuthan and Carmen~C Canavier.
\newblock Phase-resetting curves determine synchronization, phase locking, and
  clustering in networks of neural oscillators.
\newblock {\em Journal of Neuroscience}, 29(16):5218--5233, 2009.

\bibitem{acker2003synchronization}
Corey~D Acker, Nancy Kopell, and John~A White.
\newblock Synchronization of strongly coupled excitatory neurons: relating
  network behavior to biophysics.
\newblock {\em Journal of computational neuroscience}, 15(1):71--90, 2003.

\bibitem{adams2019}
N.E. Adams, C.~Teige, G.~Mollo, T.~Karapanagiotidis, P.L. Cornelissen, and
  J.~Smallwood et~al.
\newblock Theta/deltacoupling across cortical laminae contributes tosemantic
  cognition.
\newblock {\em Journal of neurophysiology}, 121(4):1150--1161, 2019.

\bibitem{albantakis2024brain}
Larissa Albantakis, Christophe Bernard, Naama Brenner, Eve Marder, and
  Rishikesh Narayanan.
\newblock The brain's best kept secret is its degenerate structure.
\newblock {\em Journal of Neuroscience}, 44(40), 2024.

\bibitem{Awal2023}
N.M. Awal, I.R. Epstein, T.J. Kaper, and T.~Vo.
\newblock Symmetry-breaking rhythms in coupled, identical fast–slow
  oscillators.
\newblock {\em Chaos: An Interdisciplinary Journal of Nonlinear Science},
  33(1), 2023.

\bibitem{Baer1989}
S.~M. Baer, T.~Erneux, and J.~Rinzel.
\newblock The slow passage through a {H}opf bifurcation: {D}elay, memory
  effects, and resonance.
\newblock {\em SIAM J. Appl. Math.}, 49(1):55--71, 1989.

\bibitem{Baldemir2020}
H.~Baldemir, D.~Avitabile, and K.~Tsaneva-Atanasova.
\newblock Pseudo-plateau bursting and mixed-mode oscillations in a model of
  developing inner hair cells.
\newblock {\em Commun Nonlinear Sci Numer Simulat}, 80:104979, 2020.

\bibitem{Bat2021}
S.~Battaglin and M.~G. Pedersen.
\newblock Geometric analysis of mixed-mode oscillations in a model of
  electrical activity in human beta-cells.
\newblock {\em Nonlinear Dyn.}, 104(4):4445--4457, 2021.

\bibitem{berger1929elektroenkephalogramm}
Hans Berger.
\newblock {\"U}ber das elektroenkephalogramm des menschen.
\newblock {\em Archiv f{\"u}r psychiatrie und nervenkrankheiten},
  87(1):527--570, 1929.

\bibitem{Brons2006}
M.~Br{\o}ns, M.~Krupa, and M.~Wechselberger.
\newblock Mixed mode oscillations due to the generalized canard phenomenon.
\newblock {\em Fields Inst. Commun.}, 49:39--63, 2006.

\bibitem{buzsaki2006rhythms}
Gy{\"o}rgy Buzs{\'a}ki.
\newblock {\em Rhythms of the Brain}.
\newblock Oxford university press, 2006.

\bibitem{buzsaki2012origin}
Gy{\"o}rgy Buzs{\'a}ki, Costas~A Anastassiou, and Christof Koch.
\newblock The origin of extracellular fields and currents—eeg, ecog, lfp and
  spikes.
\newblock {\em Nature reviews neuroscience}, 13(6):407--420, 2012.

\bibitem{buzsaki2004neuronal}
Gyorgy Buzsaki and Andreas Draguhn.
\newblock Neuronal oscillations in cortical networks.
\newblock {\em science}, 304(5679):1926--1929, 2004.

\bibitem{canavier2010pulse}
Carmen~C Canavier and Srisairam Achuthan.
\newblock Pulse coupled oscillators and the phase resetting curve.
\newblock {\em Mathematical biosciences}, 226(2):77--96, 2010.

\bibitem{cannon2015leaky}
Jonathan Cannon and Nancy Kopell.
\newblock The leaky oscillator: Properties of inhibition-based rhythms revealed
  through the singular phase response curve.
\newblock {\em SIAM Journal on Applied Dynamical Systems}, 14(4):1930--1977,
  2015.

\bibitem{Carracedo2013}
L.~M. Carracedo, H.~Kjeldsen, L.~Cunnington, A.~Jenkins, I.~Schofield, and
  et~al. M.O.~Cunningham, M.~O.
\newblock A neocortical delta rhythm facilitates reciprocal interlaminar
  interactions via nested theta rhythms.
\newblock {\em Journal of Neuroscience}, 33(26):10750--10761, 2013.

\bibitem{castejon2013phase}
Oriol Castej{\'o}n, Antoni Guillamon, and Gemma Huguet.
\newblock Phase-amplitude response functions for transient-state stimuli.
\newblock {\em The Journal of Mathematical Neuroscience}, 3(1):13, 2013.

\bibitem{chandrasekaran2009natural}
Chandramouli Chandrasekaran, Andrea Trubanova, S{\'e}bastien Stillittano, Alice
  Caplier, and Asif~A Ghazanfar.
\newblock The natural statistics of audiovisual speech.
\newblock {\em PLoS computational biology}, 5(7):e1000436, 2009.

\bibitem{Chumakov2015}
G.~A. Chumakov, N.~A. Chumakova, and E.~A. Lashina.
\newblock Modeling the complex dynamics of heterogeneous catalytic reactions
  with fast, intermediate, and slow variables.
\newblock {\em Chem. Eng. J.}, 282:11--19, 2015.

\bibitem{cui2009functional}
Jianxia Cui, Carmen~C Canavier, and Robert~J Butera.
\newblock Functional phase response curves: a method for understanding
  synchronization of adapting neurons.
\newblock {\em Journal of Neurophysiology}, 102(1):387--398, 2009.

\bibitem{Curtu2010}
R.~Curtu.
\newblock Singular {H}opf bifurcation and mixed-mode oscillations in a two-cell
  inhibitory neural network.
\newblock {\em Phys. D: Nonlinear Phenom.}, 239(9):504--514, 2010.

\bibitem{CR2011}
R.~Curtu and J.~Rubin.
\newblock Interaction of canard and singular {H}opf mechanisms in a neural
  model.
\newblock {\em SIAM J. Appl. Dyn. Syst.}, 10(4):1443--1479, 2011.

\bibitem{Maess2014}
P.~{D}e Maesschalck, E.~Kutafina, and N.~Popovi{\'c}.
\newblock Three time-scales in an extended {B}onhoeffer–van der {P}ol
  oscillator.
\newblock {\em J. Dyn. Differ. Equ.}, 26:955--987, 2014.

\bibitem{Maess2016}
P.~{D}e Maesschalck, E.~Kutafina, and N.~Popovi{\'c}.
\newblock Sector-delayed-{H}opf-type mixed-mode oscillations in a prototypical
  three-time-scale model.
\newblock {\em Appl. Math. Comput.}, 273:337--352, 2016.

\bibitem{Desroches2012}
M.~Desroches, J.~Guckenheimer, B.~Krauskopf, C.~Kuehn, H.~M. Osinga, and
  M.~Wechselberger.
\newblock Mixed-mode oscillations with multiple time scales.
\newblock {\em SIAM Rev.}, 54(2):211--288, 2012.

\bibitem{DK2018}
M.~Desroches and V.~Kirk.
\newblock Spike-adding in a canonical three-time-scale model: superslow
  explosion and folded-saddle canards.
\newblock {\em SIAM J. Appl. Dyn. Syst.}, 17(3):1989--2017, 2018.

\bibitem{ding2017temporal}
Nai Ding, Aniruddh~D Patel, Lin Chen, Henry Butler, Cheng Luo, and David
  Poeppel.
\newblock Temporal modulations in speech and music.
\newblock {\em Neuroscience \& Biobehavioral Reviews}, 81:181--187, 2017.

\bibitem{drover04}
J.~Drover, J.~Rubin, J.~Su, and B.~Erentrout.
\newblock Analysis of a canard mechanism by which excitatory synaptic coupling
  can synchronize neurons at low firing frequencies.
\newblock {\em SIAM J. APPL. MATH.}, 65(1):69--92, 2004.

\bibitem{edelman2001degeneracy}
Gerald~M Edelman and Joseph~A Gally.
\newblock Degeneracy and complexity in biological systems.
\newblock {\em Proceedings of the national academy of sciences},
  98(24):13763--13768, 2001.

\bibitem{engler2025delays}
Hans Engler, Hans Kaper, Tasso Kaper, and Theodore Vo.
\newblock Delays and advances in the onset of instability in the shishkova
  equation.
\newblock {\em Quarterly of Applied Mathematics}, 2025.

\bibitem{ermentrout96}
B.~Ermentrout.
\newblock Type i membranes, phase resetting curves, and synchrony.
\newblock {\em Neural Comput.}, 8:979--1001, 1996.

\bibitem{ermentrout1996type}
Bard Ermentrout.
\newblock Type i membranes, phase resetting curves, and synchrony.
\newblock {\em Neural computation}, 8(5):979--1001, 1996.

\bibitem{ermentrout2001effects}
Bard Ermentrout, Matthew Pascal, and Boris Gutkin.
\newblock The effects of spike frequency adaptation and negative feedback on
  the synchronization of neural oscillators.
\newblock {\em Neural computation}, 13(6):1285--1310, 2001.

\bibitem{ermentrout2010mathematical}
Bard Ermentrout and David~Hillel Terman.
\newblock {\em Mathematical foundations of neuroscience}, volume~35.
\newblock Springer, 2010.

\bibitem{ermentrout1981n}
G~Bard Ermentrout.
\newblock n: m phase-locking of weakly coupled oscillators.
\newblock {\em Journal of Mathematical Biology}, 12(3):327--342, 1981.

\bibitem{ermentrout1991multiple}
G~Bard Ermentrout and Nancy Kopell.
\newblock Multiple pulse interactions and averaging in systems of coupled
  neural oscillators.
\newblock {\em Journal of Mathematical Biology}, 29(3):195--217, 1991.

\bibitem{Fenichel1979}
N.~Fenichel.
\newblock Geometric singular perturbation theory for ordinary differential
  equations.
\newblock {\em J. Differ. Equ.}, 31(1):53--98, 1979.

\bibitem{Ghitza2014}
O.~Ghitza.
\newblock Behavioral evidence for the role of cortical theta oscillations in
  determining auditory channel capacity for speech.
\newblock {\em Frontiers in psychology}, 5:652, 2014.

\bibitem{Ghitza2009}
O.~Ghitza and S.~Greenberg.
\newblock On the possible role of brain rhythms in speech perception:
  intelligibility of time-compressed speech with periodic and aperiodic
  insertions of silence.
\newblock {\em Phonetica}, 66(1-2):113--126, 2009.

\bibitem{ghitza2011linking}
Oded Ghitza.
\newblock Linking speech perception and neurophysiology: speech decoding guided
  by cascaded oscillators locked to the input rhythm.
\newblock {\em Frontiers in psychology}, 2:130, 2011.

\bibitem{ghitza2012role}
Oded Ghitza.
\newblock On the role of theta-driven syllabic parsing in decoding speech:
  intelligibility of speech with a manipulated modulation spectrum.
\newblock {\em Frontiers in psychology}, 3:238, 2012.

\bibitem{giilmenbrain}
Gorsev Giilmen~Yenera’c’d’e.
\newblock Brain’s alpha, beta, gamma, delta, and theta oscillationsin
  neuropsychiatric diseases: proposal for biomarker strategies.

\bibitem{goaillard2021ion}
Jean-Marc Goaillard and Eve Marder.
\newblock Ion channel degeneracy, variability, and covariation in neuron and
  circuit resilience.
\newblock {\em Annual review of neuroscience}, 44(1):335--357, 2021.

\bibitem{greenberg1999speaking}
Steven Greenberg.
\newblock Speaking in shorthand--a syllable-centric perspective for
  understanding pronunciation variation.
\newblock {\em Speech Communication}, 29(2-4):159--176, 1999.

\bibitem{GW2000}
J.~Guckenheimer and A.~R. Willms.
\newblock Asymptotic analysis of subcritical {H}opf-homoclinic bifurcation.
\newblock {\em Phys. D: Nonlinear Phenom.}, 139(3-4):195--216, 2000.

\bibitem{gutfreund1995}
Y.~Gutfreund, Y.~Yarom, and I.~Segev.
\newblock Subthreshold oscillations and resonant frequency in guinea‐pig
  cortical neurons: physiology and modelling.
\newblock {\em The Journal of physiology}, 483(3):621--640, 1985.

\bibitem{hansel95}
D.~Hansel, G.~Mato, and C.~Meunier.
\newblock Synchrony in excitatory neural networks.
\newblock {\em Neural Comput.}, 7:307--337, 1995.

\bibitem{Harvey2011}
E.~Harvey, V.~Kirk, M.~Wechselberger, and J.~Sneyd.
\newblock Multiple timescales, mixed mode oscillations and canards in models of
  intracellular calcium dynamics.
\newblock {\em J. Nonlinear Sci.}, 21:639--683, 2011.

\bibitem{Hayes2016}
M.~G. Hayes, T.~J. Kaper, P.~Szmolyan, and M.~Wechselberger.
\newblock Geometric desingularization of degenerate singularities in the
  presence of fast rotation: {A} new proof of known results for slow passage
  through {H}opf bifurcations.
\newblock {\em Indag. Math.}, 27(5):1184--1203, 2016.

\bibitem{hovsepyan2020combining}
Sevada Hovsepyan, Itsaso Olasagasti, and Anne-Lise Giraud.
\newblock Combining predictive coding and neural oscillations enables online
  syllable recognition in natural speech.
\newblock {\em Nature communications}, 11(1):3117, 2020.

\bibitem{Hudson1979}
J.~L. Hudson, M.~Hart, and D.~Marinko.
\newblock An experimental study of multiple peak periodic and nonperiodic
  oscillations in the {B}elousov--{Z}habotinskii reaction.
\newblock {\em J. Chem. Phys.}, 71(4):1601--1606, 1979.

\bibitem{Hyafil2015}
A.~Hyafil, L.~Fontolan, C.~Kabdebon, B.~Gutkin, and A.L. Giraud.
\newblock Speech encoding by coupled cortical theta and gamma oscillations.
\newblock {\em Elife}, 4, 2015.

\bibitem{hyafil2015speech}
Alexandre Hyafil, Lorenzo Fontolan, Claire Kabdebon, Boris Gutkin, and
  Anne-Lise Giraud.
\newblock Speech encoding by coupled cortical theta and gamma oscillations.
\newblock {\em elife}, 4:e06213, 2015.

\bibitem{Jalics2010}
J.~Jalics, M.~Krupa, and H.~G. Rotstein.
\newblock Mixed-mode oscillations in a three time-scale system of {ODE}s
  motivated by a neuronal model.
\newblock {\em Dyn. Syst.}, 25(4):445--482, 2010.

\bibitem{Kak2023a}
P.~Kaklamanos and N.~Popovi{\'c}.
\newblock Complex oscillatory dynamics in a three-timescale {E}l {N}i{\~n}o
  {S}outhern {O}scillation model.
\newblock {\em Phys. D: Nonlinear Phenom.}, 449:133740, 2023.

\bibitem{Kak2022}
P.~Kaklamanos, N.~Popovi{\'c}, and K.~U. Kristiansen.
\newblock Bifurcations of mixed-mode oscillations in three-timescale systems:
  {A}n extended prototypical example.
\newblock {\em Chaos: An Interdisciplinary Journal of Nonlinear Science},
  32(1):013108, 2022.

\bibitem{Kak2023b}
P.~Kaklamanos, N.~Popovi{\'c}, and K.~U. Kristiansen.
\newblock Geometric singular perturbation analysis of the multiple-timescale
  {H}odgkin-{H}uxley equations.
\newblock {\em SIAM J. Appl. Dyn. Syst.}, 22(3):1552--1589, 2023.

\bibitem{Kimrey2020}
J.~Kimrey, T.~Vo, and R.~Bertram.
\newblock Big ducks in the heart: canard analysis can explain large early
  afterdepolarizations in cardiomyocytes.
\newblock {\em SIAM J. Appl. Dyn. Syst.}, 19(3):1701--1735, 2020.

\bibitem{2ndKimrey2020}
J.~Kimrey, T.~Vo, and R.~Bertram.
\newblock Canard analysis reveals why a large $\rm {C}a^{2+}$ window current
  promotes early afterdepolarizations in cardiac myocytes.
\newblock {\em PLoS Comput. Biol.}, 16(11):e1008341, 2020.

\bibitem{klinshov2017phase}
Vladimir Klinshov, Serhiy Yanchuk, Artur Stephan, and Vladimir Nekorkin.
\newblock Phase response function for oscillators with strong forcing or
  coupling.
\newblock {\em Europhysics Letters}, 118(5):50006, 2017.

\bibitem{kopell2002mechanisms}
Nancy Kopell and G~Bard Ermentrout.
\newblock Mechanisms of phase-locking and frequency control in pairs of coupled
  neural oscillators.
\newblock {\em Handbook of dynamical systems}, 2:3--54, 2002.

\bibitem{Krupa2008b}
M.~Krupa, N.~Popovi{\'c}, and N.~Kopell.
\newblock Mixed-mode oscillations in three time-scale systems: a prototypical
  example.
\newblock {\em SIAM J. Appl. Dyn. Syst.}, 7(2):361--420, 2008.

\bibitem{Krupa2008a}
M.~Krupa, N.~Popovi{\'c}, N.~Kopell, and H.~G. Rotstein.
\newblock Mixed-mode oscillations in a three time-scale model for the
  dopaminergic neuron.
\newblock {\em Chaos: An Interdisciplinary Journal of Nonlinear Science},
  18(1):015106, 2008.

\bibitem{Krupa2012}
M.~Krupa, A.~Vidal, M.~Desroches, and F.~Cl{\'e}ment.
\newblock Mixed-mode oscillations in a multiple time scale phantom bursting
  system.
\newblock {\em SIAM J. Appl. Dyn. Syst.}, 11(4):1458--1498, 2012.

\bibitem{KW2010}
M.~Krupa and M.~Wechselberger.
\newblock Local analysis near a folded saddle-node singularity.
\newblock {\em J. Differ. Equ.}, 248(12):2841--2888, 2010.

\bibitem{Kugler2018}
P.~K{\"u}gler, A.~H. Erhardt, and M.~A.~K. Bulelzai.
\newblock Early afterdepolarizations in cardiac action potentials as mixed mode
  oscillations due to a folded node singularity.
\newblock {\em PLoS One}, 13(12):e0209498, 2018.

\bibitem{lakatos2005oscillatory}
Peter Lakatos, Ankoor~S Shah, Kevin~H Knuth, Istvan Ulbert, George Karmos, and
  Charles~E Schroeder.
\newblock An oscillatory hierarchy controlling neuronal excitability and
  stimulus processing in the auditory cortex.
\newblock {\em Journal of neurophysiology}, 94(3):1904--1911, 2005.

\bibitem{Letson2017}
B.~Letson, J.~E. Rubin, and T.~Vo.
\newblock Analysis of interacting local oscillation mechanisms in
  three-timescale systems.
\newblock {\em SIAM J. Appl. Dyn. Syst.}, 77(3):1020--1046, 2017.

\bibitem{malerba2013phase}
Paola Malerba and Nancy Kopell.
\newblock Phase resetting reduces theta--gamma rhythmic interaction to a
  one-dimensional map.
\newblock {\em Journal of mathematical biology}, 66(7):1361--1386, 2013.

\bibitem{mirollo1990synchronization}
Renato~E Mirollo and Steven~H Strogatz.
\newblock Synchronization of pulse-coupled biological oscillators.
\newblock {\em SIAM Journal on Applied Mathematics}, 50(6):1645--1662, 1990.

\bibitem{ML1981}
C.~Morris and H.~Lecar.
\newblock Voltage oscillations in the barnacle giant muscle fiber.
\newblock {\em Biophys. J.}, 35(1):193--213, 1981.

\bibitem{Nan2015}
P.~Nan, Y.~Wang, V.~Kirk, and J.~E. Rubin.
\newblock Understanding and distinguishing three-time-scale oscillations:
  {C}ase study in a coupled {M}orris-{L}ecar system.
\newblock {\em SIAM J. Appl. Dyn. Syst.}, 14(3):1518--1557, 2015.

\bibitem{Neishtadt1987}
A.~Neishtadt.
\newblock On delayed stability loss under dynamical bifurcations {I}.
\newblock {\em Differ. Equ.}, 23:1385--1390, 1987.

\bibitem{Neishtadt1988}
A.~Neishtadt.
\newblock On delayed stability loss under dynamical bifurcations {II}.
\newblock {\em Differ. Equ.}, 24:171--176, 1988.

\bibitem{ohala1975temporal}
John~J Ohala.
\newblock The temporal regulation of speech.
\newblock {\em Auditory analysis and perception of speech}, pages 431--453,
  1975.

\bibitem{park2017utility}
Youngmin Park, Stewart Heitmann, and G~Bard Ermentrout.
\newblock The utility of phase models in studying neural synchronization.
\newblock {\em Computational models of brain and behavior}, pages 493--504,
  2017.

\bibitem{Pavlidis2022}
E.~Pavlidis, F.~Campillo, A.~Goldbeter, and M.~Desroches.
\newblock Multiple-timescale dynamics, mixed mode oscillations and mixed
  affective states in a model of bipolar disorder.
\newblock {\em Cognitive Neurodynamics}, 2022.

\bibitem{perez2020global}
Alberto P{\'e}rez-Cervera, Gemma Huguet, et~al.
\newblock Global phase-amplitude description of oscillatory dynamics via the
  parameterization method.
\newblock {\em Chaos: an interdisciplinary journal of nonlinear science},
  30(8), 2020.

\bibitem{perez2020phase}
Alberto P{\'e}rez-Cervera, Tere~M Seara, and Gemma Huguet.
\newblock Phase-locked states in oscillating neural networks and their role in
  neural communication.
\newblock {\em Communications in Nonlinear Science and Numerical Simulation},
  80:104992, 2020.

\bibitem{PW2014}
C.~Perryman and S.~Wieczorek.
\newblock Adapting to a changing environment: non-obvious thresholds in
  multi-scale systems.
\newblock {\em Proc. Math. Phys. Eng. Sci.}, 470(2170):20140226, 2014.

\bibitem{PW2024}
N.A. Phan and Y.~Wang.
\newblock Mixed-mode oscillations in a three-timescale coupled morris–lecar
  system.
\newblock {\em Chaos: An Interdisciplinary Journal of Nonlinear Science},
  34(5), 2024.

\bibitem{pittman21}
B.R. Pittman-Polletta, Y.~Wang, D.A. Stanley, C.E. Schroeder, M.A. Whittington,
  and N.J. Kopell.
\newblock Differential contributions of synaptic and intrinsic inhibitory
  currents to speech segmentation via flexible phase-locking in neural
  oscillators.
\newblock {\em PLoS computational biology}, 17(4):e1008783, 2021.

\bibitem{prinz2004similar}
Astrid~A Prinz, Dirk Bucher, and Eve Marder.
\newblock Similar network activity from disparate circuit parameters.
\newblock {\em Nature neuroscience}, 7(12):1345--1352, 2004.

\bibitem{reyner2022phase}
David Reyner-Parra and Gemma Huguet.
\newblock Phase-locking patterns underlying effective communication in exact
  firing rate models of neural networks.
\newblock {\em PLoS computational biology}, 18(5):e1009342, 2022.

\bibitem{RE1998}
J.~Rinzel and G.~B. Ermentrout.
\newblock Analysis of neural excitability and oscillations.
\newblock {\em Methods in neuronal modeling}, 2:251--292, 1998.

\bibitem{rinzel1987formal}
John Rinzel.
\newblock A formal classification of bursting mechanisms in excitable systems.
\newblock In {\em Mathematical Topics in Population Biology, Morphogenesis and
  Neurosciences: Proceedings of an International Symposium held in Kyoto,
  November 10--15, 1985}, pages 267--281. Springer, 1987.

\bibitem{rinzel1998analysis}
John Rinzel and G~Bard Ermentrout.
\newblock Analysis of neural excitability and oscillations.
\newblock {\em Methods in neuronal modeling}, 2:251--292, 1998.

\bibitem{Roberts2015}
K.~L. Roberts, J.~E. Rubin, and M.~Wechselberger.
\newblock Averaging, folded singularities, and torus canards: {E}xplaining
  transitions between bursting and spiking in a coupled neuron model.
\newblock {\em SIAM J. Appl. Dyn. Syst.}, 14(4):1808--1844, 2015.

\bibitem{Rostein2005}
H.~G. Rotstein, D.~D. Pervouchine, C.~D. Acker, M.J. Gillies, J.A. White, and
  E.H.~Buhl et~al.
\newblock Slow and fast inhibition and an h-current interact to create a theta
  rhythm in a model of ca1 interneuron network.
\newblock {\em Journal of neurophysiology}, 94(2):1509--1518, 2005.

\bibitem{rotstein2014frequency}
Horacio~G Rotstein and Farzan Nadim.
\newblock Frequency preference in two-dimensional neural models: a linear
  analysis of the interaction between resonant and amplifying currents.
\newblock {\em Journal of computational neuroscience}, 37(1):9--28, 2014.

\bibitem{RW2008}
J.E. Rubin and M.~Wechselberger.
\newblock The selection of mixed-mode oscillations in a hodgkin-huxley model
  with multiple timescales.
\newblock {\em Chaos: An Interdisciplinary Journal of Nonlinear Science},
  18(1):015105, 2008.

\bibitem{Sadhu2022}
S.~Sadhu.
\newblock Complex oscillatory patterns in a three-timescale model of a
  generalist predator and a specialist predator competing for a common prey.
\newblock {\em Discrete and Continuous Dynamical Systems-B}, 28(5):3014--3051,
  2023.

\bibitem{schwemmer2012theory}
Michael~A Schwemmer and Timothy~J Lewis.
\newblock The theory of weakly coupled oscillators.
\newblock {\em Phase response curves in neuroscience: theory, experiment, and
  analysis}, pages 3--31, 2012.

\bibitem{sherfey2018flexible}
Jason~S Sherfey, Salva Ardid, Joachim Hass, Michael~E Hasselmo, and Nancy~J
  Kopell.
\newblock Flexible resonance in prefrontal networks with strong feedback
  inhibition.
\newblock {\em PLoS computational biology}, 14(8):e1006357, 2018.

\bibitem{SW2001}
P.~Szmolyan and M.~Wechselberger.
\newblock Canards in $\mathbb{R}^3$.
\newblock {\em J. Differ. Equ.}, 177(2):419--453, 2001.

\bibitem{Teka2012}
W.~Teka, J.~Tabak, and R.~Bertram.
\newblock The relationship between two fast/slow analysis techniques for
  bursting oscillations.
\newblock {\em Chaos: An Interdisciplinary Journal of Nonlinear Science},
  22(4):043117, 2012.

\bibitem{Vo2010}
T.~Vo, R.~Bertram, J.~Tabak, and M.~Wechselberger.
\newblock Mixed mode oscillations as a mechanism for pseudo-plateau bursting.
\newblock {\em J. Comput. Neurosci.}, 28:443--458, 2010.

\bibitem{Vo2013}
T.~Vo, R.~Bertram, and M.~Wechselberger.
\newblock Multiple geometric viewpoints of mixed mode dynamics associated with
  pseudo-plateau bursting.
\newblock {\em SIAM J. Appl. Dyn. Syst.}, 12(2):789--830, 2013.

\bibitem{Vo2014}
T.~Vo, J.~Tabak, R.~Bertram, and M.~Wechselberger.
\newblock A geometric understanding of how fast activating potassium channels
  promote bursting in pituitary cells.
\newblock {\em J. Comput. Neurosci.}, 36:259--278, 2014.

\bibitem{VW2015}
T.~Vo and M.~Wechselberger.
\newblock Canards of folded saddle-node type {I}.
\newblock {\em SIAM J. Math. Anal.}, 47(4):3235--3283, 2015.

\bibitem{WR2016}
Y.~Wang and J.~E. Rubin.
\newblock Multiple timescale mixed bursting dynamics in a respiratory neuron
  model.
\newblock {\em J. Comput. Neurosci.}, 41:245--268, 2016.

\bibitem{WR2017}
Y.~Wang and J.~E. Rubin.
\newblock Timescales and mechanisms of sigh-like bursting and spiking in models
  of rhythmic respiratory neurons.
\newblock {\em J. Math. Neurosci.}, 7:1--39, 2017.

\bibitem{WR2020}
Y.~Wang and J.~E. Rubin.
\newblock Complex bursting dynamics in an embryonic respiratory neuron model.
\newblock {\em Chaos: An Interdisciplinary Journal of Nonlinear Science},
  30(4):043127, 2020.

\bibitem{wang2021shape}
Yangyang Wang, Jeffrey~P Gill, Hillel~J Chiel, and Peter~J Thomas.
\newblock Shape versus timing: linear responses of a limit cycle with hard
  boundaries under instantaneous and static perturbation.
\newblock {\em SIAM journal on applied dynamical systems}, 20(2):701--744,
  2021.

\bibitem{Wechselberger2005}
M.~Wechselberger.
\newblock Existence and bifurcation of canards in $\mathbb{R}^3$ in the case of
  a folded node.
\newblock {\em SIAM J. Appl. Dyn. Syst.}, 4(1):101--139, 2005.

\bibitem{wedgwood2013phase}
Kyle~CA Wedgwood, Kevin~K Lin, Ruediger Thul, and Stephen Coombes.
\newblock Phase-amplitude descriptions of neural oscillator models.
\newblock {\em The Journal of Mathematical Neuroscience}, 3(1):2, 2013.

\bibitem{wilson2020phase}
Dan Wilson.
\newblock Phase-amplitude reduction far beyond the weakly perturbed paradigm.
\newblock {\em Physical Review E}, 101(2):022220, 2020.

\bibitem{wilson2025phase}
Dan Wilson.
\newblock Phase-based reduced order models for parabolic and elliptic bursting
  neurons.
\newblock {\em SIAM Journal on Applied Dynamical Systems}, 24(1):187--218,
  2025.

\bibitem{wilson2019augmented}
Dan Wilson and Bard Ermentrout.
\newblock Augmented phase reduction of (not so) weakly perturbed coupled
  oscillators.
\newblock {\em SIAM Review}, 61(2):277--315, 2019.

\bibitem{wilson2016isostable}
Dan Wilson and Jeff Moehlis.
\newblock Isostable reduction of periodic orbits.
\newblock {\em Physical Review E}, 94(5):052213, 2016.

\bibitem{Yu2008}
N.~Yu, R.~Kuske, and Y.X. Li.
\newblock Stochastic phase dynamics and noise-induced mixed-mode oscillations
  in coupled oscillators.
\newblock {\em Chaos: An Interdisciplinary Journal of Nonlinear Science},
  18(1), 2008.

\bibitem{zhou2018m}
Yujia Zhou, Theodore Vo, Horacio~G Rotstein, Michelle~M McCarthy, and Nancy
  Kopell.
\newblock M-current expands the range of gamma frequency inputs to which a
  neuronal target entrains.
\newblock {\em The Journal of Mathematical Neuroscience}, 8(1):13, 2018.

\end{thebibliography}
\nocite{*}

\end{document}